\documentclass[amsfonts,amsmath,amssymb,aps,a4paper,british,dvips,floatfix,showkeys,showpacs,twocolumn,twoside]{revtex4}
\usepackage{babel}
\usepackage{graphicx}
\usepackage{hyperref,hypernat}

\newcommand{\abinitio}{\emph{ab initio\/}}%
\newcommand{\cm}{\ensuremath{\text{cm}^{-1}}}%
\newcommand{\degree}{\ensuremath{{^\circ}}}%
\newcommand{\eg}{e.\,g.}%
\newcommand{\etal}{\emph{et\,al.\/}}%
\newcommand{\ie}{i.\,e.}%

\newlength{\floatwidth} 

\makeatletter
\renewcommand{\@fnsymbol}[1]{%
   \ifcase#1\or\textasteriskcentered\or\textsection\or\textdagger
   \or\textdaggerdbl\or\textparagraph\or\textbardbl
   \or\textasteriskcentered\textasteriskcentered
   \or\textdagger\textdagger\or\textdaggerdbl\textdaggerdbl
   \else\@ctrerr\fi}
\makeatother

\begin{document}%

\title{Spectroscopy of free radicals and radical containing entrance-channel
   complexes in superfluid helium nano-droplets}%

\author{Jochen K\"upper}%
\email[Author to whom correspondence should be addressed; electronic address:
]{jochen@fhi-berlin.mpg.de}%
\affiliation{Fritz-Haber-Institut der MPG, Faradayweg 4--6, 14195 Berlin,
   Germany}%
\author{Jeremy M.\ Merritt}%
\email{merritjm@unc.edu}%
\affiliation{University of North Carolina, Chapel Hill, NC 27599, USA}%
\affiliation{Fritz-Haber-Institut der MPG, Faradayweg 4--6, 14195 Berlin,
   Germany}%

\author{\bigskip}%
\affiliation{\emph{Dedicated to Roger E.\ Miller}}%
\thanks{Deceased 6.~November~2005}


\begin{abstract}%
   \noindent%
   The spectroscopy of free radicals and radical containing entrance-channel
   complexes embedded in superfluid helium nano-droplets is reviewed. The
   collection of dopants inside individual droplets in the beam represents a
   micro-canonical ensemble, and as such each droplet may be considered an
   isolated cryo-reactor. The unique properties of the droplets, namely their
   low temperature (0.4~K) and fast cooling rates ($\sim10^{16}$~K\,s$^{-1}$)
   provides novel opportunities for the formation and high-resolution studies of
   molecular complexes containing one or more free radicals. The production
   methods of radicals are discussed in light of their applicability for
   embedding the radicals in helium droplets. The spectroscopic studies
   performed to date on molecular radicals and on entrance\,/\,exit-channel
   complexes of radicals with stable molecules are detailed. The observed
   complexes provide new information on the potential energy surfaces of several
   fundamental chemical reactions and on the intermolecular interactions present
   in open-shell systems. Prospects of further experiments of radicals embedded
   in helium droplets are discussed, especially the possibilities to prepare and
   study high-energy structures and their controlled manipulation, as well as
   the possibility of fundamental physics experiments.
\end{abstract}%

\date{\today}

\maketitle

\setlength{\floatwidth}{\linewidth}

\begin{widetext}
   \clearpage
   \tableofcontents
   \clearpage
\end{widetext}

\section{Introduction}
\label{sec:introduction}

Doped superfluid helium nano-droplets have emerged as a new and exciting tool
for the study of the structure and dynamics of a quantum solvent as well as the
embedded or attached atomic and molecular impurities themselves. The unique
properties of the droplets, namely their low temperature and rapid cooling, make
them a versatile tool for spectroscopic studies of metastable species. Although
there have been a number of reviews on the spectroscopy of particles embedded in
or attached to helium nano-droplets~\cite{Choi:IRPC25:15, Stienkemeier:JPB:R127,
   Toennies:ACIE43:2622, Makarov:PhysUspekhi47:217, Northby:JCP115:10065,
   Callegari:JCP115:10090, Stienkemeier:JCP115:10119}, none of these articles
covers the recent advances made in the study of open shell, free radical
metastable species. Therefore, an overview of this emerging field and the new
possibilities it provides for the understanding of chemical reaction dynamics is
timely.

Radicals and molecular ions are among the most chemically reactive species known
and play a key role in chemistry ranging from combustion processes
\cite{Warnatz:Combustion} to the upper atmosphere~\cite{Wayne:ChemAtmosphere}
and molecular synthesis in the interstellar medium~\cite{Chastaing:PCCP1:2247}.
This high reactivity comes at a considerable cost to experimentalists; such
transient species are often very difficult to maintain at sufficient
concentrations to be probed experimentally. While a great deal of progress has
been made by isolating radicals in vacuum, much less is known about the
interactions of radicals, or more generally, the features of their potential
energy surfaces (PES). A focal point in chemical reaction dynamics has been to
elucidate the properties of the transition state, the point on the PES where
bonds are broken and reformed. Already in 1884 van't Hoff proposed
\cite{Hoff:DynamiqueChimique214}, and five years later Arrhenius provided an
physical interpretation~\cite{Arrhenius:ZPC4:226}, for an empirical, analytical
expression for the rate constant of a reaction, relating it to an energy penalty
needed to produce an \emph{activated complex}, which could then go on to produce
the products. A quantum-mechanical interpretation of the activation process was
first given by London~\cite{London:ZEAPC35:552} and successively by Eyring and
Polyani~\cite{Eyring:ZPCB12:279}, who introduced the concepts of a reaction path
and the transition state as a saddle-point on a multi-dimensional PES
(\emph{Sattelgebiet des \glqq{}Resonanzgebirges\grqq}) for the H+H$_2$, H+HBr,
and H+Br$_2$ reaction systems.\footnote{It is interesting to note that these
   studies --- on systems which are quite similar to the X-HY systems described
   in section~\ref{sec:complex:atomic} -- were performed at the
   Kaiser~Wilhelm-Institut in Berlin-Dahlem: the predecessor of the
   Fritz~Haber-Institut were this review was written.} Later, the efficiency of
a reaction could be predicted based on the initial reagent energy (whether it be
translational or vibrational) by developing the concept of early or late
potential barriers~\cite{Hammond:JACS77:334, Polanyi:JCP51:1439}. Recent
advances in reactive scattering have now enabled fully (initial and final)
quantum state resolved reaction crosssections to be determined for a few
prototype systems, like $\text{F} + \text{H}_2$~\cite{Valentini:ARPC52:15,
   DerChao:JCP117:8341, Qiu:Science311:1440, Zhang:PRL96:093201}, which was once
considered to be the \emph{holy grail} of the field. The matrix of initial and
final quantum states provides a rigorous test of theoretical potential energy
surfaces. One should note, however, that even at the full state to state level,
impact parameter averaging acts to convolute the experimental results, sometimes
making it difficult to draw quantitative conclusions on the detailed features of
the potential. Furthermore, the scalability of such experiments to larger and
larger systems is prohibitive. Oriented collisions~\cite{Parker:ARPC40:561,
   Loesch:ARPC46:555, OrrEwing:JCSFT92:881, Casavecchia:RPP63:355}, and
especially the application of state-selected, decelerated molecular beams
\cite{Bethlem:PRL83:1558, Bethlem:IRPC22:73, Meerakker:ARPC57:159,
   Heiner:PCCP8:2666, Bethlem:JPB39:R263} and their use in scattering
experiments \cite{Gilijamse:Science:accepted} are likely to be important tools
enabling the experimentalist to further reduce the effects of impact parameter
averaging.

While reactive scattering is a sensitive probe of the repulsive wall of the PES,
much less is known about the long-range dispersive forces between reactive
species. Indeed, already Eyring and Polyani~\cite{Eyring:ZPCB12:279} discussed
the additional effects of dispersion forces on the PES, which predicted a
minimum for the symmetric H$_3$ molecule on their collinear PES. For the heavier
systems they concluded that their calculations were not accurate enough to
incorporate these effects quantitatively. However, the van der Waals minima
which result from the balance of attractive and repulsive forces lie at the base
of the transition state, and correspond to weakly bound clusters of two (or
more) reactants or products. These species are labeled as entrance or exit
channel complexes, respectively. Given that the barriers to chemical reaction
are typically several orders of magnitude larger than dispersion (van der Waals)
forces, the importance of these van der Waals forces to reaction dynamics has
largely been neglected. Recent experimental and theoretical work on the
$\text{Cl} + \text{HD} \rightarrow \text{HCl\,(DCl)} + \text{H\,(D)}$ reaction,
however, has shown that the corresponding orientational effects of the long
range potential can strongly influence the branching ratios and the final state
distributions by giving the reactants a torque, towards or away from the
transition state \cite{Weck:IRPC25:283, Werner:CPL328:500, Werner:CPL313:647,
   Skouteris:Science286:1713, Balakrishnan:JCP121:5563}. Therefore, the
long-range van der Waals forces can no longer be neglected for obtaining really
quantitative results. For collisions at even lower temperature
\cite{Gilijamse:Science:accepted} such effects will be even more important.
Moreover, also for the dissociation of formaldehyde
\cite{Townsend:Science306:1158, Chambreau:PS73:C89} and the hydrogen abstraction
from hydrocarbon molecules by chlorine~\cite{Murray:IRPC23:435} the effects of
long-range van der Waals interactions have clearly been observed. Therefore, it
must be concluded, that a quantitative understanding of van der Waals wells of
molecular complexes is imperative.

One of the legacies that Roger E.\ Miller has left with us is the sensitivity of
high resolution spectroscopy of such weakly bound complexes as a probe of the
surrounding PES. Here again, theory plays a crucial role in solving the
multidimensional problem standing between the PES and the observed spectral
transitions between eigenstates of the exact Hamiltonian. More recently, these
spectroscopic techniques have been applied to reactive systems, where now the
experimental methods have the added challenge of stabilizing the weakly bound
complex and preventing the reaction. Sometimes the cooling provided by a free
jet expansion is sufficient to stabilize such pre-reactive species, which can
then be studied spectroscopically. The most common examples are X-HY radicals,
where X are, for example, H$_2$ or noble gas atoms and Y=C, O, N, S, B, or CN
radicals. Many of the previous experiments have been performed using an
noble-gas atom such as argon or neon to study the effects of polarization, and
in these cases no reaction can occur~\cite{Kim:IRPC20:219, Heaven:IRPC24:375}.
Complexes have also been observed between reactive partners such as, for
example, OH$+$CO \cite{Lester:FD118:373}, OH$+$H$_2$~\cite{Loomis:JCP104:6984,
   Anderson:JCP109:3461, Loomis:ARPC48:643, Schwartz:CPL273:18,
   Hossenlopp:JCP109:10707}, and CN$+$H$_2$~\cite{Chen:JCP109:5171,
   Kaledin:CPL347:199, Chen:JCP112:7416, Wheeler:IRPC19:501}. However, in these
systems the barriers to reaction are still quite large. So far these
experimental techniques have relied on the exquisite sensitivity of laser
induced fluorescence double-resonance methods in order to measure the spectra,
somewhat limiting the choice of chromophores.

Many transient radicals have also been studied in solid noble-gas
matrices~\cite{Jacox:CP42:133, Mielke:JPC94:3519, Khriachtchev:JCP124:181101}.
Indeed, the inert matrix can be used to keep the radicals from interacting with
one another, and in general the number of radicals can be built up using long
deposition times. One can also raise the temperature of the solid matrix, which
gives the dopants some mobility, eventually initiating the reactions. Due to the
large matrix effects, however, the spectra are often strongly perturbed, thus
limiting the amount of information which can be extracted. Indeed, the highly
anisotropic interaction is found to quench the free rotation of most molecules.
In contrast to the more classical cryogenic matrices described above, cold,
solid para-hydrogen matrices have also recently been used to study molecules in
a more soft, quantum mechanical matrix~\cite{Momose:VS34:95, Momose:IRPC24:533,
   Yoshioka:IRPC25:469}, but even this special matrix does not give such great
control over the growth process as liquid helium droplets. Reactions of embedded
impurities with H$_2$, the matrix material itself, can easily be studied, but
for other purposes the matrix is not as inert as noble-gas
environments~\cite{Momose:JCP108:7334}.

Liquid helium nano-droplets combine the advantages of both types of solid matrix
environments. They are inert due to the very noble, unreactive character of the
helium atoms constituting the matrix and they perturb embedded impurities only
very weakly due to their quantum-fluid nature, which manifests itself in the
superfluidity of the droplets~\cite{Grebenev:Science279:2083}. The droplets
provide a very low ambient temperature of approximately 0.37~K, due to
evaporative cooling and the low binding energy of helium atoms to the droplet.
Molecules embedded in $^4$He droplets show free
rotation~\cite{Hartmann:PRL75:1566}. The formation of weakly bound molecular
complexes inside helium nano-droplets proceeds through unique growth dynamics
which are mainly determined by long-range forces between the complexation
partners~\cite{Choi:IRPC25:15}. Therefore, often metastable clusters are
stabilized and can be studied in the droplets~\cite{Nauta:Science283:1895,
   Nauta:Science287:293}. Additional information on the PES can be obtained
using infrared-infrared double-resonance spectroscopy, where the photo-initiated
annealing of complexes provides branching ratios between different minima and
limits on the barrier heights and allows to observe the products of
photo-initiated chemical reactions \cite{Merritt:JCP121:1309,
   Douberly:PCCP7:463}.

Extending such studies to transient, reactive species can provide novel
materials, that can be studied using high-resolution spectroscopy. It should be
possible, for example, to build linear chains from polar OH radicals just like
in the case of HCN molecules \cite{Nauta:Science283:1895}, or similar highly
reactive systems. Such systems are ideal candidates for chemical energy storage
and provide large energy-to-mass ratios. The details of such prospective
experiments are discussed in section~\ref{sec:high-energy}. Moreover, the
spectroscopic study of such highly energetic systems provides complementary
information to scattering experiments on chemical reaction dynamics.

In this review we will detail the experimental methods for embedding and
studying radicals embedded in superfluid helium nano-droplets. In the following
section a general introduction into helium droplet experiments is given,
followed by the details of radical production. Then we will describe the
experiments on radical monomers (section~\ref{sec:monomer}) and on
radical-molecule van der Waals complexes in section~\ref{sec:complex}.
Afterward, a discussion of molecule-droplet interactions, emphasizing the
additional information from spectroscopy of open-shell systems embedded in
helium droplets (section~\ref{sec:interactions}), and an outlook for further
applications (section~\ref{sec:future}) are given.

\section{Experimental methods}
\label{sec:experiment}

\subsection{Spectroscopy in superfluid helium nano-droplets}
\label{sec:experiment:spectroscopy}

\begin{figure}[b]
   \centering%
   \includegraphics[width=\floatwidth]{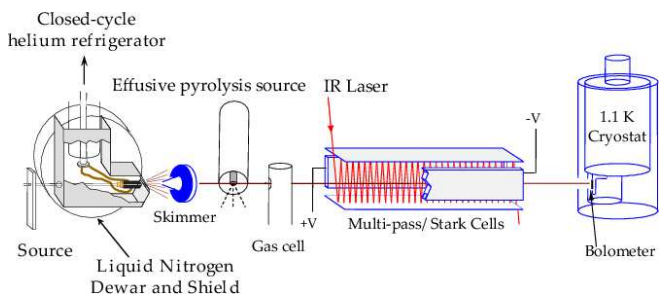}
   \caption{Experimental setup for the infrared spectroscopy of radicals in
      superfluid helium droplets; see text for details.}
   \label{fig:exp:setup}%
\end{figure}

The experimental details concerning the production of neat and doped helium
droplets have been given in previous reviews~\cite{Callegari:JCP115:10090,
   Northby:JCP115:10065, Toennies:ACIE43:2622, Makarov:PhysUspekhi47:217,
   Stienkemeier:JPB:R127, Choi:IRPC25:15} and their thermodynamic properties
have been discussed~\cite{Brink:ZPD15:257, Dalfovo:JCP115:1078,
   Lehmann:JCP119:3336}. Therefore, here we will only discuss the relevant
details and specialties for embedding radicals in the droplets. A schematic of
the first apparatus for infrared spectroscopy of radicals embedded in helium
droplets \cite{Kuepper:JCP117:647}, built at the University of North Carolina at
Chapel Hill (UNC), is shown in Figure~\ref{fig:exp:setup}.
In brief, the setup consists of a continuous helium droplet beam source and
several, differentially pumped vacuum chambers. The source chamber contains the
supersonic droplet beam source. The droplet beam passes through a skimmer into a
region with a series of load lock ports to introduce pick-up sources of varying
nature. After the droplets are doped they fly into a laser excitation region,
where a multipass cell, consisting of two plane, parallel gold-coated mirrors,
is used to increase the laser-droplet beam interaction. This spectroscopy region
also has a pair of parallel electrodes, providing homogeneous fields up to
80~kV/cm, in which Stark and pendular spectroscopy~\cite{Block:PRL68:1303} can
be performed. In all experiments reported here, the electric field is parallel
to the laser-polarization, yielding $\Delta{}m=0$ selection rules for the
associated Stark and pendular spectra. The overall kinetic energy of the beam is
then detected further downstream by a cooled semiconductor bolometer detector
\cite{Choi:IRPC25:15}. Alternatively, mass-spectrometric detection has been used
in similar ways~\cite{Stienkemeier:JPB:R127, Choi:IRPC25:15}, however, in that
case the infrared laser beam is typically sent collinear to the droplet beam to
achieve sufficient signal to noise ratios.

Droplets are typically produced by expanding high-pressure helium (20--100~bar)
through a small, cooled pinhole (5--10~$\mu$m, 5--30~K), as originally
demonstrated by Becker \etal\ in 1961~\cite{Becker:ZNA16:1259}. Droplets are
formed in the early, high-pressure portion of the expansion and then
successively cooled by evaporation. In the case of $^4$He, as used in all
spectroscopic studies on radicals embedded in helium droplets so far, the
evaporation temperature converges at $\sim0.37$~K~\cite{Brink:ZPD15:257} and
$^4$He droplets exhibit superfluidity under these conditions
\cite{Grebenev:Science279:2083}.

Atomic and molecular \emph{impurities} are embedded in the droplets by pick-up
in scattering regions~\cite{Stienkemeier:JPB:R127, Choi:IRPC25:15}. The pick-up
process is sequential, that is, individual atoms and molecules are picked up
from the scattering gas, where only low densities are necessary, because the
pick-up probability is determined by the large, geometrical crosssections of
the droplets. The cooling of dopants inside the helium droplet occurs on a
timescale considerable faster than the complexation of multiple embedded
species. For (HCN)$_n$ complexes embedded in a single droplet, for example,
about 1000~\cm{} binding energy are released upon addition of an additional HCN
molecule. If that energy would be deposited in the molecular complex, the
cluster would anneal to its global minimum structure. However, the helium
removes the energy quickly as the complex is forming, forcing it to stay at the
zero-point level while traversing the PES. Therefore, the complexes are often
trapped in local minima and metastable species can be
observed~\cite{Nauta:Science283:1895, Nauta:Science287:293}. Moreover, since the
pick-up can be performed in multiple separate scattering regions, considerable
control over the formation of larger clusters can be achieved, as, for example,
shown for HF-Ar$_n$ complexes \cite{Nauta:JCP115:10138}. This separable pick-up
also allows to produce clusters with quite complex compositions, possibly
containing multiple species, and to embed species under quite different
conditions into the droplet. For example, thermally labile bio-molecules and
metal atoms from a hot oven or radicals from an 1800~K pyrolysis source can be
embedded in the same droplet, because the two species only come together in the
cold confines of the droplet. This unique production process allows to
specifically design desired complexes inside helium droplets.

All molecular radicals or radical-molecule complexes embedded in a helium
droplet that have been studied using high-resolution spectroscopy so far have
either been stable molecules, like NO, that were picked up in an ordinary
scattering chamber, or were produced in a continuous effusive pyrolysis source
developed some years ago~\cite{Kuepper:JCP117:647}; see
section~\ref{sec:experiment:radicals} for details.

Different detection schemes used in helium droplet spectroscopy have been
described before~\cite{Choi:IRPC25:15,Stienkemeier:JPB:R127}. In all studies
described in detail in this review, infrared depletion spectroscopy is used,
where the evaporation of helium atoms from the droplet due to resonant infrared
excitation of the dopant is detected~\cite{Choi:IRPC25:15}. In most of the
studies reported here the reduced total kinetic energy of the droplet beam on a
1.1~K bolometer is measured using lock-in techniques, but mass spectrometric
techniques, albeit less sensitive, have also been employed
\cite{Choi:IRPC25:15}, also for performing electronic spectroscopy of doped
helium droplets~\cite{Stienkemeier:JCP115:10119, Stienkemeier:JPB:R127}.
However, the bolometer detection proved to be the most sensitive so far. It is
very general as it detects directly the evaporation of helium atoms from the
droplet and does not require any specialized UV laser systems. Due to its
quasi-continuous operation mode it matches the continuous droplet beams very
well and provides optimal experimental duty cycle.

In the interpretation of rotationally resolved spectra of species embedded in
helium droplets the \emph{matrix effects} of the droplet environment needs to be
considered. Vibrational frequencies of the embedded molecules are typically
shifted very little from their gas-phase values, and due to the superfluidity of
the droplets, free rotation of the embedded species can be observed
\cite{Hartmann:PRL75:1566}. The obtained rotational constants, however, are
reduced from their corresponding gas-phase values~\cite{Hartmann:PRL75:1566,
   Choi:IRPC25:15} and the influence of the droplet environment on the
electronic structure can also be considerable. These effects are discussed in
section~\ref{sec:interactions}.

\subsection{Radical production}
\label{sec:experiment:radicals}

Some molecular radicals are not transient at all and can be bought and stored
under ambient conditions. Therefore a simple pick-up cell can be used to embed
these species in the droplets. NO is an example of such a molecule that has also
been studied in helium droplets~\cite{Haeften:PRL95:215301}.

For transient species, however, one has to resort to \emph{in situ} production
of the radical in the experiment. Several different methods for producing
radicals for molecular beam studies have been developed, including pyrolysis,
radio-frequency or microwave discharges, corona discharges, and photolysis
\cite{Pauly:OtherLowEnergySources,Pauly:HighEnergySources}. The most popular
sources for use in combination with the extreme cooling provided in supersonic
jet experiments are flash pyrolysis~\cite{Kohn:RSI63:4003, Cameron:RSI67:283,
   Zhang:RSI74:3077}, electric discharge sources~\cite{Engelking:RSI57:2274,
   Engelking:CR91:399, Anderson:CPL258:207}, and photolysis sources
\cite{Monts:CP45:133, Heaven:CPL84:1, Andresen:JCP81:571}.

These methods are, however, not currently generally applicable for embedding
molecules into helium droplets, as they are mostly operated in a pulsed fashion
and require relatively high gas-pressures. While these radical sources are
efficient in producing a large number of radicals, especially the discharge and
photolysis sources often produce a large number of by-products. For gas-phase
studies this does not necessarily represent a problem as long as the background
pressure is kept low enough to prevent collisions, and that the target-radical
concentration is high enough for the experimental sensitivity. Also, if a
precursor or decomposition product absorbs in the same spectral region as the
species of interest, then it may be difficult to separate out the overlapping
bands. In a helium droplet experiment, however, all species are picked up by the
droplets without any discrimination. Therefore, extremely clean sources of
radicals are required for these experiments in order not to contaminate all
droplets. This includes buffer gas, which makes up most of the gas-phase
expansion. Moreover, most helium droplet experiments today are operated as
continuous beams and are therefore most efficiently used with continuous doping
methods. Coupling pulsed radical sources to these beams, especially photolysis
sources using intense pulsed lasers operating at 10 Hz, leads to large
reductions of duty cycle and correspondingly lower signals. Once pulsed helium
droplet sources are wider applicable~\cite{Slipchenko:RSI73:3600,
   Yang:RSI76:104102}, however, they may make such radical sources more
compatible.\footnote{Commercial atom sources provide so called \emph{clean}
   atomic beams, but for the purposes of the experiments described here the
   atom-content is still too low. Initial experiments by ourselves to use
   microwave discharge sources were not successful due to the necessary low gas
   flow and pressure, in order not to destroy the helium droplet beam by a
   gas-jet from the discharge source.}

\subsubsection*{Pyrolysis}

\begin{figure}
   \centering%
   \includegraphics[width=\floatwidth]{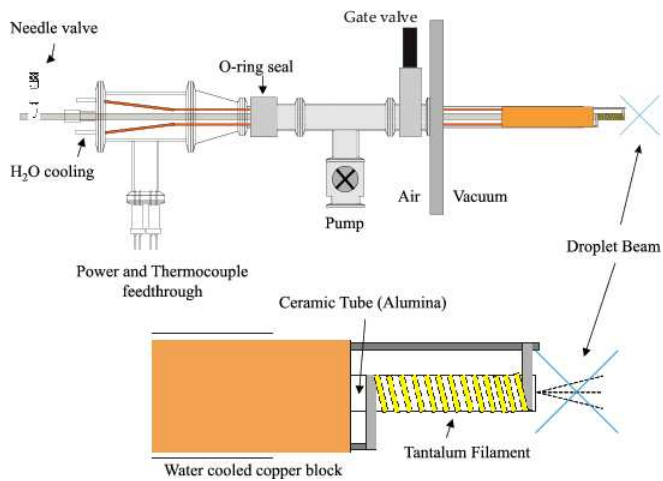}%
   \caption{A schematic diagram of a pyrolysis source that can be load-locked
      into the helium nano-droplet apparatus, to facilitate the easy change of
      pick-up sources. In this case, an alumina tube is heated by a tantalum
      filament. A needle value is used to regulate the flow of the precursor
      through the source.}%
   \label{fig:exp:pyrolysis-source}
\end{figure}

The ideas from flash pyrolysis, even if not directly applicable, can be
transferred to conditions suitable for embedding radicals in helium
droplets~\cite{Kuepper:JCP117:647}. A scheme of the low-pressure, continuous,
effusive pyrolysis source used at UNC is shown in
Figure~\ref{fig:exp:pyrolysis-source}.
The source can routinely be heated to 1800~K to produce clean effusive beams of
radicals from appropriate precursor molecules. Due to the optical transparency
of liquid helium over an extremely wide energy range (up to 21~eV), black-body
radiation from the hot source does not effect the droplet beam.

\subsubsection*{Other schemes}

While our approach has been to generate the radicals externally and dope them
into the droplets, another approach would be to produce radicals from photolysis
of a stable molecule already embedded inside a helium droplet. Near threshold
photo-dissociation of NO$_2$ has been attempted by Conjusteau \etal\
\cite{Conjusteau:thesis:2002}, with the goal of observing the NO-O
radical-radical van der Waals complex, but this has not been successful. It is
unclear whether the helium is able to quench the energy faster than the time
necessary to break the molecule apart, or the two fragments always react to give
back NO$_2$. Braun and Drabbels~\cite{Braun:PRL93:253401} have used 266~nm
photodissociation of CF$_3$I and CH$_3$I embedded in helium droplets to study
the translational dynamics of the recoiling CH$_3$ (CF$_3$) and I atom
fragments. Here, 266~nm photolysis promotes the CH$_3$I to a highly repulsive
electronic state, which is known to dissociate very rapidly in the gas-phase
\cite{Kavita:JCP112:8426}. In the helium droplet study gas-phase fragments
ejected from the droplets were detected by velocity map imaging
\cite{Parker:JCP107:2357} and their product distributions were found to be less
anisotropic and shifted to lower velocities than compared to the equivalent
gas-phase experiment~\cite{Braun:PRL93:253401}, consistent with a model of hard
sphere collisions. Since the detection method was based on imaging the gas-phase
recoil products, studies were carried out using relatively small droplets, which
allowed the fragments to escape. Photolyzing the embedding precursor in a much
larger droplet, however, could allow one or both of the products to stay within
the droplet.

Cold ion-molecule reactions have also been observed in helium droplets
\cite{Farnik:JCP122:014307}. In this study 80~eV kinetic energy electrons were
used to bombard the helium droplets, resulting in ionization of a single He atom
in the droplet. The molecular impurity may then be ionized after resonant charge
transfer~\cite{Lewis:JACS127:7235}. Although the newly formed molecular ions are
ejected from the droplet due to the energy released in the charge transfer step,
ion-molecule reactions between these secondary molecular ions and additional
molecules embedded in the droplet could be observed, yielding, for example
N$_2$D$^+$ (from N$_2^+$ and D$_2$), CH$_4$D$^+$, or CH$_3$D$_2^+$ (from
CH$_4^+$ or CH$_3^+$ and D$_2$)~\cite{Farnik:JCP122:014307}.

\section{Radical monomers in helium droplets}
\label{sec:monomer}

\subsection{Propargyl}
\label{sec:monomer:propargy}

The first spectroscopic study of a molecular radical in helium droplets was
performed on propargyl (2-propynyl, C$_3$H$_3$) just five years ago
\cite{Kuepper:JCP117:647}. In light of its role in sooting flames, the propargyl
radical has been the focus of particular experimental attention
\cite{Marinov:CombustFlame114:192, Westmoreland:JPC93:8171, Alkemade:ZPC161:19,
   Stein:SympCombust23:85}. Propargyl is one of the simplest conjugated systems
with an odd number of electrons, also making it the focus of considerable
theoretical study~\cite{Botschwina:ZPC188:29, Botschwina:JElecSpec108:109,
   Honjou:JPC91:4455, Hinchcliffe:JMolStruct37:295}. There is compelling
evidence that propargyl is the most important radical precursor in the formation
of benzene, polycyclic aromatic hydrocarbons, and soot in certain combustion
processes~\cite{Marinov:CombustFlame114:192, Westmoreland:JPC93:8171,
   Alkemade:ZPC161:19, Stein:SympCombust23:85, Wu:JPC91:6291,
   Marinov:CombustSciTech128:295, Glassman:SympCombust22:295,
   Miller:SympCombust26:461, Goodings:CombustFlame36:27,
   Olson:SympCombust18:453}. For example, the simple dimerization of two
propargyl radicals is thought to be important in the formation of benzene, as
suggested by Wu and Kern~\cite{Wu:JPC91:6291}.

The work on the transient and very reactive propargyl radical embedded in helium
droplets demonstrated how to couple our continuous, low-pressure pyrolysis
source to a helium droplet beam for pick-up of such transient species
\cite{Kuepper:JCP117:647}. From a fit to the rotationally resolved infrared
spectrum of the $\nu_1$-transition the molecular parameters given in
Table~\ref{tab:propargyl:constants} are obtained.
\begin{table}
   \centering
   \begin{tabular}{lcc}
      \hline\hline
      constant & helium droplet\footnotemark[1] & gas-phase\footnotemark[2] \\
      \hline
      $A''$ (\cm)       & 9.60847 \footnotemark[3] & 9.60847\,(18) \\
      $(B''+C'')/2$ (\cm)     & 0.1198\,(5)  & 0.312386\,(12) \\
      $B''-C''$ (\cm)         & 0.035\,(2)   & 0.0105762\,(35) \\
      $\Delta_N''$ (\cm)     & 0.00042\,(1) & $7.35\,(122)\cdot10^{-8}$ \\
      $A'$ (\cm)  & 9.60258\footnotemark[3] & 9.60258\,(11) \\
      $(B'+C')/2$ (\cm)       & 0.1185\,(5)  & 0.311641\,(7) \\
      $B'-C'$ (\cm)           & 0.035\,(2)   & 0.010496\,(13) \\
      $\Delta_N'$ (\cm)      & 0.00062\,(1) & $5.37\,(76)\cdot10^{-8}$ \\
      $\nu_0$ (\cm)           & 3322.15\,(1) & 3322.292\,(10) \\
      $\mu_a''$ (D)           & -0.150\,(5)  & --- \\
      $\Delta\mu_a$ (D)       &  0.02\,(1)   & --- \\
      \hline\hline
   \end{tabular}
   \footnotetext[1]{Reported helium droplet values are from reference
      \onlinecite{Kuepper:JCP117:647}. The values for $B-C$ were mistyped in
      the original publication and the corrected values are given here.}
   \footnotetext[2]{Reported ground state values are from
      reference~\onlinecite{Tanaka:JCP103:6450}, excited state values are from
      reference~\onlinecite{Yuan:JMolSpec187:102}.}
   \footnotetext[3]{$A"$ and $A'$ are fixed at their respective gas-phase values.}
   \caption{Summary of the molecular constants for the propargyl radical in
      superfluid helium droplets~\cite{Kuepper:JCP117:647}, compared to those
      obtained in gas-phase studies~\cite{Tanaka:JCP103:6450,
         Yuan:JMolSpec187:102, Tanaka:JCP107:2728}. Numbers in parenthesis are
      one standard deviation.}
   \label{tab:propargyl:constants}
\end{table}
We find that the $K_a=1$ states are populated due to nuclear spin statistics,
confirming the $C_{2v}$ symmetry of this radical. This shows, moreover, that the
local environment inside the helium droplet does not lower the symmetry. Indeed,
nuclear spin conversion processes are extremely slow in the gas-phase
\cite{Chapovsky:ARPC50:315}, but even the presence of the unpaired electron and
the collision rich condensed phase environment inside the droplet does not
induce any noticeable spin relaxation on the timescale of the experiment
($\sim1$~ms), demonstrating the weak perturbation of the embedded radical by the
helium droplet environment. Nuclear spin relaxation was observed in
quantum-solid para-hydrogen matrices on a much longer timescale of minutes to
hours~\cite{Fushitani:JCP116:10739} and the accelerating effect of added
ortho-hydrogen was clearly observed. Since the ensemble of complexes studied in
a helium droplet experiment is only from droplets with exactly the wanted
composition and the helium--complex interaction is generally weaker then the
solid-hydrogen--molecule interaction, one would expect the conversion even to be
considerably slower in helium droplets.

While the $A$ rotational constants cannot be determined from the experimental
data, the $(B+C)/2$ inertial parameters for ground and excited vibrational state
were about one third of the gas-phase values, a reduction that is typical for
rotational constants of this size~\cite{Choi:IRPC25:15}. The
asymmetry-splittings $B-C$ are increased in the helium droplet environment,
probably due to differential effects of the helium droplet environment on the
rotation around the different axes.

This study was the first to experimentally determine the permanent dipole moment
of propargyl radical by measuring the rotationally resolved infrared spectrum in
the presence of a strong homogeneous electric field (50~kV\,\cm). It is known
that dipole moments determined by Stark spectroscopy in helium droplets are very
comparable to the gas-phase values~\cite{Stiles:PRL90:135301}. This is confirmed
by the good agreement of the measured values for propargyl ($\mu_0=-0.150(5)$~D
and the $\Delta\mu_{v=1\leftarrow0}=0.02(1)$~D) and calculated values
($\mu_e=-0.14(3)$~D, $\mu_0=-0.124$~D, $\mu_{v=1\leftarrow0}=0.011$~D)
\cite{Botschwina:JElecSpec108:109,Botschwina:propargyl-dipole:2002}.

\subsection{Nitric oxide}
\label{sec:monomer:no}

Recently, the nitric oxide (NO) radical was studied using infrared diode laser
spectroscopy in helium droplets~\cite{Haeften:PRL95:215301}. Only the $Q(1/2)$
and $R(1/2)$ transitions of the $v=1\leftarrow0$ transition in the $^2\Pi_{1/2}$
electronic ground state were observed due to the low temperature (0.4~K) of the
droplet and the small moment of inertia of NO. From the line separation the
rotational constant of NO in helium droplets is measured to be $B=1.253$~\cm,
which is 76~\% of the gas-phase value, in good agreement with previous closed
shell systems of this size.

For the $Q(1/2)$ line sharp transitions are obtained. This reflects the fact
that vibrational relaxation is slow due the absence of low-lying vibrational
modes~\cite[and references therein]{Lindsay:JCP121:6095}. Indeed, previous
studies of HF embedded in helium droplets have shown that the monomer does not
vibrationally relax at all on the timescale of the experiment ($\sim1$~ms)
\cite{Nauta:JCP113:9466}, so instead of depleting the droplet beam intensity,
the extra vibrational energy of HF is simply carried to the bolometer. Using a
mass-spectrometer based detection scheme, such as that used in the study of NO,
one would not \emph{a priori} expect a depletion signal in this case. Molecular
absorption beam depletion can still be detected, however, due to pick-up of an
impurity, \ie\ a second NO or N$_2$, after the infrared excitation, leading to
relaxation of the vibrationally excited molecular dimer, as shown conclusively
for HF~\cite{Lindsay:JCP121:6095}. The resulting long monomer lifetime allows
the nuclear hyperfine and $\Lambda$-doubling splittings of the transition to be
resolved. The nuclear hyperfine splitting is unchanged from the isolated
gas-phase value, whereas the $\Lambda$-doubling is increased by a factor of
$1.55$ compared to the gas-phase~\cite{Haeften:PRL95:215301}. The authors
discuss this effect in terms of electronic coupling and also suggest the
possible influence of the helium density around the NO molecular axis
\cite{Haeften:PRL95:215301}. For more details of these effects see the results
on the NO-HF complex in section~\ref{sec:no-hf} and the general discussion of
molecule-droplet interactions in section~\ref{sec:interactions}.

The $R(1/2)$ line, on the other hand, shows a single unresolved band that fits
well to a Lorentzian lineshape, which is attributed to fast rotational
relaxation in the vibrationally excited state with a lifetime of
$1.52\cdot10^{-10}$~s. The large rotational constant of NO places the excited
rotational state in a region of high density of states for the fundamental
excitations of the droplet, leading to a strong coupling and the corresponding
short lifetime \cite{Choi:IRPC25:15}.

The NO molecule was also used as a sensitive probe of electronic effects of
molecule-droplet interactions~\cite{Polyakova:JCP124:214308}. In that study, NO
embedded in liquid helium droplets was photoionized and the ongoing dynamics was
found to be quite complex. NO is initially excited by a two-photon excitation
from its $^2\Pi$ electronic ground state to low-lying Rydberg states (NO$^*$).
Subsequently, the nuclear degrees of freedom in NO$^*$ and between the NO$^*$
and the helium environment relax and NO$^*$ is transported to the droplet
surface on a time-scale shorter than the laser pulse ($<10$~ns), where it can
then leave the droplet along with a modest number of helium atoms or,
alternatively, be photoionized. The formed ion can then be resolvated by
snowball formation around the ion-core or leave the droplet surrounded by helium
atoms.

\subsection{Other species}
\label{sec:monomer:other}

Several other radicals or transient species have been observed in or on liquid
helium droplets. For example the spectra of individual open shell metal atoms on
helium droplets have been obtained~\cite{Stienkemeier:ZPB98:413,
   Reho:FD1997:161, Stienkemeier:JCP115:10119, Bruehl:JCP115:10220,
   Makarov:PhysUspekhi47:217} and the chemiluminescence emission spectrum of the
$\text{BaO}^*$ product from the reaction $\text{Ba} + \text{N}_2\text{O}
\rightarrow \text{BaO}^* + \text{N}_2$ inside a helium droplet has been observed
\cite{Lugovoj:JCP112:8217}. The reaction was found to proceed very efficiently
inside the droplet, and the influence of co-embedded xenon clusters was studied,
which showed that the xenon pulls the nascent BaO, initially formed at the
surface, into the center of the droplet. Experiments on the photodissociation of
CF$_3$I to form CF$_3$ and I inside the helium
droplets~\cite{Braun:PRL93:253401} and on ion-molecule reactions
\cite{Farnik:JCP122:014307} were already discussed in
section~\ref{sec:experiment:radicals}.

\section{Radical-containing molecular complexes}
\label{sec:complex}

Whereas the effects of the moderately strong long-range electrostatic
interactions in ion-molecule reactions are well known to have a significant
influence on the associated reaction rates~\cite{Herbst:AJ185:505,
   Smith:CR92:1473}, the importance of the weaker van der Waals forces in the
entrance channels of neutral reactions have been only recently fully appreciated
\cite{Murray:IRPC23:435, Neumark:PCC5:76, Lester:FD118:373,
   Varandas:JCP125:064312}. Experimental and theoretical work on the $\text{Cl}
+ \text{HD} \rightarrow \text{HCl\,(DCl)} + \text{D\,(H)}$ reaction shows that
the torque experienced by the HD in the entrance valley of the potential has a
significant effect on the overall reaction rates and branching ratios
\cite{Werner:CPL328:500, Werner:CPL313:647, Skouteris:Science286:1713,
   Balakrishnan:JCP121:5563}. Furthermore, the rotational excitation of HCl
products formed from abstraction reactions of chlorine and organic molecules,
for example, can only be accurately described by taking into account the
features of the exit channel. The dipole-dipole interaction of the departing
fragments was found to be very important in determining the product final state
distributions~\cite{Murray:IRPC23:435}. For the dissociation of formaldehyde a
completely new dissociation mechanism (\emph{roaming hydrogen atom} mechanism)
has been proposed~\cite{Townsend:Science306:1158, Chambreau:PS73:C89}. These
results are clear examples, that entrance and exit channel complexes of
molecular reactions probe the associated PES at energies that are relevant to
the systems' chemical reaction dynamics, despite being much lower in energy than
the corresponding transition states. However, the torque on the reactants at
long range acts to deflect trajectories towards or away from the transition
state~\cite{Skouteris:Science286:1713}. Therefore, these complexes are of
considerable importance in understanding the nature of the reactions,
particularly at low translational energies (low temperatures).

The isolation and stabilization of entrance and exit channel complexes in liquid
helium droplets and their high-resolution spectroscopic study provides a novel
experimental approach for studying these important systems. In the remainder of
this section we will summarize the experiments performed on such complexes in
liquid helium droplets to date. In the first part we will discuss the
hydrogen-bound complexes of halogen atoms with small linear molecules and in the
second part of this section we will proceed to complexes of hydrocarbon radicals
with hydrogen acids.

\subsection{Complexes containing halogen atoms}
\label{sec:complex:atomic}

Recently, a large series of complexes of closed shell molecules with halogen
atoms embedded in liquid helium droplets have been studied using high-resolution
infrared spectroscopy, \ie\ the complexes of chlorine, bromine, and iodine with
HF and HCN, respectively~\cite{Merritt:PCCP:inprep,Merritt:PCCP7:67}. These
systems represent the prototypical \emph{heavy---light-heavy} X--HY (X=Cl, Br,
I; Y=F, CN) complexes which serve as prominent benchmark systems for the
understanding of elementary reaction dynamic studies. They can be studied in
great detail theoretically~\cite{Dubernet:JCP101:1939, Dubernet:JPC98:5844,
   Maierle:JCSFT93:709, Jungwirth:JPCA102:7241, Bittererova:CPL299:145,
   Meuwly:PCCP2:441, Meuwly:JCP112:592, Meuwly:JCP119:8873, Klos:JCP115:3085,
   Zeimen:JPCA107:5110, Neumark:ARPC43:153, Deskevich:JCP124:224303,
   Fishchuk:JPCA110:5273,Fishchuk:JPCA110:5280}, as they contain only two heavy
atoms (X-HF) and only a few nuclear degrees of freedom. Some stationary points
of such systems have also been studied experimentally, for example, by
spectroscopy of the transition states~\cite{Neumark:ARPC43:153, Liu:CPL299:374,
   Imura:JMolStruct552:137, Mestdagh:IRPC22:285} and using infrared spectroscopy
in noble-gas matrices~\cite{Andrews:JCP89:3502, Hunt:JCP88:3599,
   Hunt:JPC92:3769}. These studies, however, did not reveal detailed information
on the associated entrance or exit channel regions of the reactions.

The X-HF~\cite{Ding:FDCS55:252, Wuerzberg:CPL57:373, Wuerzberg:JCP72:5915,
   Moore:IJCK26:813, Tamagake:JCP73:2203, Zolot:F+HCl:inprep} and
X-HCN~\cite{Metz:CPL221:347, Kreher:JCP104:4481, Decker:JPCA105:5759} PES have
also been studied in collision experiments. Recently, strong rotational
enhancement has been described for the F + HCl reaction \cite{Hayes:JPCA110:436,
   Zolot:F+HCl:inprep}. The chemiluminescence from HF produced in chemical
lasers, \ie\ from reactions of ClF with hydrogen, has been
observed~\cite{Krogh:JCP67:2993}. However, so far no spectra of the stationary
points on the PES have been obtained for the free molecular complexes.

\begin{figure*}
   \centering
   \includegraphics[width=\linewidth]{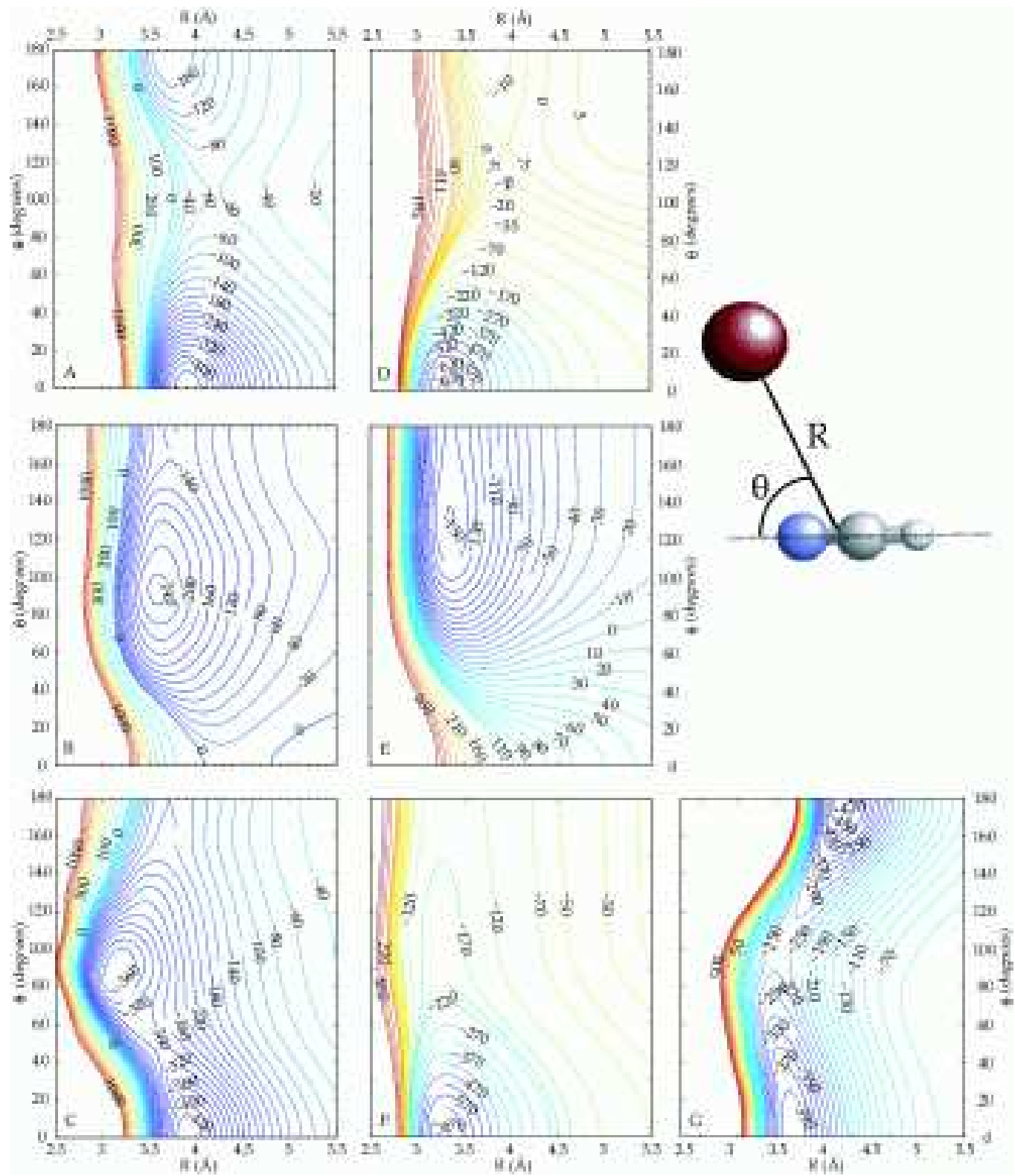}
   \caption{(A, D) Diabatic potential energy surfaces for Cl-HCl and Cl-HF
      $j_A=3/2 \; \omega=3/2$ and (B, E) $j_A=3/2 \; \omega=1/2$, respectively.
      (C--F) The lowest adiabatic potential energy surface including spin-orbit
      coupling for Cl-HCl (C), Cl-HF (F), and Br-HCN (G), respectively. Jacobi
      coordinates as shown in the inset have been used throughout and
      $\theta=0\,\degree$ corresponds to the hydrogen bound arrangement. See
      text for details.}
   \label{fig:hcn-x:pes}
\end{figure*}

For these X-HY systems, the interaction of the quadrupole moment of the halogen
atom with the dipole moment of the molecule suggests two linear structures with
the atom sitting on either end of the linear molecule. However, T-shaped minima
are also predicted due to the considerable quadrupole moments of the molecules
\cite{Dubernet:JCP101:1939, Dubernet:JPC98:5844, Meuwly:JCP112:592,
   Meuwly:JCP119:8873, Zeimen:JPCA107:5110}. The weak interaction between a
halogen atom and the molecules in a van der Waals complex suggests, that, to a
first approximation, the orbital angular momentum $j_A$ of the atom will be
conserved in the complex. This interaction removes the degeneracy of the $^2P$
atomic states, which in $C_s$ symmetry gives rise to two states of $A'$ and one
state of $A''$ symmetry. These correspond to the three relative orientations of
the singly occupied orbital with respect to the complexation partner, \ie\ the
unpaired electron being in the $p_x$, $p_y$, and $p_z$ orbital, respectively.
Solving for the eigenvalues of the electronic Schr\"odinger equation under the
Born-Oppenheimer approximation yields the adiabatic PES. However, since these
surfaces are coupled by non-adiabatic coupling terms originating from the
nuclear kinetic energy operator, it is often helpful to solve for the bound
states in a diabatic representation \cite{Dubernet:JPC98:5844,
   Zeimen:JPCA107:5110}. The electronic anisotropy is further complicated due to
electron spin, and spin-orbit coupling can also have an important effect on
shaping the PES. For comparison we reproduce the lowest diabatic and adiabatic
PES including spin-orbit coupling for the Cl-HCl, Cl-HF, and Br-HCN complexes in
Figure~\ref{fig:hcn-x:pes}(A--C), (D--F), and (G), respectively.
The results for Cl-HCl \cite{Klos:JCP115:3085, Zeimen:JPCA107:5110} and Cl-HF
\cite{Fishchuk:JPCA110:5273, Fishchuk:JPCA110:5280} have been published
previously, while preliminary results for Br-HCN have been provided by A.\ van
der Avoird. The general topology of the Cl-HCl and Cl-HF surfaces are quite
similar. However, the larger dipole of HF, compared to HCl, makes the linear
hydrogen bound minimum deeper for Cl-HF and the T-shaped minimum has almost
completely disappeared in the lowest adiabatic surface. Indeed, for Cl-HF only
the linear hydrogen bound isomer was observed experimentally in helium droplets,
suggesting that the T-shaped or linear HF-X isomers are not formed. Detailed
comparisons of the experimental and theoretical results for the X-HF systems
have been given elsewhere \cite{Merritt:PCCP7:67, Fishchuk:JPCA110:5273,
   Fishchuk:JPCA110:5280, Meuwly:JCP119:8873}. In contrast, the lowest adiabatic
surface for the Br-HCN system shows two deep minima corresponding to the
hydrogen-bound and nitrogen-bound linear isomers, respectively, which are also
both observed experimentally \cite{Merritt:PCCP:inprep}.

The $1\,^2A'$ and $1\,^2A''$ surfaces are degenerate at the linear H-bound
configuration ($\theta=0$), whereas the $1\,^2A'$ and $2\,^2A'$ surfaces are
degenerate at the linear N-bound configuration ($\theta=180$). Therefore, the
electronic symmetry is $^2\Sigma$ for HCN-X, with the unpaired electron along
the axis of the molecule, and $^2\Pi$ for X-HCN, with the unpaired electron
perpendicular to the axis of the molecule.

\begin{figure}
   \centering%
   \includegraphics[width=\floatwidth]{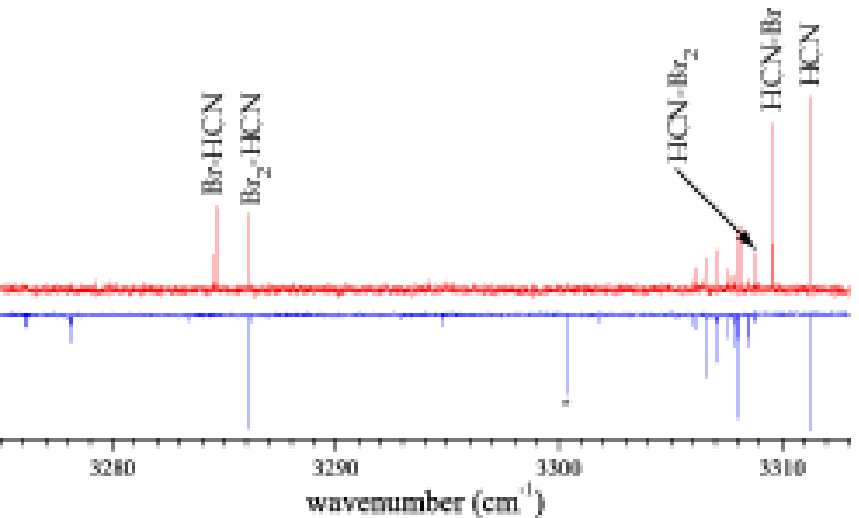}%
   \caption{Pendular survey scans revealing the pyrolysis source temperature
      dependence (upward peaks are with a hot source, downward peaks are with a
      cold source) for the HCN + bromine experiment. The scans show peaks which
      we identify as molecular and atomic bromine complexed with HCN. The peak
      labeled with an asterisk is a known impurity.}
   \label{fig:halogen:X-HCN:survey}%
\end{figure}

In Figure~\ref{fig:halogen:X-HCN:survey} survey scans for the Br-HCN systems
with cold (room temperature) and heated ($\sim1600$~K) pyrolysis source are
shown as an example.
Such overview scans are infrared depletion spectra obtained in the presence of
strong homogeneous electric fields ($\sim60$~kV\,\cm) in which the complete
rotational structure is collapsed into a single \emph{pendular} transition
\cite{Loesch:JCP93:4779, Rost:PRL68:1299, Block:PRL68:1303} due to the large
degree of orientation of the polar molecules in the cold environment.

When the pyrolysis source is cold, only complexes of HCN with halogen molecules
are formed. When the pyrolysis nozzle is heated, however, the signals associated
with the precursor complexes are significantly reduced, due to the dissociation
into atoms, and a new set of signals is observed to grow in, which are readily
assigned to binary complexes of bromine atoms with HCN. Based on harmonic
vibrational frequency calculations as well as physical intuition, the largely
different frequency shifts of the $\nu_\text{CH}$ stretching vibrations result
from the two different structures of the complexes, namely the peak shifted
furthest to the red corresponds to the Br-HCN complex, and the peak which
exhibits only a small shift is related to the HCN-Br complex. This assignment is
confirmed by analysis of the field-free spectra of each of these transitions
\cite{Merritt:thesis:2006, Merritt:PCCP:inprep}.

\begin{figure}[t]
   \centering
   \includegraphics[width=\floatwidth]{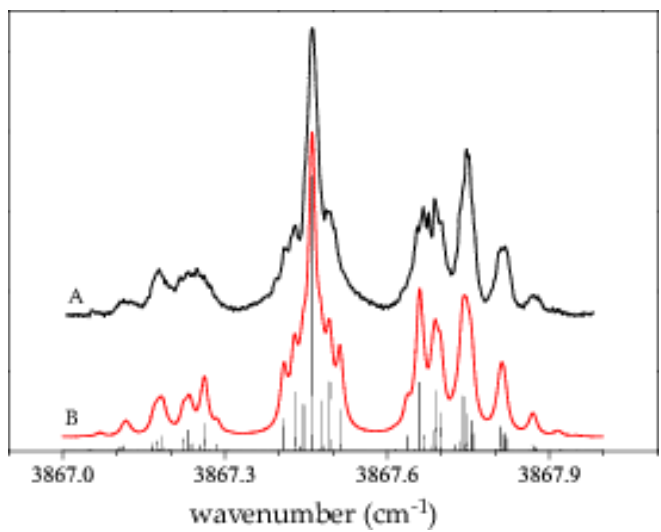}
   \caption{(A) The field-free infrared spectrum of Br-HF shown together with
      (B) a simulation that includes nuclear magnetic hyperfine interactions due
      to the bromine nucleus ($I=3/2$). The stick spectrum shows all of the
      transitions that underlie the observed features.}%
   \label{fig:halogen:Br-HF-spectrum:field-free}%
\end{figure}

\begin{figure}
   \centering
   \includegraphics[width=\floatwidth]{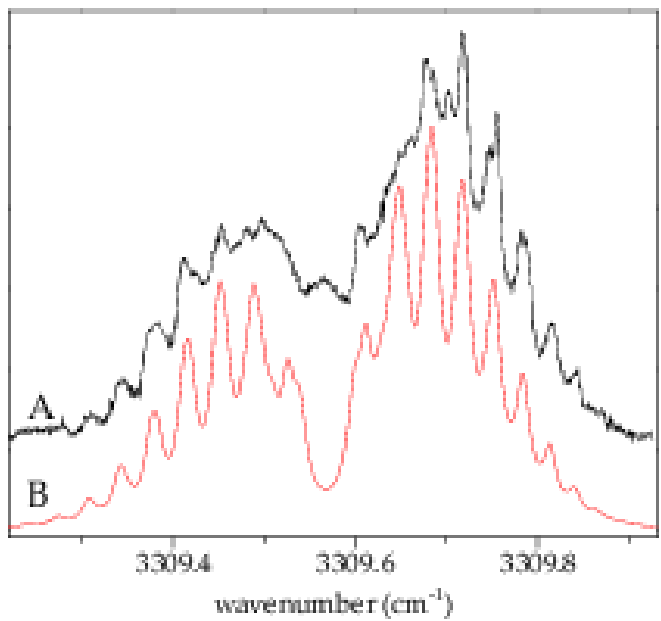}
   \caption{(A) The field-free infrared spectrum of HCN-Br shown together with
      (B) a simulation that includes parity splitting induced by spin-orbit
      coupling of the $^2\Sigma_{1/2}$ and the $^2\Pi_{1/2}$ states of the
      complex.}%
   \label{fig:halogen:HCN-Br-spectrum:field-free}%
\end{figure}

In Figure~\ref{fig:halogen:Br-HF-spectrum:field-free} the field free spectrum of
Br-HF is shown as an example.
The vibrational frequency is shifted by 131~\cm{} to lower energy compared to
the HF monomer transition~\cite{Nauta:JCP113:9466}, indicating the binding of Br
to the hydrogen end of HF. The simulation of the rotationally resolved spectrum,
constituted by all the individual simulated transitions shown at the bottom of
the figure, reproduces all of the important features of the spectrum. Some
deviations are observed especially for the lowest-$J$ transitions, an effect
that is regularly observed for molecules embedded inside helium droplets; see
section~\ref{sec:interactions} for details. The overall structure and all
features are clearly reproduced by the simulation. Therefore, these spectra
confirm the $^2\Pi_{3/2}$ ground states of these hydrogen bound complexes,
consistent with the $^2P_{3/2}$ ground state of atomic bromine. Spectra and
simulations of the same quality are obtained for all other hydrogen bound
X-HF/HCN complexes studied. The molecular parameters obtained for the ground
states for all complexes are given in
Table~\ref{tab:halogen:molecular-constants}.
\begin{table*}
   \centering
   \scriptsize
   \begin{minipage}{\linewidth}
      \begin{tabular}{lccccccc}
         \hline\hline
         species & $\nu_0$ (\cm) & $B$ \footnotemark[1] (\cm)
         & $D$ ($10^{-4}\,\cm$) & $\mu$ (D)
         & $a\Lambda+(b+c)\Sigma$ (\cm) & isotope splitting \footnotemark[2] (\cm)
         & $B_\text{gas}/B_\text{He}$\footnotemark[3] \\
         \hline
         $^{35}$Cl-HF & 3887.54 & 0.055  & 1.2  & 1.9  & 0.005 & 0.038 & 2.3 \\
         $^{79}$Br-HF & 3867.46 & 0.043  & 0.95 & 2.1  & 0.045 & 0.010 & 2.2 \\
         I-HF         & 3847.82 & 0.037  & 0.3  & 2.2  & 0.035 & --    & 2.2 \\
         Br-HCN       & 3284.61 & 0.0151 & 1.5  &      & 0.04  &\footnotemark[4]       & 2.7 \\
         I-HCN        & 3277.79 & 0.0120 & 1.0  &      & 0.04  & --    & 2.83 \\
         HCN-Cl       & 3309.33 & 0.032  & 0.50 & 3.0  & --    & 0.001\footnotemark[4] & 2.7 \\
         HCN-Br       & 3309.55 & 0.019  & 0.12 & 3.78 & --    & \footnotemark[4]      & 2.75\\
         HCN-I        & 3309.37 & 0.016  & 0.10 & (3.91)\footnotemark[5]
         & --         & --      & 2.75 \\
         \hline\hline
      \end{tabular}
      \footnotetext[1]{From references~\onlinecite{Merritt:PCCP7:67,
            Merritt:PCCP:inprep}}%
      \footnotetext[2]{The frequency of the heavier complex subtracted from the
         corresponding one of the lighter isotopologue.}%
      \footnotetext[3]{Gas-phase values are from \abinitio{}
         calculations~\cite{Merritt:PCCP7:67, Merritt:PCCP:inprep}. For the
         HCN-X complexes the gas-phase values are $B_0$ values from
         one-dimensionally spin-orbit corrected adiabatic
         PES~\cite{Merritt:PCCP:inprep}.}%
      \footnotetext[4]{No isotope splitting is observed experimentally, which is
         consistent with bound state calculations on an one-dimensional
         \abinitio{} potential energy surface for HCN-Cl; splittings for the
         bromine complexes are expected to be even
         smaller~\cite{Merritt:PCCP:inprep}.}%
      \footnotetext[5]{\abinitio{} value from
         reference~\onlinecite{Merritt:PCCP:inprep}}%
   \end{minipage}
   \caption{Ground states molecular parameters from infrared spectroscopy for
      all complexes of HF and HCN with halogen atoms observed so
      far~\cite{Merritt:PCCP7:67, Merritt:PCCP:inprep}. For chlorine and bromine
      the constants of the lighter isotopologue are given (iodine has only a
      single isotope) and the isotope splittings are the frequency of the
      lighter isotopologue subtracted from the frequency of the heavier one.} 
   \label{tab:halogen:molecular-constants}
\end{table*}
In Figure~\ref{fig:halogen:HCN-Br-spectrum:field-free} the field-free spectrum
of the nitrogen-bound HCN-Br complex is shown for a comparison and the molecular
constants of all corresponding HCN-X complexes are also given in
Table~\ref{tab:halogen:molecular-constants}.
These nitrogen-bound complexes exhibit quite different features than the
hydrogen bound complexes. Whereas the $\nu_\text{CH}$ vibrational frequencies
are shifted by tens of wavenumbers for the hydrogen bound complexes, these
N-bound complexes are shifted only by a few \cm{} compared to the free HCN
molecule. The rotationally resolved spectra of the HCN-Br and HCN-I complexes
also show a very different fine-structure, which is assigned to parity-splitting
due to the spin-orbit interaction between the $^2\Sigma_{1/2}$ and the
$^2\Pi_{1/2}$ states of the complex~\cite{Merritt:PCCP:inprep,
   Merritt:thesis:2006}. For the HCN-Cl complex no such fine-structure could be
observed, what is consistent with the much smaller spin-orbit coupling constant
of atomic chlorine than for bromine and iodine an the correspondingly small
splittings of the transitions that are not visible at the resolution of our
experiment.

It had been proposed that Cl-HCl complexes may posses T-shaped structures, but
display several of the properties of a linear open-shell molecule
\cite{Zeimen:JPCA107:5110}. Given the frequency shift from the HF monomer,
however, this is unlikely the case for the X-HF complexes observed here; see
reference~\onlinecite{Merritt:PCCP7:67} for a detailed discussion. Nevertheless,
the fact that this point is subject of considerable debate illustrates how
poorly we still understand these species, calling out for more experiments and
theory on these entrance channel complexes.

No complexes of atomic fluorine with HF or HCN have been observed, despite
extensive searches for such complexes and a previous assignment for the
$\nu_\text{HF}$ stretching vibrations of F-HF in argon matrices
\cite{Hunt:JPC92:3769}. Moreover, from \abinitio\ calculations it is clear, that
the reaction of F + HCN to form HFCN via an insertion mechanism is quite
exothermic and exhibits only a small barrier, if at all
\cite{Merritt:PCCP:inprep, Merritt:thesis:2006}. The fact that the fluorine
containing F-HCN complexes cannot be observed, could indeed be an indication
that the insertion reaction takes place, even at 0.37~K. This reasoning is also
supported by argon matrix work where the insertion product has been observed,
whereas no pre-reactive (van der Waals) complexes were found
\cite{Hunt:IC26:3051, Goldschleger:MC2001:43}. Guided by the experimental
vibrational frequency of 3018~\cm\ for HFCN in solid argon, we also searched for
the HFCN reaction product in the helium droplets. However, since this frequency
is at the edge of the F-center laser tuning range and the transition strength is
about 50 times weaker than for the HCN monomer, no manifestations of the HFCN
product could be observed. Revisiting the experiment using newer technology OPO
lasers~\cite{Schneider:OL22:1293, Herpen:OL27:640, Merritt:thesis:2006}, which
have enhanced tuning ranges and improved power, might show the corresponding
spectral signatures.

Similar studies to the ones described here have also been performed on the
pre-reactive complexes of open-shell metal atoms (Al, Ga, In)
\cite{Merritt:thesis:2006} and semiconductor atoms (Ge) with HCN and preliminary
results were presented in an earlier review~\cite{Choi:IRPC25:15}. Upon complex
formation in liquid helium droplets the same linear van der Waals complexes as
for the halogen atom HCN complexes are observed. However, these atom-HCN
complexes can undergo an exothermic insertion reaction and the corresponding
barriers are low enough that the reactions do actually proceed even at the low
ambient temperatures of the helium droplet, depending on the metal, either
directly or initiated by a single-quantum excitation of the $\nu_\text{CH}$
stretching vibration in the HCN moiety. Details on the van der Waals complexes
and the insertion products will be published elsewhere
\cite{Merritt:inprep:HCN-Ga}. Furthermore, the complexes of HCN with several
alkaline (Na, K, Rb, Cs) and earth alkaline (Mg, Ca, Sr) atoms
\cite{Douberly:metals, Douberly:thesis:2006}, with Zn atoms
\cite{Stiles:JPCA110:5620}, and with Cu and Au atoms have been observed. The
complexes of one to six magnesium atoms with several small molecules have been
studied extensively~\cite{Nauta:Science292:481, Stiles:JCP121:3130,
   Moore:JPCA108:9908, Dong:JPCA108:2181}. Due to the electronic quasi-closed
shell structure of magnesium and high barriers to reaction in these systems,
these complexes exhibit the simple spectra of closed shell van der Waals
complexes. For the heavier atoms from these series it is known that the atom
resides on the surface of the droplet, forming a dimple~\cite[and references
therein]{Stienkemeier:JPB:R127}. It is interesting to study the effects of
embedding an atom that wants to stay on the surface of the droplet, and a
molecule that wants to be inside the droplet, which, in addition, strongly
attract each other. For Na-HCN, one of these systems, a very strong absorption
line is observed~\cite{Douberly:metals, Douberly:thesis:2006}. The remaining
rotational structure, however, is the spectrum of a linear or spherical top with
a very much larger inertial moment than that expected for a Na-HCN binary
complex, even inside a helium droplet. The spectrum is actually consistent with
the rotation of the Na-HCN mass around the droplet and the observed rotational
constant scales nicely with the droplet radius. Therefore, it is concluded, that
the sodium pulls the HCN into a dimple on the surface of the droplet and the
complex can freely rotate over the surface of the helium droplet
\cite{Douberly:metals, Douberly:thesis:2006}. However, no rotational
fine-structure can be observed due to the unresolved features in the extremely
small $B$ spectra. The spectra of the heavier atoms (K, Rb, Cs, Ca, Sr) show
similar effects; for some of them even both species can be observed
simultaneously -- an embedded van der Waals complex as for Mg-HCN and a
surface-bound complex as for Na-HCN~\cite{Douberly:metals,
   Douberly:thesis:2006}.

\subsection{Complexes of hydrocarbon radicals}
\label{sec:complex:hydrocarbon}

\subsubsection{Complexes of methyl radical with HF and HCN}

The reactions of hydrocarbons are of particular interest due to the wide range
of reactions for which they play a key role, for example combustion processes
\cite{Warnatz:Combustion}. The hydrogen exchange reaction with fluorine atoms
form a basis of our understanding of much larger systems. The most simple
fluorine + hydrocarbon system, $\text{F} + \text{CH}_4 \rightarrow \text{HF} +
\text{CH}_3$, has been extensively studied both experimentally
\cite{Shiu:PRL92:103201} and theoretically~\cite{Chu:CPL424:243}, however,
high-resolution spectroscopic studies of the entrance and exit channel regions
of the PES are still lacking. Previous studies have identified the CH$_3$-HF
complex in argon matrices~\cite{Jacox:CP42:133, Johnson:JACS102:5736,
   Misochko:JACS117:11997, Misochko:JCP106:3146}.

In order to probe the exit channel region of the PES, methyl radicals have been
produced via pyrolysis, while a second pick-up cell was used to dope the droplet
with an HF molecule. Azomethane, di-tert-butyl peroxide (DTBP), and CH$_3$I were
all used as precursors to confirm that the signals genuinely arose from CH$_3$.
In addition, several transitions of the CH$_3$ monomer were also observed in
helium.
\begin{figure}
   \centering%
   \includegraphics[width=\floatwidth]{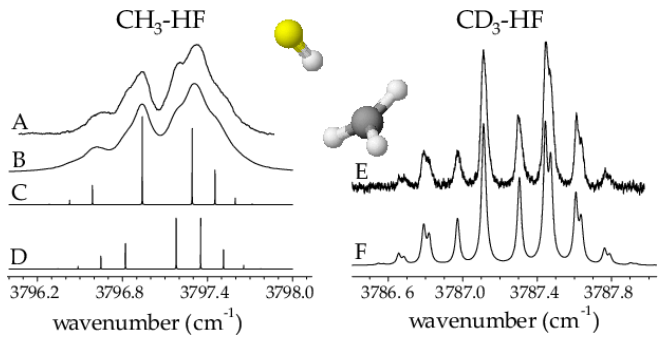}%
   \caption{(A) An experimental spectrum corresponding to the HF stretching
      vibration of the CH$_3$-HF molecular complex, along with a symmetric top
      simulation (B) which clearly shows the C$_{3v}$ symmetry of this complex.
      In (C) and (D) we plot the contributions from the K=1 and K=0 bands
      respectively, illustrating that the nuclear spin statistics prevent
      interconversion of these two manifolds of states. Scan (E) shows the
      corresponding spectrum for CD$_3$ radicals, illustrating a dramatic
      increase in the lifetime of the complex and (F) the corresponding
      simulation.}%
   \label{fig:CH3-HF}%
\end{figure}
Figure~\ref{fig:CH3-HF}\,(A) shows a partially rotationally resolved
experimental spectrum which is assigned to the HF stretching vibration of the
CH$_3$-HF molecular complex. The C$_{3v}$ symmetry of the complex can be
uniquely identified from the large Q branch present in the spectrum. Indeed, at
the temperature of the droplets, we would not normally expect population of the
excited K states since the rotation about the $a$-axis only moves the three
hydrogen atoms, giving rise to a large energy spacing between different $K$
states. Assuming nuclear spin conversion is slow compared with the timescale of
the experiment, the nuclear spin statistics of the ro-vibrational wavefunction
treat the $K=0,3,6\ldots$ and $K=1,2,4,5,7\ldots$ states as separate identities
of $A$ and $E$ symmetry, respectively, and thus they do not interconvert upon
cooling to 0.37~K. At the temperature of the droplets, all of the population is
cooled into the $K=0$ and $K=1$ states, and only for the $K=1$ states
$\Delta{}J=0$ transitions are allowed. The intensities reflect the population of
$A:E$ states of CH$_3$ in the high-temperature source, yielding a $\sim1:1$
population ration for the K=0 and K=1 bands.

The linewidths in the CH$_3$-HF spectrum are rather broad, somewhat limiting the
accuracy of the molecular parameters determined by the fit to the spectrum,
shown in Figure~\ref{fig:CH3-HF}\,(B)--(D). Considering the cause of the
broadening we concluded that a vibration-vibration resonance was the likely
cause of the shortening of the excited state lifetime, coupling the HF stretch
excited state to the CH$_3$ stretches of the methyl radical. To test this idea,
we also recorded the spectrum of CD$_3$-HF, which is shown in
Figure~\ref{fig:CH3-HF}\,(E).
\begin{figure}
   \centering%
   \includegraphics[width=0.5\textwidth]{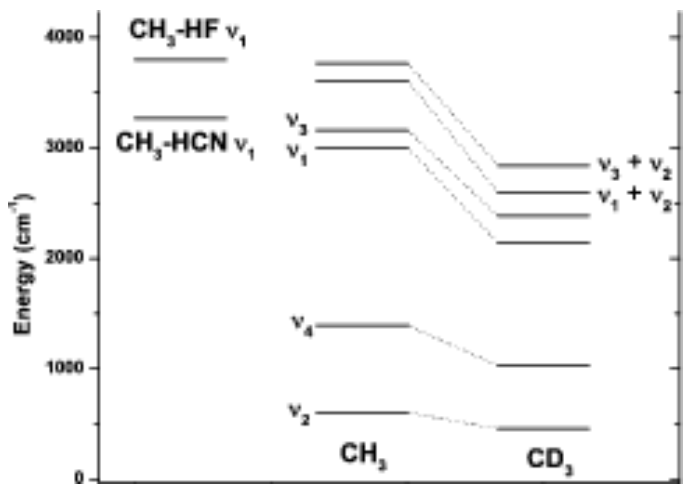}%
   \caption{A plot of the corresponding energy levels of the CH$_3$ and CD$_3$
      monomers. The effect of deuteration in the CD$_3$-HF complex is to lower
      the vibrational frequencies of the CH stretches, effectively quenching the
      vibration-vibration resonance that occurs with the excited HF stretching
      state of the complex. See text for details.}%
   \label{fig:CH3-CD3-energies}%
\end{figure}
CD$_3$ radicals were produced by pyrolysis of CD$_3$I. Deuteration of CH$_3$
lowers the frequencies of the vibrations as shown in
Figure~\ref{fig:CH3-CD3-energies}, detuning the resonance, and resulting in much
narrower lines, as expected. The molecular parameters derived from the fits to
the spectra are summarized in Table~\ref{tab:hydrocarbon-constants}.
\begin{table*}
   \centering%
   \begin{tabular}{lcccccc}
      \hline\hline
      species & $\nu_0$ (\cm) & $A$ (\cm) & $\Delta{}A$ (\cm)
      & $(B+C)/2$ (\cm) & $D_J$ (\cm) & $\mu$ (D) \\
      \hline
      CH$_3$-HF      & 3797.00 & & 0.06\footnotemark[1]  & 0.09  & $2.5\cdot10^{-4}$ & 2.6 \\
      CD$_3$-HF      & 3787.14 & & 0.027\footnotemark[1] & 0.083 & $2.6\cdot10^{-4}$ & 2.6 \\
      C$_2$H$_5$-HF  & 3774.45 & 0.30  & & 0.059 & $4.8\cdot10^{-5}$ & 2.7 \\
      C$_3$H$_5$-HF  & 3810.10 & 0.095 & & 0.040 & $1.0\cdot10^{-4}$ & 2.4 \\
      CH$_3$-HCN     & 3265.70 & & 0.04\footnotemark[1]  & 0.030 & $3.7\cdot10^{-5}$ & 3.1 \\
      CD$_3$-HCN     & 3262.09 & & 0.018\footnotemark[1] & 0.027 & $2.6\cdot10^{-5}$ & 3.1 \\
      C$_2$H$_5$-HCN & 3260.29 & 0.30  & & 0.15  & $2.9\cdot10^{-5}$ & 4.1 \\
      C$_3$H$_5$-HCN & 3260.14 & 0.09  & & 0.016 & $8.0\cdot10^{-6}$ & 3.2 \\
      \hline\hline
   \end{tabular}%
   \footnotetext[1]{These parameters are taken from gas-phase
      work~\cite{Jacox:CP42:133, Johnson:JACS102:5736}.} 
   \caption{Ground-state experimental molecular constants of
      hydrocarbon-HF and -HCN complexes. Comparison to \abinitio{} inertial
      parameters shows reduction factors for the rotational constants like
      closed shell systems of the same size~\cite{Choi:IRPC25:15}.}
   \label{tab:hydrocarbon-constants}%
\end{table*}

In addition to the importance of the long-range van der Waals forces in giving
the reactants a torque, which may be towards or away from the transition state
\cite{Weck:IRPC25:283, Werner:CPL328:500, Werner:CPL313:647,
   Skouteris:Science286:1713, Balakrishnan:JCP121:5563}, the stabilization of
weakly bound reactive complexes may represent an ideal starting point for the
study of reactive resonances~\cite{Qiu:Science311:1440}. The experimental
observation of reactive, or Feshbach, resonances in chemical reactions are
highly sought after because they result from purely quantum mechanical effects
and probe regions of the PES which are typically not sampled by current
experimental techniques. Such resonances result from a coupling of the
vibrationally adiabatic PES near the transition state, giving rise to dynamical
trapping. To date, reactive resonances have been confirmed in the crossed
molecular beam reactive scattering of
$\text{F}+\text{H}_2\rightarrow\text{HF}+\text{H}$ \cite{Qiu:Science311:1440}
and the photo-electron spectrum of $\text{IHI}^-$ \cite{Liu:CPL332:65}.
Resonances of this type are not expected to be constrained to such simple
systems, it is only our ability to detect or recognize them, which has limited
current experiments to such simple systems. Only recently has a reactive
resonance been implicated in a true polyatomic reaction, namely $\text{F} +
\text{CH}_4 \rightarrow \text{HF} + \text{CH}_3$~\cite{Shiu:PRL92:103201}. At
low collision energies the transient resonance in this reaction can proceed via
predissociation into $\text{HF}(v=2,j')+\text{CH}_3(v=0)$, similar to F + HD
\cite{Qiu:Science311:1440}. The extra degrees of freedom of the CH$_3$, however,
may allow intramolecular vibrational energy relaxation (IVR) which would shorten
the resonance lifetime, allowing the $\text{HF}(v=2) + \text{CH}_3(v_1=1)$
channel to compete effectively with the pre-dissociative decay described above.
At higher collision energies the decay of the resonance state into
$\text{HF}(v=3) + \text{CH}_3(v=0)$ dominates~\cite{Shiu:PRL92:103201}.

From our experiments on the CH$_3$-HF and CD$_3$-HF complexes we can confirm a
similar coupling between the $\text{HF}(v=1)+\text{CH}_3(v_1=0)$ and
$\text{HF}(v=0)+\text{CH}_3(v_1=1)$ states, as indicated by the
vibration-vibration resonance lifetime, which is considerably shortened in the
case of CD$_3$-HF. Overtone pumping of the CH$_3$-HF van der Waals complex may
allow access to the region of the PES directly involved in the reactive
resonance, as proposed for $\text{F} + \text{H}_2$~\cite{Qiu:Science311:1440}.

In order to further experimentally explore the V-V coupling mechanism observed
for the CH$_3$-HF complex, one can also alter the energy of the bright state by
changing the identity of the chromophore. Rotationally resolved infrared spectra
have been recorded for the CH$_3$-HCN and CD$_3$-HCN complexes
\cite{Rudic:JCP124:104305}. As seen from the energy level diagram in
Figure~\ref{fig:CH3-CD3-energies}, the CH stretching frequency of HCN lies below
the combination levels of the CH$_3$ monomer, implicated in the relaxation of
CH$_3$-HF, so a more direct comparison of the coupling can be made with just the
$\nu_1$ and $\nu_3$ levels of CH$_3$. Interestingly for CD$_3$-HCN, the
combination band for CD$_3$ is a viable relaxation channel. The peaks in the
CH$_3$-HCN spectrum are well fit by a Lorentzian lineshape, just as in
CH$_3$-HF, suggesting that the broadening results from the finite lifetime of
the excited state. Comparing the observed linewidths for CH$_3$-HCN and
CH$_3$-HF we find that the lifetime of CH$_3$-HCN is approximately two times
longer than for CH$_3$-HF, suggesting that the coupling to the combination band
is indeed the important relaxation channel for CH$_3$-HF. The CD$_3$-HCN
molecular complex exhibited linewidths which were only slightly smaller than for
CH$_3$-HCN, clearly showing that the effect of coupling to the CH stretches of
the CH$_3$ directly is much smaller.

To our knowledge no studies have investigated the possibility that combination
modes are excited in the F + CH$_4$ reaction, other than pure overtones. Based
on our observed coupling of the $\text{HF}(v=1) + \text{CH}_3(v=0)$ to
$\text{HF}(v=0) + \text{CH}_3(v_1=1\;\text{or}\;v_3=1+v_2=1)$ it may be possible
to observe this reaction channel in an crossed molecular beam reactive
scattering experiment similar to the ones reported by the group of Kopin
Liu~\cite{Shiu:PRL92:103201,Liu:ARPC52:139}, where the experimental conditions
favor the production of methyl in the CH-stretch excited state, \ie\ near
threshold.

\subsubsection{Larger hydrocarbon radicals}

\begin{figure}
   \centering%
   \includegraphics[width=\floatwidth]{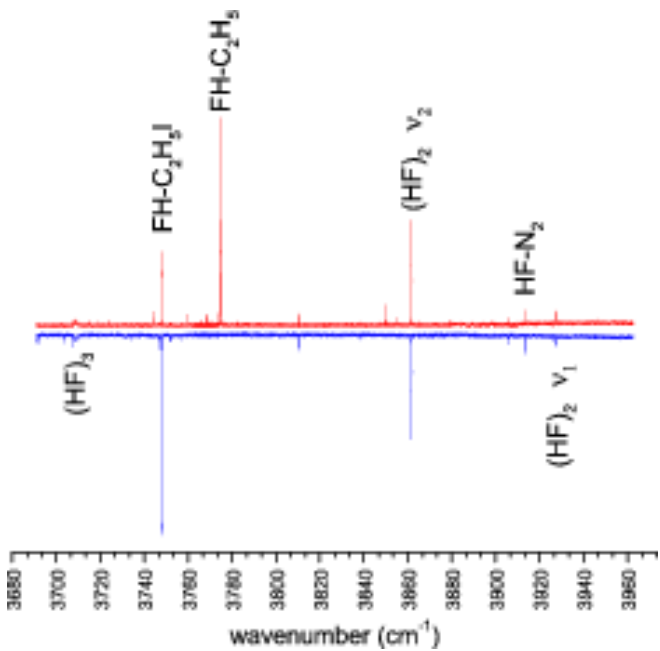}%
   \caption{Pendular survey scans corresponding to flowing ethyl iodide through
      a cold (downward going peaks) or hot (upward going peaks) pyrolysis
      source. HF is then picked up downstream in a separate pick-up cell. We
      assign the peak at 3747 cm$^{-1}$ to an HF-C$_2$H$_5$I complex, the
      intensity of which decreases substantially at temperatures appropriate for
      pyrolysis. At pyrolysis temperatures of about 1000 K, the peak at 3774
      cm$^{-1}$ appears which is assigned to the HF-Ethyl radical complex.}%
   \label{fig:HF-ethyl-survey}%
\end{figure}
\begin{figure}
   \centering
   \includegraphics[width=\floatwidth]{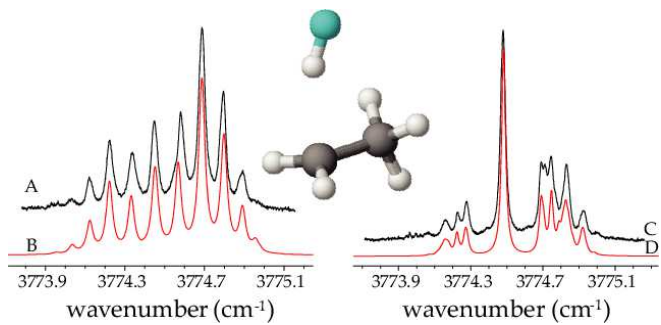}%
   \caption{Rotationally resolved field-free (A) and Stark (C) spectra
      corresponding to the HF stretching vibration of the HF-C$_2$H$_5$ complex
      and the corresponding simulations, (B) and (D), respectively.}%
   \label{fig:HF-ethyl}%
\end{figure}
\begin{figure}[t]
   \centering%
   \includegraphics[width=\floatwidth]{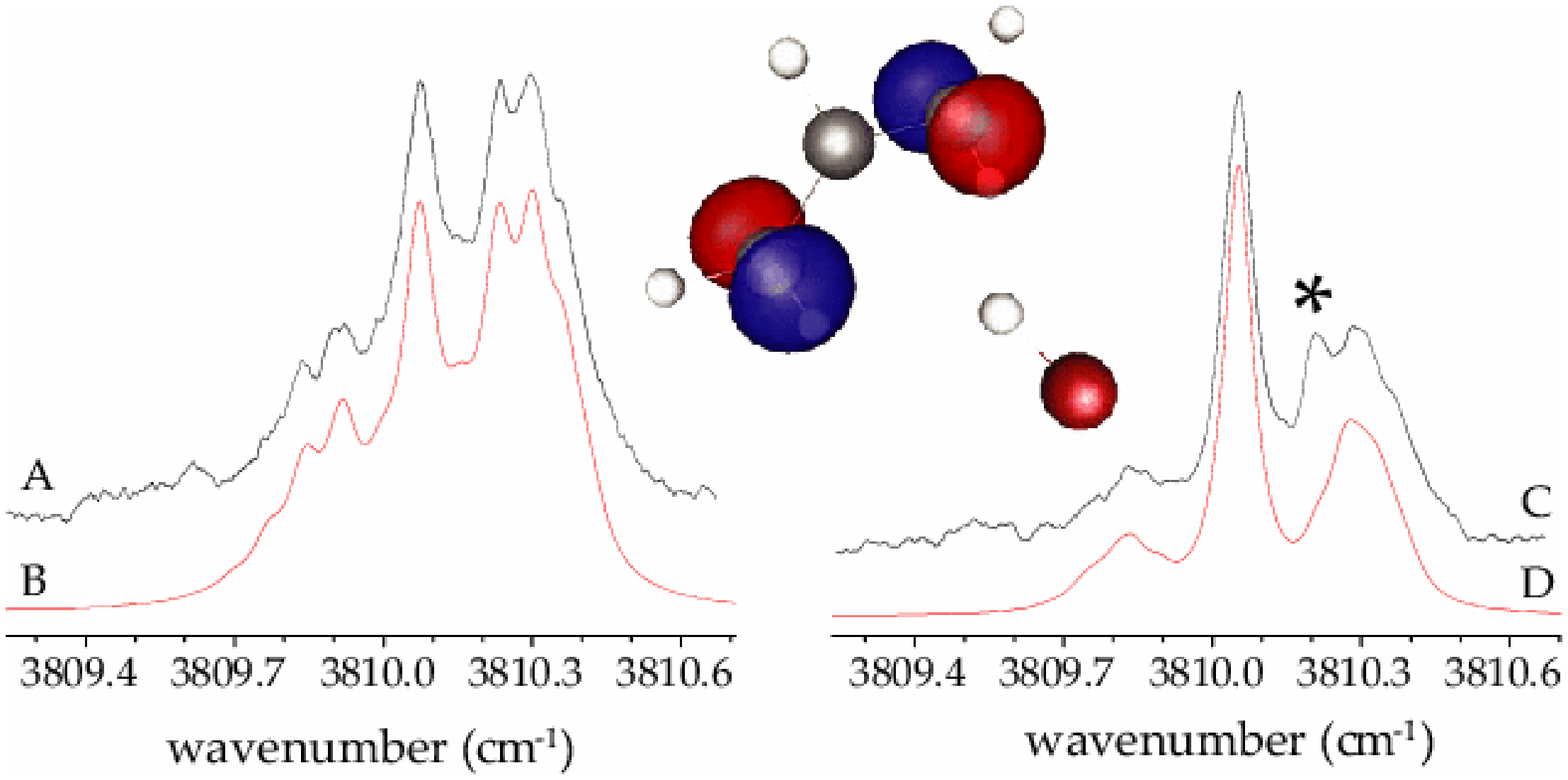}
   \caption{Field-free (A) and Stark (C) spectra for the HF-Allyl radical
      complex, and the corresponding simulations, (B) and (D), respectively.
      The peak marked with an asterisk corresponds to a known impurity which
      overlaps at this frequency.}
   \label{fig:HF-allyl}%
\end{figure}

Larger hydrocarbon radicals, including ethyl (C$_2$H$_5$) and allyl (C$_3$H$_5$)
radicals, may be readily formed by pyrolysis of their corresponding iodine
precursors~\cite{Kohn:RSI63:4003, Gilbert:JCP110:5485, Castiglioni:JPCA109:962,
   Blanksby:ACR36:255}, and spectra have been recorded for their complexes with
HCN and HF embedded in helium droplets. These systems are of interest since they
may be directly compared with the methyl radical complexes observed previously.
Figure~\ref{fig:HF-ethyl-survey} shows a pendular survey scan for the HF
stretching region. In the bottom trace C$_2$H$_5$I is flowed through a room
temperature pyrolysis source, and the peak at 3747 cm$^{-1}$ is assigned to a
C$_2$H$_5$I-HF complex. Upon heating the source, this peak is observed to
decrease in intensity, and a new strong peak is observed at 3774 cm$^{-1}$ which
we assign to a C$_2$H$_5$-HF complex. Turning off the electric field we recover
the rotationally resolved spectrum shown in Figure~\ref{fig:HF-ethyl}\,(A).
Figure~\ref{fig:HF-ethyl}\,C) shows the Stark spectrum for this band recorded at
an electric field strength of 3.86 kV cm$^{-1}$. The molecular parameters
determined from the fits to the spectra are given in
Table~\ref{tab:hydrocarbon-constants}, together with the values for the
ethyl-HCN complex. In Figure~\ref{fig:HF-allyl}\,(A) and (C) the field free
spectrum and the Stark spectrum of the allyl radical-HF complex are shown,
respectively. Still for a molecule of that size clear rotational structure is
observed and the obtained molecular parameters are summarized in
Table~\ref{tab:hydrocarbon-constants}, together with the values for the
allyl-HCN complex.
The results on these larger hydrocarbon radicals are very promising for studying
radicals that are even more important in radical chain reactions in combustion
processes, such as, for example, C$_n$H$_{2n+1}$OO peroxy-radicals, which are
some of the most important intermediates in these
reactions~\cite{Curran:CombustFlame114:149}. High resolution spectra of these
molecules have been difficult to obtain in free-jet expansions,
though~\cite{Nielsen:PeroxyRadicals:UV, Pushkarsky:JCP112:10695,
   Zalyubovsky:JPCA107:7704, Zalyubovsky:JPCA109:1308, Glover:JPCA109:11191,
   Tarczay:CPL406:81}.

Moreover, given that we have shown that pure effusive beams of hydrocarbon
radicals can be made under suitable conditions for helium droplet pick-up, the
possibility of studying radical-radical and radical-molecule reactions inside
the helium droplets is reasonable. A very interesting and simple experiment
would be to dope a single oxygen molecule into a droplet containing a
hydrocarbon radical. Theoretical calculations predict that the reaction to form
hydroperoxy radicals (ROO$^\bullet$) is barrierless
\cite{Rienstra-Kiracofe:JPCA104:9823}, however, at low temperatures this might
only apply if the oxygen directly approaches the radical center. Instead, the
helium might trap the complex in a local van der Waals minimum, forming a
pre-reactive complex, similar to what is discussed in
section~\ref{sec:high-energy}. Vibrational excitation could then be used to
initiate the corresponding reaction.

\subsection{NO-HF}
\label{sec:no-hf}

Given the recent observation that the $\Lambda$ doubling of NO was increased by
55~\% upon solvation in a helium droplet, there is considerable need to further
study this phenomenon, since exciting new molecules and clusters might only be
formed in helium, with no gas-phase data available for comparison. This
statement is already particularly true given our preliminary results on open
shell clusters with Br, I, Ga, and In atoms which have revealed large parity
splittings. Indeed, recent high-level theoretical calculations on the HCN-Br
complex indicate that the parity splitting is much smaller in a helium droplet
than one might expect in the gas-phase~\cite{Fishchuk:HCN-Br:inprep}. The
electronic angular momentum couplings, which are responsible for such open-shell
effects, are very sensitive probes of excited electronic states. Moreover, there
is relatively little information available so far on how the helium droplet
effects these.

To further our understanding of such open shell species in helium, we have
recorded the infrared spectrum of the NO-HF complex, which has been observed
previously in the gas-phase by Fawzy \etal~\cite{Fawzy:JCP93:2992}. While NO-HF
is a bent complex, we can use a diatomic, or more generally, a linear molecule
Hamiltonian to describe the $|\Omega|=1/2$ levels just as one can use a closed
shell linear rotor Hamiltonian for the $K=0$ levels of a symmetric top. If we
take $P$ as the projection of the total angular momentum on the $a$-axis of the
complex, then we are considering only the $|P|=1/2$ levels. Since NO-HF is only
slightly bent away from linearity, these excited $P$ states are of significantly
higher energy and are not populated at the rotational temperatures typical of
molecular beams or helium droplets. The observed spectrum in the gas phase is
dominated by pairs of transitions, reminiscent of $\Lambda$-type doubling. In
their original model, the rotational energies were obtained by fitting the
spectrum using
\begin{widetext}
   \begin{equation}
      F(J) = BJ(J+1)-DJ^2(J+1)^2
      \pm \frac{1}{2}[p(J+\frac{1}{2})-p_J(J+\frac{1}{2})J(J+1)]
   \end{equation}
\end{widetext}
where $J$ is half-integer, starting at $J=1/2$, and the $-$ and $+$ signs in
front of the square bracketed term are defined for the $e$ and $f$ parity
sub-states, respectively. $B$ and $p$ are the rotational and parity doubling
constants, respectively, and $D$ and $p_J$ are the corresponding centrifugal
distortion parameters. Fitting the helium droplet spectrum to the same model
Hamiltonian resulted in the molecular constants listed in
Table~\ref{tab:NO-HF-constants}, which are compared with the gas-phase
values~\cite{Fawzy:JCP93:2992}.
\begin{table}
   \centering
   \begin{tabular}{lll}
      \hline\hline
      constant (cm$^{-1}$) & helium droplet & gas-phase\\
      \hline
      $\nu_0$ & 3871.76 & 3877.47269(12)\\
      $B''$ & 0.0473 & 0.111320(17)\\
      $B'$ & 0.0479 & 0.116167(19)\\
      $p''$ & 0.0676 & 0.15274(19)\\
      $p'$ & 0.0802 & 0.19652(18)\\
      D$_J''$ & 1.25$\times$10$^{-4}$ & 2.56(31)$\times$10$^{-6}$\\
      D$_J'$ & 1.43$\times$10$^{-4}$ & 2.56(31)$\times$10$^{-6}$\\
      p$_J''$ & 3.8$\times$10$^{-4}$ & 0.932(54)$\times$10$^{-4}$\\
      p$_J'$ & 4.2$\times$10$^{-4}$ & 0.503(38)$\times$10$^{-4}$\\
      \hline\hline
   \end{tabular}
   \caption{Molecular constants of the NO-HF complex derived by fitting the
      experimentally observed spectrum for the HF stretching vibration in both
      helium droplets and in the gas-phase~\cite{Fawzy:JCP93:2992}. The fit to
      the field free spectrum was performed using the contour fitting routines
      in pgopher~\cite{Western:pgopher}. Note that in pgopher the signs of p$_J$
      must be reversed when compared to the origin model of
      Fawzy~\cite{Fawzy:JCP93:2992}.}
   \label{tab:NO-HF-constants}
\end{table}
In contrast to NO in helium, we find that for NO-HF, the parity constant $p$, is
reduced from the gas-phase value by exactly the same amount (within experimental
uncertainty) as the rotational constant $B$. As pointed out for NO
\cite{Haeften:PRL95:215301}, the expression for the parity splitting does have a
linear relationship with $B$, and this is exactly what we see for NO-HF. For NO
monomer, however, the parity splitting was increased by 55~\% despite the fact
that the rotational constant was 76~\% of the gas-phase value, illustrating that
the mechanism with which the helium interacts with the molecule is
different~\cite{Haeften:PRL95:215301}.

\section{Interactions between helium droplets and embedded molecules}
\label{sec:interactions}

Since the first spectroscopic investigation of SF$_6$ embedded in helium
droplets in 1992~\cite{Goyal:PRL69:933} and the first observation of free
rotation thereof a few years later~\cite{Hartmann:PRL75:1566}, a large number of
molecules and cluster systems have been studied and trends are now being
established. The most easily recognized and well publicized trend is the effect
of the helium droplet on the associated rotational constants of the embedded
cluster. Two dynamical regimes have been established. Light rotors with their
large rotational constants do not couple efficiently to the helium, resulting in
only a slight decrease in the observed rotational constant compared to the gas
phase. In contrast, heavier rotors typically exhibit a much more anisotropic
interaction with the helium, which allows some of the helium to follow the
rotational motion, thus adding to its moment of inertia. Typically rotational
constants of \emph{heavy} rotors ($B<1$~\cm) are reduced by a factor of
$2.5\pm0.5$ when compared to the gas-phase~\cite{Choi:IRPC25:15}. The scatter in
the reduction factors of the rotational constants results from the unique
dopant-helium interactions for the individual systems, and, for example, two
similar molecules such as CO$_2$ and N$_2$O, which have nearly the same
gas-phase rotational constant, have very different helium droplet rotational
constants due to the differences in the interaction
potentials~\cite{Nauta:JCP115:10254}. Since inside helium droplets metastable
structures can be formed, which may be impossible to be observed in the
gas-phase, there is certainly a desire to be able to extract quantitative bond
lengths and angles from the rotationally resolved spectra. Fully quantum
mechanical calculations have been able to reproduce the experimentally
determined rotational constants for a few prototype systems, which is certainly
a first step~\cite{Lee:PRL83:3812, Viel:JCP115:10186, Blinov:JCP120:5916}.
Despite the ambiguity in rotational constants, the overall symmetry of the
system is not affected, what can aid in structural determinations. Indeed the
nuclear spin statistical weights for propargyl and all CH$_3$-HX-type complexes
have confirmed the $C_{2v}$ and $C_{3v}$ symmetries of the isolated molecules,
respectively.

As can be seen from previous reviews on the subject, almost all of our knowledge
on the interaction of the droplet with a dopant has come from closed shell
molecules. From the admittedly small group of open-shell molecules studied,
which are all summarized in this review, we find that the interactions are
generally the same for radicals. The biggest exception would be the fact that
open shell alkali and the quasi closed shell heavier earth alkali metal atoms
are not solvated by the droplets and instead reside on the surface, which
generally results from a stronger He-He interaction than the corresponding
He-metal interaction. It was first unclear as to whether the interaction with an
open shell molecule would be favorable at all, \ie\ whether the radical would go
into the droplet, based on the known repulsive interaction between lone
electrons and helium droplets. For open shell atoms in non-$S$ states, \ie\
halogen atoms ($^2P$), the electrostatic interactions with the quadrupole moment
of the atom are quite strong, thus in general allowing for solvation.

Even if the species goes into the droplet, a localized unpaired electron may
cause density distortions of the surrounding helium, which could lower the
overall symmetry. Indeed such effects were postulated for NO, which would give
rise to a first order splitting of the two parity components. In that light the
different effects of the helium droplet environment on the parity splitting of
bare NO and the NO-HF complex show how such effects depend on details of the
molecular system and the molecule-helium interactions. Theoretical support is
urgently needed to describe and understand these detailed experimental results.

\section{Future directions}
\label{sec:future}

\subsection{High energy structures for chemical energy storage}
\label{sec:high-energy}

The possibilities to build metastable molecular cluster structures inside liquid
helium nano-droplets was demonstrated, for example, by observing a ring of six
water molecules~\cite{Nauta:Science287:293} or by the self-assembly of long
linear chains of HCN~\cite{Nauta:Science283:1895}. These chains are built up due
to the specific kinetic control of the complex formation inside the helium
droplets. Since individual molecules are picked up successively by the droplet
and are rapidly cooled inside with a rate of $\sim10^{16}$~K\,s$^{-1}$, they
approach each other isothermally at low temperature (0.4~K). Already at
relatively long distance the molecules are oriented in the field produced by the
dipole of the complexation partner. The ground states are high-field seeking and
therefore the molecular dipoles of both partners are oriented along the
intermolecular axis while they approach each other. Once they are close to each
other, they find themself in a linear, hydrogen bonded geometry. Due to the low
temperature they cannot rearrange to form the thermodynamically more stable
cyclic structures observed in free jets~\cite{Jucks:JCP88:2196,
   Anex:JPC92:2913}.

In the same way as for HCN it should be possible to assemble chains of polar
radicals in helium droplets, \eg\ chains of CN or OH radicals. Considering two
ground-state OH radicals approaching one another from long range at very low
temperatures, the hydrogen peroxide structure is the global minimum product.
Nevertheless, there is also a hydrogen bonded minimum at long range analogous to
the HF dimer~\cite{Howard:JCP81:5417, Quack:JCP95:28}. At long range the ground
state OH--OH interactions are essentially the same as for the HF dimer. Since
the length scale for the hydrogen-bond is very different from that of the
chemical well, it is expected that after reaching the hydrogen-bonded well there
is a barrier to further bond-compression, before the O--O distance is decreased
sufficiently to access the chemical bonding region. In systems with such
barriers the possibility exists for stabilizing the pre-reactive complex in the
hydrogen bonded well inside a helium droplet, in analogy to the physisorbed
state in surface science. The challenge is to remove the condensation energy
from the system as it forms, before it can surmount any barriers, thus trapping
it in the pre-reactive form. The natural extension to larger OH clusters
suggests that one might be able to make a whole new class of pre-reactive
radical solids.

The qualitative description given above is supported by extensive theoretical
studies of the hydrogen peroxide system. Particularly interesting is a
high-level theoretical study by Kuhn \etal, in which a minimum energy path
analysis shows how the bonding character changes from long range hydrogen
bonding to covalent bonding at decreased O--O distances~\cite{Kuhn:JCP111:2565}.
These two regions are separated by a significant barrier -- some hundred
wavenumbers for the highest quality potential -- that is presumably high enough
to exhibit bound states in the hydrogen bonded well. Moreover, the minimum
energy path from these calculations shows a cusp at the maximum of the barrier,
corresponding to a sudden change in the angular geometry of the complex at that
distance. This is due to the need for switching from a geometry with the
molecular dipoles oriented parallel (OH--OH) to an head-to-head structure
(HO--OH).

With this theoretical work supporting the existence of a barrier between the
hydrogen bonded and the chemical bonded minima, there is good reason to think
that the pre-reactive dimer, as well as larger nanoclusters, can be stabilized
by following the minimum energy approach path analogous to the production of HCN
chains~\cite{Nauta:Science283:1895}, cyclic water hexamer
\cite{Nauta:Science287:293}, and polymers of HF~\cite{Blume:JCP105:8666,
   Huisken:CPL245:319, Huisken:JCP103:5366, Douberly:JPCB107:4500}.
$(\text{H}_2)_n$-HF clusters, which are a good model for the OH/H$_2$ metastable
system, have also been studied in helium droplets~\cite{Moore:JCP118:9629,
   Moore:JCP119:4713, Moore:JPCA107:10805, Moore:JPCA108:1930}. These studies
show that there is considerable control on the number of molecules picked up by
the droplets, allowing the study of a wide range of stoichiometries. Generally,
supermolecular structures with large dipoles are favored by the formation
process. It has to be pointed out that all systems mentioned form different,
metastable, structures in liquid helium droplets, than obtained from gas-phase
nucleation. The implication is that liquid helium provides a new growth medium
for creating and stabilizing higher energy isomers of clusters that are not
normally observed in the gas-phase. It is also encouraging that different
clusters of quite large size using can be differentiated using infrared pendular
state spectroscopy \cite{Nauta:Science283:1895}.

To illustrate what might be achieved by producing analogous metastable
structures of radicals, consider the hydrogen-oxygen
reactions~\cite{CRC:HandbookChemPhys71, Strazisar:Science290:958,
   Smith:PCCP4:3543}:
%
%
\begin{align*}
   2\;\text{H}_2 + \text{O}_2 & \longrightarrow 2\;\text{H}_2\text{O}
   & \mathrm{\Delta_f H^\circ} & = \text{-484~kJ\,mol}^{-1} \\
   \text{H}_2 + \text{O}_2   & \longrightarrow \text{H}_2\text{O}_2
   & \mathrm{\Delta_f H^\circ} & = \text{-273~kJ\,mol}^{-1} \\
   \intertext{and for comparison, consider the following reactions:}
   2\;\text{OH} + \text{H}_2 & \longrightarrow 2\;\text{H}_2\text{O}
   & \mathrm{\Delta_f H^\circ} & = \text{-562~kJ\,mol}^{-1} \\
   2\;\text{OH} & \longrightarrow \text{H}_2\text{O}_2
   & \mathrm{\Delta_f H^\circ} & = \text{-351~kJ\,mol}^{-1}
\end{align*}
Thus, if solids of pure OH or with a OH/H$_2$ ration of 2:1 could be stabilized,
these would exhibit exothermicities that are considerably higher than that of
the hydrogen-oxygen system, which is widely used as fuel. As such they would be
ideal combustibles for high energy consumptions. From the discussion given
above, there is plenty of evidence to suggest this could be achieved, both in
nano-droplets and in bulk helium.

Other systems that could potentially provide an even larger amount of chemical
energy storage are nitrogen (oxygen) oligomers, which react very exothermally to
form stable N$_2$ (O$_2$) molecules. The N$_3$ and N$_4$ molecules have been
predicted~\cite{Bittererova:CPL347:220, Bittererova:JCP116:9740,
   Zhang:JCP122:014106} and observed experimentally~\cite{Cacace:Science295:480,
   Hansen:JPCA107_10608}. The cluster formation mechanism in helium droplets
described above should allow one to produce a wide variety of such metastable
N$_n$ systems from nitrogen atoms, where $n$ could be much larger than 3 and is
solely determined by the N-atom density in the pick-up region.

Gordon and co-workers have shown that similar radical solids can be stabilized
in bulk liquid helium.\footnote{It was recently proposed that similar effects
   should also be obtainable in atomic Bose-Einstein condensates, where ionic
   impurities could lead to the formation of \emph{mesoscopic molecular
      ions}~\cite{Cote:PRL89:093001}.} They directed a helium beam containing
atomic nitrogen into a dewar of liquid helium, to find that a highly energetic,
snow-like precipitate was formed~\cite{Gordon:CPL54:282, Gordon:CP170:411,
   Boltnev:CP189:367}. Upon warming that material thermally detonated. Electron
spin measurements have shown an unexpectedly high concentration of N atoms with
more than 10~\% of the N$_2$ concentration in the original experiments. Later
the experiments were optimized to obtain atomic concentrations as high as 50~\%
\cite{Gordon:CPL155:301}. Although a number of techniques have been used to
study the properties of these cryo-solids, detailed structural information is
still lacking. Nevertheless the authors point out that the liquid helium is only
necessary during the formation process and have shown that the resulting solids
are stable when the liquid helium is completely evaporated, as long as the
temperature remains below approximately 8~K~\cite{Gordon:CP170:411}. Most
remarkably the authors were able to use a plunger to compact the solid
precipitate into a pellet without causing it to react
\cite{Gordon:Plunger:2000}. They suggest that the method should be applicable to
a wide range of species and we suspect that it will be even more effective for
\emph{molecular} radicals, where there are steric considerations that also
inhibit recombination. A number of fascinating observations have been made by
this group, including the appearance of an \emph{intense green glow} when the
sample was irradiated with an helium neon laser. More recently they showed that
captured N$(^2D)$ atoms show thermoluminescence upon slight heating by 0.1~K or
when irradiating them using microwaves~\cite{Boltnev:OSU2000:WI13,
   Popov:OSU2000:WI14}. Details of the impurity condensation in liquid and solid
helium and the interpretation of their X-ray and IR spectroscopic studies have
recently been discussed \cite{Gordon:LTP30:756}.

Structural information on such materials is clearly desirable and can be
obtained from vibrational spectroscopy on small (nanoscale) samples of these
materials. Studies in liquid helium droplets will provide better control of the
growth, as well as the ability to use high resolution vibrational and rotational
spectroscopy to characterize the resulting structures. For smaller clusters
rotational resolution will be possible, providing detailed information on the
corresponding structures.

The nanosolids we are proposing to make in these studies are highly energetic
and it is interesting to consider what will happen when they are vibrationally
excited. If the vibrational energy deposited in the molecules is sufficient to
overcome the barriers that are responsible for the stability of the clusters,
and the energy becomes redistributed into the reaction coordinate by
intramolecular vibrational energy redistribution (IVR), it is expected to
observe strong responses to the laser excitation. This laser-induced detonation
(LID) will result in complete evaporation of the helium droplet resulting in
very strong depletion signals Alternatively, the clusters might dissipate the
vibrational energy to the helium so quickly that no reaction does occur and the
system simply cools back down by evaporation of liquid helium atoms. These
effects could be studied using IR-IR pump-probe experiments, similar to the ones
demonstrated for stable species embedded in helium droplets
\cite{Merritt:JCP121:1309, Douberly:PCCP7:463}. Considering the OH--OH system,
it is interesting to consider excitation of the OH--OH intermolecular
vibrations, \ie\ using combination modes of this vibration. This will put energy
into the reaction coordinate and may push the system over the barrier and into
the reactive well. Such tests of the stability of these solid materials are
important in accessing the usefulness of such materials as specialty fuels.
Understanding the growth of such new classes of high energy density materials in
liquid helium can be used to make and study the properties and structures of
these systems on the nanometer scale. What is exciting abut this approach is
that it has the potential to be scaled up to macroscopic sizes, as demonstrated
by the experiments of Gordon and co-workers~\cite{Boltnev:CP189:367,
   Gordon:CP170:411, Gordon:CPL155:301, Gordon:CPL54:282}.

Considering the Br-HCN and HCN-Br complexes described above, it seems reasonable
to allow for the formation of a metastable Br-HCN-Br complex as such a
high-energy structure. Whereas we were not able to find this HCN complex, first
experiments using the equivalent cyanoacetylene (HCCCN) chromophore reveal the
existence of such complexes.
\begin{figure}[b]
   \centering%
   \includegraphics[width=\floatwidth]{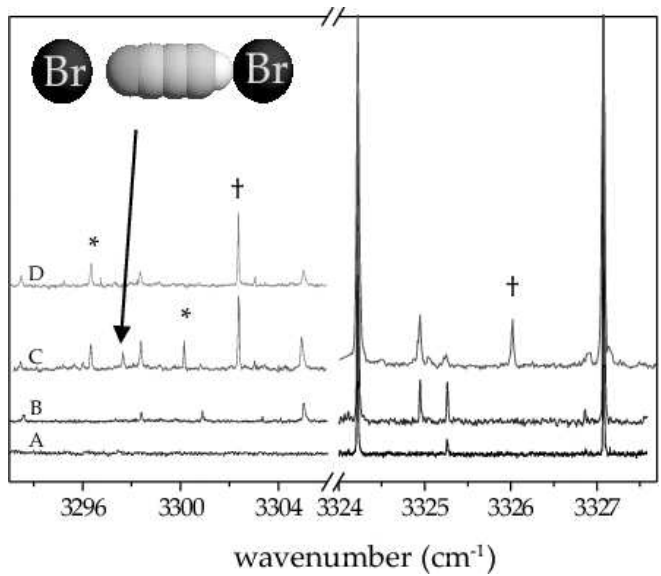}%
   \caption{Pendular survey scans for the Br + HCCCN system, where the bromine
      pressure is intentionally high to facilitate forming larger clusters. In
      (A) only HCCCN is picked up by the droplets and in (B) Br$_2$ is flowing
      through the cold pyrolysis source. Scans (C) and (D) were both recorded
      with the pyrolysis source at the appropriate temperature for bromine atom
      production. Whereas in (C) the HCCCN is picked up first, in (D) the
      bromine atoms are picked up first. See text for details.
   }
   \label{fig:br-hcccn:spectrum}%
\end{figure}
In Figure~\ref{fig:br-hcccn:spectrum} pendular survey scans for the
bromine-cyanoacetylene system are shown for different experimental conditions.
It is important to note that scans \ref{fig:br-hcccn:spectrum}\,(A) - (C) were
recorded by picking up the HCCCN first, but in \ref{fig:br-hcccn:spectrum}\,(D)
the order has been reversed, and we instead pick-up from the pyrolysis source
first. In scan~\ref{fig:br-hcccn:spectrum}\,(A), only HCCCN is added to the
droplets, while in \ref{fig:br-hcccn:spectrum}\,(B), Br$_{2}$ is flowing through
a room temperature pyrolysis source. The result of heating the pyrolysis to the
appropriate temperatures for bromine atom pick-up are shown in
scans~\ref{fig:br-hcccn:spectrum}\,(C) and (D). In good agreement with the
results of X + HCN, we find two peaks at 3326.1 and 3302.4~\cm, marked by a
$\dagger$, which we assign to the Br-HCCCN and HCCCN-Br complexes, respectively,
based on their frequency shifts and signal strengths. The peaks labeled with an
$*$ were found to optimize at higher HCCCN pressure, therefore they correspond
to complexes containing more than one HCCCN. The peak at 3297.65~\cm\ optimizes
at the same HCCCN pressure as the 1:1 complexes, however, it is found to
optimize at higher bromine pressure, suggesting that it is a complex containing
two bromine atoms~\cite{Merritt:PCCP:inprep, Merritt:thesis:2006}. Since it only
appears in the spectrum when the HCCCN is picked up first, it is a good
candidate for being Br-HCCCN-Br, a metastable van der Waals complex containing
two spatially separated bromine atoms. The dependence on pick-up order is to be
expected because two bromine atoms will likely recombine to form Br$_{2}$ in the
absence of the molecular spacer.

To aid in this preliminary assignment we performed bromine pressure dependence
measurements~\cite{Merritt:PCCP:inprep}. This new peak is found to optimize at
higher bromine pressures than those peaks assigned to the 1:1 complexes, and
agrees well with a simulation of a dimer using a Poisson distribution for the
pick-up statistics. Although we cannot rule out that a Br-HCCCN-Br$_{2}$, or
similar complex might have such a pick-up cell pressure dependence, a
double-resonance population transfer experiment~\cite{Douberly:PCCP7:463} in
which this new peak is pumped and the HCCCN-Br$_{2}$ complex is recovered would
be definitive.

It is also interesting to note, that the the original interest in stabilizing
spin-polarized hydrogen arose because of the same reasons, namely searching for
novel high-energy structures \cite{Etters:JCP62:313, Silvera:PRL44:164,
   Meyer:PRB50:9339}. The reaction $\text{H} + \text{H} \rightarrow \text{H}_2$
is exothermic by more than 400~kJ\,\cm, yielding an very large energy-to-mass
ratio if samples of atomic hydrogen could be stabilized. For two polarized
hydrogen atoms direct molecule formation is not possible, as their
$^3\Sigma_u^+$ potential is purely repulsive, and a spin-flip is necessary for
the binding energy of the hydrogen molecule to be released. Whereas the
preparation of dense samples of spin-polarized hydrogen for energy storage has
not been successful yet, these studies emerged into the quest for the
experimental realization of Bose-Einstein condensation and yielded a whole new
field in atomic and molecular physics \cite{Cornell:RMP74:875,
   Ketterle:RMP74:1131}. Nowadays similar ideas are also discussed in the
context of deep space propulsion systems based on antihydrogen
\cite{Nieto:JOB5:S547}.

\subsection{Other applications}
\label{sec:outlook:other}

Recently a new experiment for a sensitive search for the electric dipole moment
(EDM) of the electron using matrix-isolated dipolar radicals was proposed
\cite{Kozlov:arXiv:physics/0602111}. It was suggested to measure the
magnetization of an oriented sample of diatomic radicals embedded in a solid
noble-gas matrix. While the solid state matrix will provide a larger number
density of radicals, it is not clear to what degree molecular orientation can be
achieved, since neither the break-down voltages of the matrix nor the details of
the cage effects in the matrix sites are know
\cite{Kozlov:arXiv:physics/0602111}. As shown for several examples in this
review, pendular states with a high degree of orientation can easily be achieved
for polar molecules embedded in liquid helium droplet. Therefore, the techniques
of strong field orientation of radicals embedded in liquid helium droplets, as
described in this review, might provide a well-defined environment for sensitive
EDM tests.

The OH radical receives also considerable attention in the field of ultracold
polar molecules and it has been successfully confined in electrostatic traps for
times $>1$~s and at temperatures of $\sim50$~mK~\cite{Meerakker:PRL94:023004,
   Meerakker:ARPC57:159}, and there is considerable interest to cool such
trapped molecules further down to even lower temperatures. In that context the
interaction of two OH molecules at very long range has been considered and for
two rotational ground state OH radicals in their low-field seeking state,
so-called \emph{field-linked} states have been predicted: bound states of
long-range OH-OH dipole aligned dimers that are stabilized by an external
electric field on the order of 10~kV/cm~\cite{Avdeenkov:PRL90:043006}. Such
field linked states provide a similar degree of control over the chemical
reaction as the van der Waals-dimers described above. As long as the field is
on, they are safely separated by several nanometers in their long-range well,
but when the field is ramped down they can react --- assuming that, for example,
the exothermic reaction 2\,OH$\rightarrow$H$_2$O$+$O proceeds at ultracold
temperatures~\cite{Avdeenkov:PRL90:043006}. However, a gas of OH radicals in
low-field seeking states is subject to large collisional losses and is therefore
not suitable for evaporative cooling~\cite{Avdeenkov:PRA66:052718}. The
high-field seeking state is collisionally stable, being the lowest energy state
of OH, however, it does not show such long-range states and the collisional
parameters are strongly depending on the short-range potential and must be
determined by a ``suitable experiment''~\cite{Avdeenkov:PRL90:043006}.
Experiments on entrance-channel complexes in helium droplets, like the ones
described in this paper, would provide detailed information on this
PES.\footnote{In a similar approach groups from the field of ultracold atomic
   physics and helium droplet spectroscopy have successfully collaborated to
   obtain information on the PES and spectroscopy of mixed alkali
   dimers~\cite{Mudrich:EPJD31:291}.}

\subsection{Experimental improvements}
\label{sec:exp-improvement}

The main hurdle for studying transient species embedded in superfluid helium
droplets is the generation of the necessary clean samples of radicals. Albeit
the continuous pyrolysis source~\cite{Kuepper:JCP117:647} proves to be a very
successful and versatile tool for the production of a wide range of radicals,
the types of radicals that can be produced is still quite limited. It seems
impossible, for example, to use it to create clean sources of oxygen or nitrogen
atoms, as necessary for the experiments proposed above, as that would require to
thermally break the extremely strong $O_2$ or N$_2$ bonds.

Alternative techniques that have been used so successfully in free jet
experiments, such as photolysis or discharge sources, can not easily be coupled
to the continuous droplet beam setups, which up to date in the vast majority are
operated in a cw fashion. Once pulsed droplet beams mature they will provide
much larger droplet densities, what might allow to couple them to pulsed radical
sources. Nevertheless, the droplet property of embedding any species in the
scattering region requires extremely clean sources that would still need to be
developed, but the possibility of using photodissociation should help
considerably to break strong bonds.

It would be extremely interesting to perform time-resolved studies on the
processes of molecules embedded in helium droplets in order to study reactions
and relaxation dynamics in real time. Femtosecond pump-probe experiments on
alkali-doped helium droplets~\cite{Stienkemeier:PRL83:2320, Schulz:PRL87:153401,
   Droppelmann:PRL93:023402} show the feasibility of such experiments. In these
experiments the dynamics of the helium-alkali atom interaction could clearly be
observed. In a similar way it should be possible to observe relaxation dynamics
after vibrational excitation of molecules like HCN embedded \emph{inside} the
helium droplets. Once metastable radical complexes, as described above, can be
produced inside the helium droplets one could even photoinitiate their reaction
and follow the complete chemical dynamics from the educts to the final products
due to the confining environment of the droplets.

\section{Summary}
\label{sec:summary}

Superfluid helium droplets provide a novel medium for the study of transient and
reactive species. A continuous pyrolysis source to produce extremely clean,
effusive beams of radicals suitable for pick-up by helium droplet beams was
developed and the radicals embedded in helium droplets were studied using
high-resolution infrared spectroscopy. Using this setup, halogen atoms and
hydrocarbon radicals were produced and their van der Waals complexes with HF or
HCN were studied. In related studies NO and also a few other transient molecules
were studied as well. Many of the systems studied provide new, detailed
information on stationary points of PES that were previously only studied using
reactive scattering experiments. Even further information and the possibility
for optically controlled chemistry can be obtained using double-resonance
techniques.

The results from these studies also provide a very sensitive probe of the
quantum fluid the helium droplet environment provides, due to the influence of
the droplet not only on the inertial parameters, as for closed shell molecules,
but also on the electronic properties, as manifested in parity splitting or
$\Lambda$-doubling parameters.

The ability to design complexes of multiple, potentially different stable and
transient species inside these cryo-reactors promises a wide variety of
possibilities, from studying stationary points on the PES of these reactive
species, developing novel high-energy species, studying reaction dynamics under
cold and defined conditions, to optically induced chemical reactions.

However, the extraction of quantitative structural data from high-resolution
spectra of molecules embedded in helium droplets is still quite limited, due to
missing accurate models for the molecule-helium droplet interaction. However,
there are progresses in the theoretical descriptions of these effects and they
might once allow to determine accurate structural information, \ie\ geometric
rotational constants for the isolated molecular system from helium droplet
experiments.

\begin{acknowledgments}
   Most of the experimental work described herein was performed at the
   University of North Carolina at Chapel Hill in the group of Roger E.\ Miller,
   who deceased on 6.~November 2005. The authors are very grateful for the
   opportunity to work with him and for the stimulating environment and support
   he provided to his group.
   
   J.\,M.\,M.\ thanks Tom Baer for supporting him during the final stage of his
   graduate work. We thank Ad van der Avoird for providing the potential energy
   surfaces in Figure~\ref{fig:hcn-x:pes}.
   
   Financial support by NSF and AFOSR is acknowledged. J.\,K.\ gratefully
   acknowledges a Feodor Lynen fellowship of the Alexander von Humboldt
   Foundation and J.\,M.\,M.\ a scholarship by the Max Planck Society.
\end{acknowledgments}

\bibliography{string,mp}%

\begin{thebibliography}{244}
\expandafter\ifx\csname natexlab\endcsname\relax\def\natexlab#1{#1}\fi
\expandafter\ifx\csname bibnamefont\endcsname\relax
  \def\bibnamefont#1{#1}\fi
\expandafter\ifx\csname bibfnamefont\endcsname\relax
  \def\bibfnamefont#1{#1}\fi
\expandafter\ifx\csname citenamefont\endcsname\relax
  \def\citenamefont#1{#1}\fi
\expandafter\ifx\csname url\endcsname\relax
  \def\url#1{\texttt{#1}}\fi
\expandafter\ifx\csname urlprefix\endcsname\relax\def\urlprefix{URL }\fi
\providecommand{\bibinfo}[2]{#2}
\providecommand{\eprint}[2][]{\url{#2}}

\bibitem[{\citenamefont{Choi et~al.}(2006)\citenamefont{Choi, Douberly,
  Falconer, Lewis, Lindsay, Merritt, Stiles, and Miller}}]{Choi:IRPC25:15}
\bibinfo{author}{\bibfnamefont{M.~Y.} \bibnamefont{Choi}},
  \bibinfo{author}{\bibfnamefont{G.~E.} \bibnamefont{Douberly}},
  \bibinfo{author}{\bibfnamefont{T.~M.} \bibnamefont{Falconer}},
  \bibinfo{author}{\bibfnamefont{W.~K.} \bibnamefont{Lewis}},
  \bibinfo{author}{\bibfnamefont{C.~M.} \bibnamefont{Lindsay}},
  \bibinfo{author}{\bibfnamefont{J.~M.} \bibnamefont{Merritt}},
  \bibinfo{author}{\bibfnamefont{P.~L.} \bibnamefont{Stiles}},
  \bibnamefont{and} \bibinfo{author}{\bibfnamefont{R.~E.}
  \bibnamefont{Miller}}, \bibinfo{journal}{Int. Rev. Phys. Chem.}
  \textbf{\bibinfo{volume}{25}}, \bibinfo{pages}{15} (\bibinfo{year}{2006}).

\bibitem[{\citenamefont{Stienkemeier and
  Lehmann}(2006)}]{Stienkemeier:JPB:R127}
\bibinfo{author}{\bibfnamefont{F.}~\bibnamefont{Stienkemeier}}
  \bibnamefont{and} \bibinfo{author}{\bibfnamefont{K.~K.}
  \bibnamefont{Lehmann}}, \bibinfo{journal}{J. Phys. B}
  \textbf{\bibinfo{volume}{39}}, \bibinfo{pages}{R127} (\bibinfo{year}{2006}).

\bibitem[{\citenamefont{Toennies and Vilesov}(2004)}]{Toennies:ACIE43:2622}
\bibinfo{author}{\bibfnamefont{J.~P.} \bibnamefont{Toennies}} \bibnamefont{and}
  \bibinfo{author}{\bibfnamefont{A.~F.} \bibnamefont{Vilesov}},
  \bibinfo{journal}{Angew. Chem. Int. Ed.} \textbf{\bibinfo{volume}{43}},
  \bibinfo{pages}{2622} (\bibinfo{year}{2004}).

\bibitem[{\citenamefont{Makarov}(2004)}]{Makarov:PhysUspekhi47:217}
\bibinfo{author}{\bibfnamefont{G.~N.} \bibnamefont{Makarov}},
  \bibinfo{journal}{Physics-Uspekhi} \textbf{\bibinfo{volume}{47}},
  \bibinfo{pages}{217} (\bibinfo{year}{2004}).

\bibitem[{\citenamefont{Northby}(2001)}]{Northby:JCP115:10065}
\bibinfo{author}{\bibfnamefont{J.~A.} \bibnamefont{Northby}},
  \bibinfo{journal}{J. Chem. Phys.} \textbf{\bibinfo{volume}{115}},
  \bibinfo{pages}{10065} (\bibinfo{year}{2001}).

\bibitem[{\citenamefont{Callegari et~al.}(2001)\citenamefont{Callegari,
  Lehmann, Schmied, and Scoles}}]{Callegari:JCP115:10090}
\bibinfo{author}{\bibfnamefont{C.}~\bibnamefont{Callegari}},
  \bibinfo{author}{\bibfnamefont{K.~K.} \bibnamefont{Lehmann}},
  \bibinfo{author}{\bibfnamefont{R.}~\bibnamefont{Schmied}}, \bibnamefont{and}
  \bibinfo{author}{\bibfnamefont{G.}~\bibnamefont{Scoles}},
  \bibinfo{journal}{J. Chem. Phys.} \textbf{\bibinfo{volume}{115}},
  \bibinfo{pages}{10090} (\bibinfo{year}{2001}).

\bibitem[{\citenamefont{Stienkemeier and
  Vilesov}(2001)}]{Stienkemeier:JCP115:10119}
\bibinfo{author}{\bibfnamefont{F.}~\bibnamefont{Stienkemeier}}
  \bibnamefont{and} \bibinfo{author}{\bibfnamefont{A.~F.}
  \bibnamefont{Vilesov}}, \bibinfo{journal}{J. Chem. Phys.}
  \textbf{\bibinfo{volume}{115}}, \bibinfo{pages}{10119}
  (\bibinfo{year}{2001}).

\bibitem[{\citenamefont{Warnatz et~al.}(1996)\citenamefont{Warnatz, Maas, and
  Dibble}}]{Warnatz:Combustion}
\bibinfo{author}{\bibfnamefont{J.}~\bibnamefont{Warnatz}},
  \bibinfo{author}{\bibfnamefont{U.}~\bibnamefont{Maas}}, \bibnamefont{and}
  \bibinfo{author}{\bibfnamefont{W.}~\bibnamefont{Dibble}},
  \emph{\bibinfo{title}{Combustion: Physical and Chemical Fundamentals,
  Modeling and Simulation, Experiments: Pollutant Formation}}
  (\bibinfo{publisher}{Springer}, \bibinfo{address}{Berlin},
  \bibinfo{year}{1996}).

\bibitem[{\citenamefont{Wayne}(1991)}]{Wayne:ChemAtmosphere}
\bibinfo{author}{\bibfnamefont{R.~P.} \bibnamefont{Wayne}},
  \emph{\bibinfo{title}{Chemistry of Atmospheres: An introduction to the
  Chemistry of the Atmospheres of Earth, the Planets, and their satellites}}
  (\bibinfo{publisher}{Oxford University Press}, \bibinfo{address}{Oxford, GB},
  \bibinfo{year}{1991}), \bibinfo{edition}{2nd} ed.

\bibitem[{\citenamefont{Chastaing et~al.}(1999)\citenamefont{Chastaing, James,
  Sims, and Smith}}]{Chastaing:PCCP1:2247}
\bibinfo{author}{\bibfnamefont{D.}~\bibnamefont{Chastaing}},
  \bibinfo{author}{\bibfnamefont{P.~L.} \bibnamefont{James}},
  \bibinfo{author}{\bibfnamefont{I.~R.} \bibnamefont{Sims}}, \bibnamefont{and}
  \bibinfo{author}{\bibfnamefont{I.~W.~M.} \bibnamefont{Smith}},
  \bibinfo{journal}{Phys. Chem. Chem. Phys.} \textbf{\bibinfo{volume}{1}},
  \bibinfo{pages}{2247} (\bibinfo{year}{1999}).

\bibitem[{\citenamefont{Van't~Hoff}(1884)}]{Hoff:DynamiqueChimique214}
\bibinfo{author}{\bibfnamefont{J.~H.} \bibnamefont{Van't~Hoff}},
  \emph{\bibinfo{title}{\'Etudes de dynamique chimique}}
  (\bibinfo{publisher}{Muller}, \bibinfo{address}{Amsterdam},
  \bibinfo{year}{1884}).

\bibitem[{\citenamefont{Arrhenius}(1889)}]{Arrhenius:ZPC4:226}
\bibinfo{author}{\bibfnamefont{S.}~\bibnamefont{Arrhenius}},
  \bibinfo{journal}{Z. Phys. Chem.} \textbf{\bibinfo{volume}{4}},
  \bibinfo{pages}{226} (\bibinfo{year}{1889}).

\bibitem[{\citenamefont{London}(1929)}]{London:ZEAPC35:552}
\bibinfo{author}{\bibfnamefont{F.}~\bibnamefont{London}},
  \bibinfo{journal}{Zeitschrift f\"ur Elektrochemie und Angewandte
  Physikalische Chemie} \textbf{\bibinfo{volume}{35}}, \bibinfo{pages}{552}
  (\bibinfo{year}{1929}).

\bibitem[{\citenamefont{Eyring and Polanyi}(1931)}]{Eyring:ZPCB12:279}
\bibinfo{author}{\bibfnamefont{H.}~\bibnamefont{Eyring}} \bibnamefont{and}
  \bibinfo{author}{\bibfnamefont{M.}~\bibnamefont{Polanyi}},
  \bibinfo{journal}{Z. Phys. Chem. B} \textbf{\bibinfo{volume}{12}},
  \bibinfo{pages}{279} (\bibinfo{year}{1931}).

\bibitem[{\citenamefont{Hammond}(1955)}]{Hammond:JACS77:334}
\bibinfo{author}{\bibfnamefont{G.~S.} \bibnamefont{Hammond}},
  \bibinfo{journal}{J. Am. Chem. Soc.} \textbf{\bibinfo{volume}{77}},
  \bibinfo{pages}{334} (\bibinfo{year}{1955}).

\bibitem[{\citenamefont{Polanyi and Wong}(1969)}]{Polanyi:JCP51:1439}
\bibinfo{author}{\bibfnamefont{J.~C.} \bibnamefont{Polanyi}} \bibnamefont{and}
  \bibinfo{author}{\bibfnamefont{W.~H.} \bibnamefont{Wong}},
  \bibinfo{journal}{J. Chem. Phys.} \textbf{\bibinfo{volume}{51}},
  \bibinfo{pages}{1439} (\bibinfo{year}{1969}).

\bibitem[{\citenamefont{Qiu et~al.}(2006)\citenamefont{Qiu, Ren, Che, Dai,
  Harich, Wang, Yang, Xu, Xie, Gustafsson et~al.}}]{Qiu:Science311:1440}
\bibinfo{author}{\bibfnamefont{M.~H.} \bibnamefont{Qiu}},
  \bibinfo{author}{\bibfnamefont{Z.~F.} \bibnamefont{Ren}},
  \bibinfo{author}{\bibfnamefont{L.}~\bibnamefont{Che}},
  \bibinfo{author}{\bibfnamefont{D.~X.} \bibnamefont{Dai}},
  \bibinfo{author}{\bibfnamefont{S.~A.} \bibnamefont{Harich}},
  \bibinfo{author}{\bibfnamefont{X.~Y.} \bibnamefont{Wang}},
  \bibinfo{author}{\bibfnamefont{X.~M.} \bibnamefont{Yang}},
  \bibinfo{author}{\bibfnamefont{C.~X.} \bibnamefont{Xu}},
  \bibinfo{author}{\bibfnamefont{D.~Q.} \bibnamefont{Xie}},
  \bibinfo{author}{\bibfnamefont{M.}~\bibnamefont{Gustafsson}},
  \bibnamefont{et~al.}, \bibinfo{journal}{Science}
  \textbf{\bibinfo{volume}{311}}, \bibinfo{pages}{1440} (\bibinfo{year}{2006}).

\bibitem[{\citenamefont{Chao et~al.}(2002)\citenamefont{Chao, Harich, Dai,
  Wang, Yang, and Skodje}}]{DerChao:JCP117:8341}
\bibinfo{author}{\bibfnamefont{S.~D.} \bibnamefont{Chao}},
  \bibinfo{author}{\bibfnamefont{S.~A.} \bibnamefont{Harich}},
  \bibinfo{author}{\bibfnamefont{D.~X.} \bibnamefont{Dai}},
  \bibinfo{author}{\bibfnamefont{C.~C.} \bibnamefont{Wang}},
  \bibinfo{author}{\bibfnamefont{X.~M.} \bibnamefont{Yang}}, \bibnamefont{and}
  \bibinfo{author}{\bibfnamefont{R.~T.} \bibnamefont{Skodje}},
  \bibinfo{journal}{J. Chem. Phys.} \textbf{\bibinfo{volume}{117}},
  \bibinfo{pages}{8341} (\bibinfo{year}{2002}).

\bibitem[{\citenamefont{Zhang et~al.}(2006)\citenamefont{Zhang, Dai, Wang,
  Harich, Wang, Yang, Gustafsson, and Skodje}}]{Zhang:PRL96:093201}
\bibinfo{author}{\bibfnamefont{J.~Y.} \bibnamefont{Zhang}},
  \bibinfo{author}{\bibfnamefont{D.~X.} \bibnamefont{Dai}},
  \bibinfo{author}{\bibfnamefont{C.~C.} \bibnamefont{Wang}},
  \bibinfo{author}{\bibfnamefont{S.~A.} \bibnamefont{Harich}},
  \bibinfo{author}{\bibfnamefont{X.~Y.} \bibnamefont{Wang}},
  \bibinfo{author}{\bibfnamefont{X.~M.} \bibnamefont{Yang}},
  \bibinfo{author}{\bibfnamefont{M.}~\bibnamefont{Gustafsson}},
  \bibnamefont{and} \bibinfo{author}{\bibfnamefont{R.~T.}
  \bibnamefont{Skodje}}, \bibinfo{journal}{Phys. Rev. Lett.}
  \textbf{\bibinfo{volume}{96}}, \bibinfo{pages}{093201}
  (\bibinfo{year}{2006}).

\bibitem[{\citenamefont{Valentini}(2001)}]{Valentini:ARPC52:15}
\bibinfo{author}{\bibfnamefont{J.~J.} \bibnamefont{Valentini}},
  \bibinfo{journal}{Ann. Rev. Phys. Chem.} \textbf{\bibinfo{volume}{52}},
  \bibinfo{pages}{15} (\bibinfo{year}{2001}).

\bibitem[{\citenamefont{Parker and Bernstein}(1989)}]{Parker:ARPC40:561}
\bibinfo{author}{\bibfnamefont{D.~H.} \bibnamefont{Parker}} \bibnamefont{and}
  \bibinfo{author}{\bibfnamefont{R.~B.} \bibnamefont{Bernstein}},
  \bibinfo{journal}{Ann. Rev. Phys. Chem.} \textbf{\bibinfo{volume}{40}},
  \bibinfo{pages}{561} (\bibinfo{year}{1989}).

\bibitem[{\citenamefont{Loesch}(1995)}]{Loesch:ARPC46:555}
\bibinfo{author}{\bibfnamefont{H.~J.} \bibnamefont{Loesch}},
  \bibinfo{journal}{Ann. Rev. Phys. Chem.} \textbf{\bibinfo{volume}{46}},
  \bibinfo{pages}{555} (\bibinfo{year}{1995}).

\bibitem[{\citenamefont{OrrEwing}(1996)}]{OrrEwing:JCSFT92:881}
\bibinfo{author}{\bibfnamefont{A.~J.} \bibnamefont{OrrEwing}},
  \bibinfo{journal}{J. Chem. Soc. -- Faraday Trans.}
  \textbf{\bibinfo{volume}{92}}, \bibinfo{pages}{881} (\bibinfo{year}{1996}).

\bibitem[{\citenamefont{Casavecchia}(2000)}]{Casavecchia:RPP63:355}
\bibinfo{author}{\bibfnamefont{P.}~\bibnamefont{Casavecchia}},
  \bibinfo{journal}{Rep. Prog. Phys.} \textbf{\bibinfo{volume}{63}},
  \bibinfo{pages}{355} (\bibinfo{year}{2000}).

\bibitem[{\citenamefont{Bethlem et~al.}(2006)\citenamefont{Bethlem, Tarbutt,
  K\"upper, Carty, Wohlfart, Hinds, and Meijer}}]{Bethlem:JPB39:R263}
\bibinfo{author}{\bibfnamefont{H.~L.} \bibnamefont{Bethlem}},
  \bibinfo{author}{\bibfnamefont{M.~R.} \bibnamefont{Tarbutt}},
  \bibinfo{author}{\bibfnamefont{J.}~\bibnamefont{K\"upper}},
  \bibinfo{author}{\bibfnamefont{D.}~\bibnamefont{Carty}},
  \bibinfo{author}{\bibfnamefont{K.}~\bibnamefont{Wohlfart}},
  \bibinfo{author}{\bibfnamefont{E.~A.} \bibnamefont{Hinds}}, \bibnamefont{and}
  \bibinfo{author}{\bibfnamefont{G.}~\bibnamefont{Meijer}},
  \bibinfo{journal}{J. Phys. B} \textbf{\bibinfo{volume}{39}},
  \bibinfo{pages}{R263} (\bibinfo{year}{2006}).

\bibitem[{\citenamefont{Bethlem et~al.}(1999)\citenamefont{Bethlem, Berden, and
  Meijer}}]{Bethlem:PRL83:1558}
\bibinfo{author}{\bibfnamefont{H.~L.} \bibnamefont{Bethlem}},
  \bibinfo{author}{\bibfnamefont{G.}~\bibnamefont{Berden}}, \bibnamefont{and}
  \bibinfo{author}{\bibfnamefont{G.}~\bibnamefont{Meijer}},
  \bibinfo{journal}{Phys. Rev. Lett.} \textbf{\bibinfo{volume}{83}},
  \bibinfo{pages}{1558} (\bibinfo{year}{1999}).

\bibitem[{\citenamefont{Bethlem and Meijer}(2003)}]{Bethlem:IRPC22:73}
\bibinfo{author}{\bibfnamefont{H.~L.} \bibnamefont{Bethlem}} \bibnamefont{and}
  \bibinfo{author}{\bibfnamefont{G.}~\bibnamefont{Meijer}},
  \bibinfo{journal}{Int. Rev. Phys. Chem.} \textbf{\bibinfo{volume}{22}},
  \bibinfo{pages}{73} (\bibinfo{year}{2003}).

\bibitem[{\citenamefont{van~de Meerakker et~al.}(2006)\citenamefont{van~de
  Meerakker, Vanhaecke, and Meijer}}]{Meerakker:ARPC57:159}
\bibinfo{author}{\bibfnamefont{S.~Y.~T.} \bibnamefont{van~de Meerakker}},
  \bibinfo{author}{\bibfnamefont{N.}~\bibnamefont{Vanhaecke}},
  \bibnamefont{and} \bibinfo{author}{\bibfnamefont{G.}~\bibnamefont{Meijer}},
  \bibinfo{journal}{Ann. Rev. Phys. Chem.} \textbf{\bibinfo{volume}{57}},
  \bibinfo{pages}{159} (\bibinfo{year}{2006}).

\bibitem[{\citenamefont{Heiner et~al.}(2006)\citenamefont{Heiner, Bethlem, and
  Meijer}}]{Heiner:PCCP8:2666}
\bibinfo{author}{\bibfnamefont{C.~E.} \bibnamefont{Heiner}},
  \bibinfo{author}{\bibfnamefont{H.~L.} \bibnamefont{Bethlem}},
  \bibnamefont{and} \bibinfo{author}{\bibfnamefont{G.}~\bibnamefont{Meijer}},
  \bibinfo{journal}{Phys. Chem. Chem. Phys.} \textbf{\bibinfo{volume}{8}},
  \bibinfo{pages}{2666} (\bibinfo{year}{2006}).

\bibitem[{\citenamefont{Gilijamse et~al.}(2006)\citenamefont{Gilijamse,
  Hoekstra, van~de Meerakker, Groeneboom, and
  Meijer}}]{Gilijamse:Science:accepted}
\bibinfo{author}{\bibfnamefont{J.~J.} \bibnamefont{Gilijamse}},
  \bibinfo{author}{\bibfnamefont{S.}~\bibnamefont{Hoekstra}},
  \bibinfo{author}{\bibfnamefont{S.~Y.~T.} \bibnamefont{van~de Meerakker}},
  \bibinfo{author}{\bibfnamefont{G.~C.} \bibnamefont{Groeneboom}},
  \bibnamefont{and} \bibinfo{author}{\bibfnamefont{G.}~\bibnamefont{Meijer}},
  \bibinfo{journal}{Science}  (\bibinfo{year}{2006}), \bibinfo{note}{accepted
  for publication}.

\bibitem[{\citenamefont{Werner et~al.}(2000)\citenamefont{Werner, Bian,
  Menendez, Aoiz, Casavecchia, Cartechini, and Balucani}}]{Werner:CPL328:500}
\bibinfo{author}{\bibfnamefont{H.~J.} \bibnamefont{Werner}},
  \bibinfo{author}{\bibfnamefont{W.}~\bibnamefont{Bian}},
  \bibinfo{author}{\bibfnamefont{M.}~\bibnamefont{Menendez}},
  \bibinfo{author}{\bibfnamefont{F.~J.} \bibnamefont{Aoiz}},
  \bibinfo{author}{\bibfnamefont{P.}~\bibnamefont{Casavecchia}},
  \bibinfo{author}{\bibfnamefont{L.}~\bibnamefont{Cartechini}},
  \bibnamefont{and} \bibinfo{author}{\bibfnamefont{N.}~\bibnamefont{Balucani}},
  \bibinfo{journal}{Chem. Phys. Lett.} \textbf{\bibinfo{volume}{328}},
  \bibinfo{pages}{500} (\bibinfo{year}{2000}).

\bibitem[{\citenamefont{Balakrishnan}(2004)}]{Balakrishnan:JCP121:5563}
\bibinfo{author}{\bibfnamefont{N.}~\bibnamefont{Balakrishnan}},
  \bibinfo{journal}{J. Chem. Phys.} \textbf{\bibinfo{volume}{121}},
  \bibinfo{pages}{5563} (\bibinfo{year}{2004}).

\bibitem[{\citenamefont{Skouteris et~al.}(1999)\citenamefont{Skouteris,
  Manolopoulos, Bian, Werner, Lai, and Liu}}]{Skouteris:Science286:1713}
\bibinfo{author}{\bibfnamefont{D.}~\bibnamefont{Skouteris}},
  \bibinfo{author}{\bibfnamefont{D.~E.} \bibnamefont{Manolopoulos}},
  \bibinfo{author}{\bibfnamefont{W.}~\bibnamefont{Bian}},
  \bibinfo{author}{\bibfnamefont{H.~J.} \bibnamefont{Werner}},
  \bibinfo{author}{\bibfnamefont{L.~H.} \bibnamefont{Lai}}, \bibnamefont{and}
  \bibinfo{author}{\bibfnamefont{K.}~\bibnamefont{Liu}},
  \bibinfo{journal}{Science} \textbf{\bibinfo{volume}{286}},
  \bibinfo{pages}{1713} (\bibinfo{year}{1999}).

\bibitem[{\citenamefont{Werner et~al.}(1999)\citenamefont{Werner, Bian, and
  Manthe}}]{Werner:CPL313:647}
\bibinfo{author}{\bibfnamefont{H.~J.} \bibnamefont{Werner}},
  \bibinfo{author}{\bibfnamefont{W.}~\bibnamefont{Bian}}, \bibnamefont{and}
  \bibinfo{author}{\bibfnamefont{U.}~\bibnamefont{Manthe}},
  \bibinfo{journal}{Chem. Phys. Lett.} \textbf{\bibinfo{volume}{313}},
  \bibinfo{pages}{647} (\bibinfo{year}{1999}).

\bibitem[{\citenamefont{Weck and Balakrishnan}(2006)}]{Weck:IRPC25:283}
\bibinfo{author}{\bibfnamefont{P.~F.} \bibnamefont{Weck}} \bibnamefont{and}
  \bibinfo{author}{\bibfnamefont{N.}~\bibnamefont{Balakrishnan}},
  \bibinfo{journal}{Int. Rev. Phys. Chem.} \textbf{\bibinfo{volume}{25}},
  \bibinfo{pages}{283} (\bibinfo{year}{2006}).

\bibitem[{\citenamefont{Townsend et~al.}(2004)\citenamefont{Townsend, Lahankar,
  Lee, Chambreau, Suits, Zhang, Rheinecker, Harding, and
  Bowman}}]{Townsend:Science306:1158}
\bibinfo{author}{\bibfnamefont{D.}~\bibnamefont{Townsend}},
  \bibinfo{author}{\bibfnamefont{S.~A.} \bibnamefont{Lahankar}},
  \bibinfo{author}{\bibfnamefont{S.~K.} \bibnamefont{Lee}},
  \bibinfo{author}{\bibfnamefont{S.~D.} \bibnamefont{Chambreau}},
  \bibinfo{author}{\bibfnamefont{A.~G.} \bibnamefont{Suits}},
  \bibinfo{author}{\bibfnamefont{X.}~\bibnamefont{Zhang}},
  \bibinfo{author}{\bibfnamefont{J.}~\bibnamefont{Rheinecker}},
  \bibinfo{author}{\bibfnamefont{L.~B.} \bibnamefont{Harding}},
  \bibnamefont{and} \bibinfo{author}{\bibfnamefont{J.~M.}
  \bibnamefont{Bowman}}, \bibinfo{journal}{Science}
  \textbf{\bibinfo{volume}{306}}, \bibinfo{pages}{1158} (\bibinfo{year}{2004}).

\bibitem[{\citenamefont{Chambreau et~al.}(2006)\citenamefont{Chambreau,
  Townsend, Lahankar, Lee, and Suits}}]{Chambreau:PS73:C89}
\bibinfo{author}{\bibfnamefont{S.~D.} \bibnamefont{Chambreau}},
  \bibinfo{author}{\bibfnamefont{D.}~\bibnamefont{Townsend}},
  \bibinfo{author}{\bibfnamefont{S.~A.} \bibnamefont{Lahankar}},
  \bibinfo{author}{\bibfnamefont{S.~K.} \bibnamefont{Lee}}, \bibnamefont{and}
  \bibinfo{author}{\bibfnamefont{A.~G.} \bibnamefont{Suits}},
  \bibinfo{journal}{Physica Scripta} \textbf{\bibinfo{volume}{73}},
  \bibinfo{pages}{C89} (\bibinfo{year}{2006}).

\bibitem[{\citenamefont{Murray and Orr-Ewing}(2004)}]{Murray:IRPC23:435}
\bibinfo{author}{\bibfnamefont{C.}~\bibnamefont{Murray}} \bibnamefont{and}
  \bibinfo{author}{\bibfnamefont{A.~J.} \bibnamefont{Orr-Ewing}},
  \bibinfo{journal}{Int. Rev. Phys. Chem.} \textbf{\bibinfo{volume}{23}},
  \bibinfo{pages}{435} (\bibinfo{year}{2004}).

\bibitem[{\citenamefont{Heaven}(2005)}]{Heaven:IRPC24:375}
\bibinfo{author}{\bibfnamefont{M.~C.} \bibnamefont{Heaven}},
  \bibinfo{journal}{Int. Rev. Phys. Chem.} \textbf{\bibinfo{volume}{24}},
  \bibinfo{pages}{375} (\bibinfo{year}{2005}).

\bibitem[{\citenamefont{Kim and Meyer}(2001)}]{Kim:IRPC20:219}
\bibinfo{author}{\bibfnamefont{Y.}~\bibnamefont{Kim}} \bibnamefont{and}
  \bibinfo{author}{\bibfnamefont{H.}~\bibnamefont{Meyer}},
  \bibinfo{journal}{Int. Rev. Phys. Chem.} \textbf{\bibinfo{volume}{20}},
  \bibinfo{pages}{219} (\bibinfo{year}{2001}).

\bibitem[{\citenamefont{Lester et~al.}(2001)\citenamefont{Lester, Pond,
  Marshall, Anderson, Harding, and Wagner}}]{Lester:FD118:373}
\bibinfo{author}{\bibfnamefont{M.~I.} \bibnamefont{Lester}},
  \bibinfo{author}{\bibfnamefont{B.~V.} \bibnamefont{Pond}},
  \bibinfo{author}{\bibfnamefont{M.~D.} \bibnamefont{Marshall}},
  \bibinfo{author}{\bibfnamefont{D.~T.} \bibnamefont{Anderson}},
  \bibinfo{author}{\bibfnamefont{L.~B.} \bibnamefont{Harding}},
  \bibnamefont{and} \bibinfo{author}{\bibfnamefont{A.~F.}
  \bibnamefont{Wagner}}, \bibinfo{journal}{Faraday Disc.}
  \textbf{\bibinfo{volume}{118}}, \bibinfo{pages}{373} (\bibinfo{year}{2001}).

\bibitem[{\citenamefont{Loomis et~al.}(1996)\citenamefont{Loomis, Schwartz, and
  Lester}}]{Loomis:JCP104:6984}
\bibinfo{author}{\bibfnamefont{R.~A.} \bibnamefont{Loomis}},
  \bibinfo{author}{\bibfnamefont{R.~L.} \bibnamefont{Schwartz}},
  \bibnamefont{and} \bibinfo{author}{\bibfnamefont{M.~I.}
  \bibnamefont{Lester}}, \bibinfo{journal}{J. Chem. Phys.}
  \textbf{\bibinfo{volume}{104}}, \bibinfo{pages}{6984} (\bibinfo{year}{1996}).

\bibitem[{\citenamefont{Anderson et~al.}(1998)\citenamefont{Anderson, Schwartz,
  Todd, and Lester}}]{Anderson:JCP109:3461}
\bibinfo{author}{\bibfnamefont{D.~T.} \bibnamefont{Anderson}},
  \bibinfo{author}{\bibfnamefont{R.~L.} \bibnamefont{Schwartz}},
  \bibinfo{author}{\bibfnamefont{M.~W.} \bibnamefont{Todd}}, \bibnamefont{and}
  \bibinfo{author}{\bibfnamefont{M.~I.} \bibnamefont{Lester}},
  \bibinfo{journal}{J. Chem. Phys.} \textbf{\bibinfo{volume}{109}},
  \bibinfo{pages}{3461} (\bibinfo{year}{1998}).

\bibitem[{\citenamefont{Loomis and Lester}(1997)}]{Loomis:ARPC48:643}
\bibinfo{author}{\bibfnamefont{R.~A.} \bibnamefont{Loomis}} \bibnamefont{and}
  \bibinfo{author}{\bibfnamefont{M.~I.} \bibnamefont{Lester}},
  \bibinfo{journal}{Ann. Rev. Phys. Chem.} \textbf{\bibinfo{volume}{48}},
  \bibinfo{pages}{643} (\bibinfo{year}{1997}).

\bibitem[{\citenamefont{Schwartz et~al.}(1997)\citenamefont{Schwartz, Anderson,
  Todd, and Lester}}]{Schwartz:CPL273:18}
\bibinfo{author}{\bibfnamefont{R.~L.} \bibnamefont{Schwartz}},
  \bibinfo{author}{\bibfnamefont{D.~T.} \bibnamefont{Anderson}},
  \bibinfo{author}{\bibfnamefont{M.~W.} \bibnamefont{Todd}}, \bibnamefont{and}
  \bibinfo{author}{\bibfnamefont{M.~I.} \bibnamefont{Lester}},
  \bibinfo{journal}{Chem. Phys. Lett.} \textbf{\bibinfo{volume}{273}},
  \bibinfo{pages}{18} (\bibinfo{year}{1997}).

\bibitem[{\citenamefont{Hossenlopp et~al.}(1998)\citenamefont{Hossenlopp,
  Anderson, Todd, and Lester}}]{Hossenlopp:JCP109:10707}
\bibinfo{author}{\bibfnamefont{J.~M.} \bibnamefont{Hossenlopp}},
  \bibinfo{author}{\bibfnamefont{D.~T.} \bibnamefont{Anderson}},
  \bibinfo{author}{\bibfnamefont{M.~W.} \bibnamefont{Todd}}, \bibnamefont{and}
  \bibinfo{author}{\bibfnamefont{M.~I.} \bibnamefont{Lester}},
  \bibinfo{journal}{J. Chem. Phys.} \textbf{\bibinfo{volume}{109}},
  \bibinfo{pages}{10707} (\bibinfo{year}{1998}).

\bibitem[{\citenamefont{Chen and Heaven}(1998)}]{Chen:JCP109:5171}
\bibinfo{author}{\bibfnamefont{Y.~L.} \bibnamefont{Chen}} \bibnamefont{and}
  \bibinfo{author}{\bibfnamefont{M.~C.} \bibnamefont{Heaven}},
  \bibinfo{journal}{J. Chem. Phys.} \textbf{\bibinfo{volume}{109}},
  \bibinfo{pages}{5171} (\bibinfo{year}{1998}).

\bibitem[{\citenamefont{Kaledin and Heaven}(2001)}]{Kaledin:CPL347:199}
\bibinfo{author}{\bibfnamefont{A.~L.} \bibnamefont{Kaledin}} \bibnamefont{and}
  \bibinfo{author}{\bibfnamefont{M.~C.} \bibnamefont{Heaven}},
  \bibinfo{journal}{Chem. Phys. Lett.} \textbf{\bibinfo{volume}{347}},
  \bibinfo{pages}{199} (\bibinfo{year}{2001}).

\bibitem[{\citenamefont{Chen and Heaven}(2000)}]{Chen:JCP112:7416}
\bibinfo{author}{\bibfnamefont{Y.~L.} \bibnamefont{Chen}} \bibnamefont{and}
  \bibinfo{author}{\bibfnamefont{M.~C.} \bibnamefont{Heaven}},
  \bibinfo{journal}{J. Chem. Phys.} \textbf{\bibinfo{volume}{112}},
  \bibinfo{pages}{7416} (\bibinfo{year}{2000}).

\bibitem[{\citenamefont{Wheeler et~al.}(2000)\citenamefont{Wheeler, Anderson,
  and Lester}}]{Wheeler:IRPC19:501}
\bibinfo{author}{\bibfnamefont{M.~D.} \bibnamefont{Wheeler}},
  \bibinfo{author}{\bibfnamefont{D.~T.} \bibnamefont{Anderson}},
  \bibnamefont{and} \bibinfo{author}{\bibfnamefont{M.~I.}
  \bibnamefont{Lester}}, \bibinfo{journal}{Int. Rev. Phys. Chem.}
  \textbf{\bibinfo{volume}{19}}, \bibinfo{pages}{501} (\bibinfo{year}{2000}).

\bibitem[{\citenamefont{Mielke and Andrews}(1990)}]{Mielke:JPC94:3519}
\bibinfo{author}{\bibfnamefont{Z.}~\bibnamefont{Mielke}} \bibnamefont{and}
  \bibinfo{author}{\bibfnamefont{L.}~\bibnamefont{Andrews}},
  \bibinfo{journal}{J. Phys. Chem.} \textbf{\bibinfo{volume}{94}},
  \bibinfo{pages}{3519} (\bibinfo{year}{1990}).

\bibitem[{\citenamefont{Jacox}(1979)}]{Jacox:CP42:133}
\bibinfo{author}{\bibfnamefont{M.~E.} \bibnamefont{Jacox}},
  \bibinfo{journal}{Chem. Phys.} \textbf{\bibinfo{volume}{42}},
  \bibinfo{pages}{133} (\bibinfo{year}{1979}).

\bibitem[{\citenamefont{Khriachtchev et~al.}(2006)\citenamefont{Khriachtchev,
  Tanskanen, and Rasanen}}]{Khriachtchev:JCP124:181101}
\bibinfo{author}{\bibfnamefont{L.}~\bibnamefont{Khriachtchev}},
  \bibinfo{author}{\bibfnamefont{H.}~\bibnamefont{Tanskanen}},
  \bibnamefont{and} \bibinfo{author}{\bibfnamefont{M.}~\bibnamefont{Rasanen}},
  \bibinfo{journal}{J. Chem. Phys.} \textbf{\bibinfo{volume}{124}},
  \bibinfo{pages}{181101} (\bibinfo{year}{2006}).

\bibitem[{\citenamefont{Yoshioka et~al.}(2006)\citenamefont{Yoshioka, Raston,
  and Anderson}}]{Yoshioka:IRPC25:469}
\bibinfo{author}{\bibfnamefont{K.}~\bibnamefont{Yoshioka}},
  \bibinfo{author}{\bibfnamefont{P.~L.} \bibnamefont{Raston}},
  \bibnamefont{and} \bibinfo{author}{\bibfnamefont{D.~T.}
  \bibnamefont{Anderson}}, \bibinfo{journal}{Int. Rev. Phys. Chem.}
  \textbf{\bibinfo{volume}{25}}, \bibinfo{pages}{469} (\bibinfo{year}{2006}).

\bibitem[{\citenamefont{Momose et~al.}(2004)\citenamefont{Momose, Hoshina,
  Fushitani, and Katsuki}}]{Momose:VS34:95}
\bibinfo{author}{\bibfnamefont{T.}~\bibnamefont{Momose}},
  \bibinfo{author}{\bibfnamefont{H.}~\bibnamefont{Hoshina}},
  \bibinfo{author}{\bibfnamefont{M.}~\bibnamefont{Fushitani}},
  \bibnamefont{and} \bibinfo{author}{\bibfnamefont{H.}~\bibnamefont{Katsuki}},
  \bibinfo{journal}{Vibrational Spectroscopy} \textbf{\bibinfo{volume}{34}},
  \bibinfo{pages}{95} (\bibinfo{year}{2004}).

\bibitem[{\citenamefont{Momose et~al.}(2005)\citenamefont{Momose, Fushitani,
  and Hoshina}}]{Momose:IRPC24:533}
\bibinfo{author}{\bibfnamefont{T.}~\bibnamefont{Momose}},
  \bibinfo{author}{\bibfnamefont{M.}~\bibnamefont{Fushitani}},
  \bibnamefont{and} \bibinfo{author}{\bibfnamefont{H.}~\bibnamefont{Hoshina}},
  \bibinfo{journal}{Int. Rev. Phys. Chem.} \textbf{\bibinfo{volume}{24}},
  \bibinfo{pages}{533} (\bibinfo{year}{2005}).

\bibitem[{\citenamefont{Momose et~al.}(1998)\citenamefont{Momose, Hoshina,
  Sogoshi, Katsuki, Wakabayashi, and Shida}}]{Momose:JCP108:7334}
\bibinfo{author}{\bibfnamefont{T.}~\bibnamefont{Momose}},
  \bibinfo{author}{\bibfnamefont{H.}~\bibnamefont{Hoshina}},
  \bibinfo{author}{\bibfnamefont{N.}~\bibnamefont{Sogoshi}},
  \bibinfo{author}{\bibfnamefont{H.}~\bibnamefont{Katsuki}},
  \bibinfo{author}{\bibfnamefont{T.}~\bibnamefont{Wakabayashi}},
  \bibnamefont{and} \bibinfo{author}{\bibfnamefont{T.}~\bibnamefont{Shida}},
  \bibinfo{journal}{J. Chem. Phys.} \textbf{\bibinfo{volume}{108}},
  \bibinfo{pages}{7334} (\bibinfo{year}{1998}).

\bibitem[{\citenamefont{Grebenev et~al.}(1998)\citenamefont{Grebenev, Toennies,
  and Vilesov}}]{Grebenev:Science279:2083}
\bibinfo{author}{\bibfnamefont{S.}~\bibnamefont{Grebenev}},
  \bibinfo{author}{\bibfnamefont{J.~P.} \bibnamefont{Toennies}},
  \bibnamefont{and} \bibinfo{author}{\bibfnamefont{A.~F.}
  \bibnamefont{Vilesov}}, \bibinfo{journal}{Science}
  \textbf{\bibinfo{volume}{279}}, \bibinfo{pages}{2083} (\bibinfo{year}{1998}).

\bibitem[{\citenamefont{Hartmann et~al.}(1995)\citenamefont{Hartmann, Miller,
  Toennies, and Vilesov}}]{Hartmann:PRL75:1566}
\bibinfo{author}{\bibfnamefont{M.}~\bibnamefont{Hartmann}},
  \bibinfo{author}{\bibfnamefont{R.~E.} \bibnamefont{Miller}},
  \bibinfo{author}{\bibfnamefont{J.~P.} \bibnamefont{Toennies}},
  \bibnamefont{and} \bibinfo{author}{\bibfnamefont{A.~F.}
  \bibnamefont{Vilesov}}, \bibinfo{journal}{Phys. Rev. Lett.}
  \textbf{\bibinfo{volume}{75}}, \bibinfo{pages}{1566} (\bibinfo{year}{1995}).

\bibitem[{\citenamefont{Nauta and Miller}(1999)}]{Nauta:Science283:1895}
\bibinfo{author}{\bibfnamefont{K.}~\bibnamefont{Nauta}} \bibnamefont{and}
  \bibinfo{author}{\bibfnamefont{R.~E.} \bibnamefont{Miller}},
  \bibinfo{journal}{Science} \textbf{\bibinfo{volume}{283}},
  \bibinfo{pages}{1895} (\bibinfo{year}{1999}).

\bibitem[{\citenamefont{Nauta and
  Miller}(2000{\natexlab{a}})}]{Nauta:Science287:293}
\bibinfo{author}{\bibfnamefont{K.}~\bibnamefont{Nauta}} \bibnamefont{and}
  \bibinfo{author}{\bibfnamefont{R.~E.} \bibnamefont{Miller}},
  \bibinfo{journal}{Science} \textbf{\bibinfo{volume}{287}},
  \bibinfo{pages}{293} (\bibinfo{year}{2000}{\natexlab{a}}).

\bibitem[{\citenamefont{Merritt et~al.}(2004)\citenamefont{Merritt, Douberly,
  and Miller}}]{Merritt:JCP121:1309}
\bibinfo{author}{\bibfnamefont{J.~M.} \bibnamefont{Merritt}},
  \bibinfo{author}{\bibfnamefont{G.~E.} \bibnamefont{Douberly}},
  \bibnamefont{and} \bibinfo{author}{\bibfnamefont{R.~E.}
  \bibnamefont{Miller}}, \bibinfo{journal}{J. Chem. Phys.}
  \textbf{\bibinfo{volume}{121}}, \bibinfo{pages}{1309} (\bibinfo{year}{2004}).

\bibitem[{\citenamefont{Douberly et~al.}(2005)\citenamefont{Douberly, Merritt,
  and Miller}}]{Douberly:PCCP7:463}
\bibinfo{author}{\bibfnamefont{G.~E.} \bibnamefont{Douberly}},
  \bibinfo{author}{\bibfnamefont{J.~M.} \bibnamefont{Merritt}},
  \bibnamefont{and} \bibinfo{author}{\bibfnamefont{R.~E.}
  \bibnamefont{Miller}}, \bibinfo{journal}{Phys. Chem. Chem. Phys.}
  \textbf{\bibinfo{volume}{7}}, \bibinfo{pages}{463} (\bibinfo{year}{2005}).

\bibitem[{\citenamefont{Brink and Stringari}(1990)}]{Brink:ZPD15:257}
\bibinfo{author}{\bibfnamefont{D.~M.} \bibnamefont{Brink}} \bibnamefont{and}
  \bibinfo{author}{\bibfnamefont{S.}~\bibnamefont{Stringari}},
  \bibinfo{journal}{Z. Phys. D} \textbf{\bibinfo{volume}{15}},
  \bibinfo{pages}{257} (\bibinfo{year}{1990}).

\bibitem[{\citenamefont{Dalfovo and Stringari}(2001)}]{Dalfovo:JCP115:1078}
\bibinfo{author}{\bibfnamefont{F.}~\bibnamefont{Dalfovo}} \bibnamefont{and}
  \bibinfo{author}{\bibfnamefont{S.}~\bibnamefont{Stringari}},
  \bibinfo{journal}{J. Chem. Phys.} \textbf{\bibinfo{volume}{115}},
  \bibinfo{pages}{1078} (\bibinfo{year}{2001}).

\bibitem[{\citenamefont{Lehmann}(2003)}]{Lehmann:JCP119:3336}
\bibinfo{author}{\bibfnamefont{K.~K.} \bibnamefont{Lehmann}},
  \bibinfo{journal}{J. Chem. Phys.} \textbf{\bibinfo{volume}{119}},
  \bibinfo{pages}{3336} (\bibinfo{year}{2003}).

\bibitem[{\citenamefont{K\"upper et~al.}(2002)\citenamefont{K\"upper, Merritt,
  and Miller}}]{Kuepper:JCP117:647}
\bibinfo{author}{\bibfnamefont{J.}~\bibnamefont{K\"upper}},
  \bibinfo{author}{\bibfnamefont{J.~M.} \bibnamefont{Merritt}},
  \bibnamefont{and} \bibinfo{author}{\bibfnamefont{R.~E.}
  \bibnamefont{Miller}}, \bibinfo{journal}{J. Chem. Phys.}
  \textbf{\bibinfo{volume}{117}}, \bibinfo{pages}{647} (\bibinfo{year}{2002}).

\bibitem[{\citenamefont{Block et~al.}(1992)\citenamefont{Block, Bohac, and
  Miller}}]{Block:PRL68:1303}
\bibinfo{author}{\bibfnamefont{P.~A.} \bibnamefont{Block}},
  \bibinfo{author}{\bibfnamefont{E.~J.} \bibnamefont{Bohac}}, \bibnamefont{and}
  \bibinfo{author}{\bibfnamefont{R.~E.} \bibnamefont{Miller}},
  \bibinfo{journal}{Phys. Rev. Lett.} \textbf{\bibinfo{volume}{68}},
  \bibinfo{pages}{1303} (\bibinfo{year}{1992}).

\bibitem[{\citenamefont{Becker et~al.}(1961)\citenamefont{Becker, Klingelhofer,
  and Lohse}}]{Becker:ZNA16:1259}
\bibinfo{author}{\bibfnamefont{E.~W.} \bibnamefont{Becker}},
  \bibinfo{author}{\bibfnamefont{R.}~\bibnamefont{Klingelhofer}},
  \bibnamefont{and} \bibinfo{author}{\bibfnamefont{P.}~\bibnamefont{Lohse}},
  \bibinfo{journal}{Z. Naturforsch. A} \textbf{\bibinfo{volume}{16}},
  \bibinfo{pages}{1259} (\bibinfo{year}{1961}).

\bibitem[{\citenamefont{Nauta and
  Miller}(2001{\natexlab{a}})}]{Nauta:JCP115:10138}
\bibinfo{author}{\bibfnamefont{K.}~\bibnamefont{Nauta}} \bibnamefont{and}
  \bibinfo{author}{\bibfnamefont{R.~E.} \bibnamefont{Miller}},
  \bibinfo{journal}{J. Chem. Phys.} \textbf{\bibinfo{volume}{115}},
  \bibinfo{pages}{10138} (\bibinfo{year}{2001}{\natexlab{a}}).

\bibitem[{\citenamefont{von Haeften et~al.}(2005)\citenamefont{von Haeften,
  Metzelthin, Rudolph, Staemmler, and Havenith}}]{Haeften:PRL95:215301}
\bibinfo{author}{\bibfnamefont{K.}~\bibnamefont{von Haeften}},
  \bibinfo{author}{\bibfnamefont{A.}~\bibnamefont{Metzelthin}},
  \bibinfo{author}{\bibfnamefont{S.}~\bibnamefont{Rudolph}},
  \bibinfo{author}{\bibfnamefont{V.}~\bibnamefont{Staemmler}},
  \bibnamefont{and} \bibinfo{author}{\bibfnamefont{M.}~\bibnamefont{Havenith}},
  \bibinfo{journal}{Phys. Rev. Lett.} \textbf{\bibinfo{volume}{95}},
  \bibinfo{pages}{215301} (\bibinfo{year}{2005}).

\bibitem[{\citenamefont{Pauly}(1988{\natexlab{a}})}]{Pauly:OtherLowEnergySourc%
es}
\bibinfo{author}{\bibfnamefont{H.}~\bibnamefont{Pauly}}, in
  \cite{_Scoles:MolBeam:1}, chap.~\bibinfo{chapter}{4}, pp.
  \bibinfo{pages}{83--123}.

\bibitem[{\citenamefont{Pauly}(1988{\natexlab{b}})}]{Pauly:HighEnergySources}
\bibinfo{author}{\bibfnamefont{H.}~\bibnamefont{Pauly}}, in
  \cite{_Scoles:MolBeam:1}, chap.~\bibinfo{chapter}{5}, pp.
  \bibinfo{pages}{124--152}.

\bibitem[{\citenamefont{Kohn et~al.}(1992)\citenamefont{Kohn, Clauberg, and
  Chen}}]{Kohn:RSI63:4003}
\bibinfo{author}{\bibfnamefont{D.~W.} \bibnamefont{Kohn}},
  \bibinfo{author}{\bibfnamefont{H.}~\bibnamefont{Clauberg}}, \bibnamefont{and}
  \bibinfo{author}{\bibfnamefont{P.}~\bibnamefont{Chen}},
  \bibinfo{journal}{Rev. Sci. Instrum.} \textbf{\bibinfo{volume}{63}},
  \bibinfo{pages}{4003} (\bibinfo{year}{1992}).

\bibitem[{\citenamefont{Cameron and Kable}(1996)}]{Cameron:RSI67:283}
\bibinfo{author}{\bibfnamefont{M.~R.} \bibnamefont{Cameron}} \bibnamefont{and}
  \bibinfo{author}{\bibfnamefont{S.~H.} \bibnamefont{Kable}},
  \bibinfo{journal}{Rev. Sci. Instrum.} \textbf{\bibinfo{volume}{67}},
  \bibinfo{pages}{283} (\bibinfo{year}{1996}).

\bibitem[{\citenamefont{Zhang et~al.}(2003)\citenamefont{Zhang, Friderichsen,
  Nandi, Ellison, David, McKinnon, Lindeman, Dayton, and
  Nimlos}}]{Zhang:RSI74:3077}
\bibinfo{author}{\bibfnamefont{X.}~\bibnamefont{Zhang}},
  \bibinfo{author}{\bibfnamefont{A.~V.} \bibnamefont{Friderichsen}},
  \bibinfo{author}{\bibfnamefont{S.}~\bibnamefont{Nandi}},
  \bibinfo{author}{\bibfnamefont{G.~B.} \bibnamefont{Ellison}},
  \bibinfo{author}{\bibfnamefont{D.~E.} \bibnamefont{David}},
  \bibinfo{author}{\bibfnamefont{J.~T.} \bibnamefont{McKinnon}},
  \bibinfo{author}{\bibfnamefont{T.~G.} \bibnamefont{Lindeman}},
  \bibinfo{author}{\bibfnamefont{D.~C.} \bibnamefont{Dayton}},
  \bibnamefont{and} \bibinfo{author}{\bibfnamefont{M.~R.}
  \bibnamefont{Nimlos}}, \bibinfo{journal}{Rev. Sci. Instrum.}
  \textbf{\bibinfo{volume}{74}}, \bibinfo{pages}{3077} (\bibinfo{year}{2003}).

\bibitem[{\citenamefont{Engelking}(1986)}]{Engelking:RSI57:2274}
\bibinfo{author}{\bibfnamefont{P.~C.} \bibnamefont{Engelking}},
  \bibinfo{journal}{Rev. Sci. Instrum.} \textbf{\bibinfo{volume}{57}},
  \bibinfo{pages}{2274} (\bibinfo{year}{1986}).

\bibitem[{\citenamefont{Engelking}(1991)}]{Engelking:CR91:399}
\bibinfo{author}{\bibfnamefont{P.~C.} \bibnamefont{Engelking}},
  \bibinfo{journal}{Chem. Rev.} \textbf{\bibinfo{volume}{91}},
  \bibinfo{pages}{399} (\bibinfo{year}{1991}).

\bibitem[{\citenamefont{Anderson et~al.}(1996)\citenamefont{Anderson, Davis,
  Zwier, and Nesbitt}}]{Anderson:CPL258:207}
\bibinfo{author}{\bibfnamefont{D.~T.} \bibnamefont{Anderson}},
  \bibinfo{author}{\bibfnamefont{S.}~\bibnamefont{Davis}},
  \bibinfo{author}{\bibfnamefont{T.~S.} \bibnamefont{Zwier}}, \bibnamefont{and}
  \bibinfo{author}{\bibfnamefont{D.~J.} \bibnamefont{Nesbitt}},
  \bibinfo{journal}{Chem. Phys. Lett.} \textbf{\bibinfo{volume}{258}},
  \bibinfo{pages}{207} (\bibinfo{year}{1996}).

\bibitem[{\citenamefont{Monts et~al.}(1980)\citenamefont{Monts, Dietz, Duncan,
  and Smalley}}]{Monts:CP45:133}
\bibinfo{author}{\bibfnamefont{D.~L.} \bibnamefont{Monts}},
  \bibinfo{author}{\bibfnamefont{T.~G.} \bibnamefont{Dietz}},
  \bibinfo{author}{\bibfnamefont{M.~A.} \bibnamefont{Duncan}},
  \bibnamefont{and} \bibinfo{author}{\bibfnamefont{R.~E.}
  \bibnamefont{Smalley}}, \bibinfo{journal}{Chem. Phys.}
  \textbf{\bibinfo{volume}{45}}, \bibinfo{pages}{133} (\bibinfo{year}{1980}).

\bibitem[{\citenamefont{Heaven et~al.}(1981)\citenamefont{Heaven, Miller, and
  Bondybey}}]{Heaven:CPL84:1}
\bibinfo{author}{\bibfnamefont{M.}~\bibnamefont{Heaven}},
  \bibinfo{author}{\bibfnamefont{T.~A.} \bibnamefont{Miller}},
  \bibnamefont{and} \bibinfo{author}{\bibfnamefont{V.~E.}
  \bibnamefont{Bondybey}}, \bibinfo{journal}{Chem. Phys. Lett.}
  \textbf{\bibinfo{volume}{84}}, \bibinfo{pages}{1} (\bibinfo{year}{1981}).

\bibitem[{\citenamefont{Andresen et~al.}(1984)\citenamefont{Andresen, Hausler,
  and Lulf}}]{Andresen:JCP81:571}
\bibinfo{author}{\bibfnamefont{P.}~\bibnamefont{Andresen}},
  \bibinfo{author}{\bibfnamefont{D.}~\bibnamefont{Hausler}}, \bibnamefont{and}
  \bibinfo{author}{\bibfnamefont{H.~W.} \bibnamefont{Lulf}},
  \bibinfo{journal}{J. Chem. Phys.} \textbf{\bibinfo{volume}{81}},
  \bibinfo{pages}{571} (\bibinfo{year}{1984}).

\bibitem[{\citenamefont{Slipchenko et~al.}(2002)\citenamefont{Slipchenko, Kuma,
  Momose, and Vilesov}}]{Slipchenko:RSI73:3600}
\bibinfo{author}{\bibfnamefont{M.~N.} \bibnamefont{Slipchenko}},
  \bibinfo{author}{\bibfnamefont{S.}~\bibnamefont{Kuma}},
  \bibinfo{author}{\bibfnamefont{T.}~\bibnamefont{Momose}}, \bibnamefont{and}
  \bibinfo{author}{\bibfnamefont{A.~F.} \bibnamefont{Vilesov}},
  \bibinfo{journal}{Rev. Sci. Instrum.} \textbf{\bibinfo{volume}{73}},
  \bibinfo{pages}{3600} (\bibinfo{year}{2002}).

\bibitem[{\citenamefont{Yang et~al.}(2005)\citenamefont{Yang, Brereton, and
  Ellis}}]{Yang:RSI76:104102}
\bibinfo{author}{\bibfnamefont{S.}~\bibnamefont{Yang}},
  \bibinfo{author}{\bibfnamefont{S.~M.} \bibnamefont{Brereton}},
  \bibnamefont{and} \bibinfo{author}{\bibfnamefont{A.~M.} \bibnamefont{Ellis}},
  \bibinfo{journal}{Rev. Sci. Instrum.} \textbf{\bibinfo{volume}{76}},
  \bibinfo{pages}{104102} (\bibinfo{year}{2005}).

\bibitem[{\citenamefont{Conjusteau}(2002)}]{Conjusteau:thesis:2002}
\bibinfo{author}{\bibfnamefont{A.}~\bibnamefont{Conjusteau}}, Ph.D. thesis,
  \bibinfo{school}{Princeton University}, \bibinfo{address}{Princeton, NJ, USA}
  (\bibinfo{year}{2002}).

\bibitem[{\citenamefont{Braun and Drabbels}(2004)}]{Braun:PRL93:253401}
\bibinfo{author}{\bibfnamefont{A.}~\bibnamefont{Braun}} \bibnamefont{and}
  \bibinfo{author}{\bibfnamefont{M.}~\bibnamefont{Drabbels}},
  \bibinfo{journal}{Phys. Rev. Lett.} \textbf{\bibinfo{volume}{93}},
  \bibinfo{pages}{253401} (\bibinfo{year}{2004}).

\bibitem[{\citenamefont{Kavita and Das}(2000)}]{Kavita:JCP112:8426}
\bibinfo{author}{\bibfnamefont{K.}~\bibnamefont{Kavita}} \bibnamefont{and}
  \bibinfo{author}{\bibfnamefont{P.~K.} \bibnamefont{Das}},
  \bibinfo{journal}{J. Chem. Phys.} \textbf{\bibinfo{volume}{112}},
  \bibinfo{pages}{8426} (\bibinfo{year}{2000}).

\bibitem[{\citenamefont{Parker and Eppink}(1997)}]{Parker:JCP107:2357}
\bibinfo{author}{\bibfnamefont{D.~H.} \bibnamefont{Parker}} \bibnamefont{and}
  \bibinfo{author}{\bibfnamefont{A.}~\bibnamefont{Eppink}},
  \bibinfo{journal}{J. Chem. Phys.} \textbf{\bibinfo{volume}{107}},
  \bibinfo{pages}{2357} (\bibinfo{year}{1997}).

\bibitem[{\citenamefont{F\'arnik and Toennies}(2005)}]{Farnik:JCP122:014307}
\bibinfo{author}{\bibfnamefont{M.}~\bibnamefont{F\'arnik}} \bibnamefont{and}
  \bibinfo{author}{\bibfnamefont{J.~P.} \bibnamefont{Toennies}},
  \bibinfo{journal}{J. Chem. Phys.} \textbf{\bibinfo{volume}{122}},
  \bibinfo{pages}{014307} (\bibinfo{year}{2005}).

\bibitem[{\citenamefont{Lewis et~al.}(2005)\citenamefont{Lewis, Lindsay,
  Bemish, and Miller}}]{Lewis:JACS127:7235}
\bibinfo{author}{\bibfnamefont{W.~K.} \bibnamefont{Lewis}},
  \bibinfo{author}{\bibfnamefont{C.~M.} \bibnamefont{Lindsay}},
  \bibinfo{author}{\bibfnamefont{R.~J.} \bibnamefont{Bemish}},
  \bibnamefont{and} \bibinfo{author}{\bibfnamefont{R.~E.}
  \bibnamefont{Miller}}, \bibinfo{journal}{J. Am. Chem. Soc.}
  \textbf{\bibinfo{volume}{127}}, \bibinfo{pages}{7235} (\bibinfo{year}{2005}).

\bibitem[{\citenamefont{Marinov et~al.}(1998)\citenamefont{Marinov, Pitz,
  Westbrook, Vincitore, Castaldi, Senkan, and
  Melius}}]{Marinov:CombustFlame114:192}
\bibinfo{author}{\bibfnamefont{N.~M.} \bibnamefont{Marinov}},
  \bibinfo{author}{\bibfnamefont{W.~J.} \bibnamefont{Pitz}},
  \bibinfo{author}{\bibfnamefont{C.~K.} \bibnamefont{Westbrook}},
  \bibinfo{author}{\bibfnamefont{A.~M.} \bibnamefont{Vincitore}},
  \bibinfo{author}{\bibfnamefont{M.~J.} \bibnamefont{Castaldi}},
  \bibinfo{author}{\bibfnamefont{S.~M.} \bibnamefont{Senkan}},
  \bibnamefont{and} \bibinfo{author}{\bibfnamefont{C.~F.}
  \bibnamefont{Melius}}, \bibinfo{journal}{Combust. Flame}
  \textbf{\bibinfo{volume}{114}}, \bibinfo{pages}{192} (\bibinfo{year}{1998}).

\bibitem[{\citenamefont{Westmoreland et~al.}(1989)\citenamefont{Westmoreland,
  Dean, Howard, and Longwell}}]{Westmoreland:JPC93:8171}
\bibinfo{author}{\bibfnamefont{P.~R.} \bibnamefont{Westmoreland}},
  \bibinfo{author}{\bibfnamefont{A.~M.} \bibnamefont{Dean}},
  \bibinfo{author}{\bibfnamefont{J.~B.} \bibnamefont{Howard}},
  \bibnamefont{and} \bibinfo{author}{\bibfnamefont{J.~P.}
  \bibnamefont{Longwell}}, \bibinfo{journal}{J. Phys. Chem.}
  \textbf{\bibinfo{volume}{93}}, \bibinfo{pages}{8171} (\bibinfo{year}{1989}).

\bibitem[{\citenamefont{Alkemade and Homann}(1989)}]{Alkemade:ZPC161:19}
\bibinfo{author}{\bibfnamefont{U.}~\bibnamefont{Alkemade}} \bibnamefont{and}
  \bibinfo{author}{\bibfnamefont{K.~H.} \bibnamefont{Homann}},
  \bibinfo{journal}{Z. Phys. Chem.} \textbf{\bibinfo{volume}{161}},
  \bibinfo{pages}{19} (\bibinfo{year}{1989}).

\bibitem[{\citenamefont{Stein et~al.}(1990)\citenamefont{Stein, Walker, Suryan,
  and Fahr}}]{Stein:SympCombust23:85}
\bibinfo{author}{\bibfnamefont{S.~E.} \bibnamefont{Stein}},
  \bibinfo{author}{\bibfnamefont{J.~A.} \bibnamefont{Walker}},
  \bibinfo{author}{\bibfnamefont{M.~M.} \bibnamefont{Suryan}},
  \bibnamefont{and} \bibinfo{author}{\bibfnamefont{A.}~\bibnamefont{Fahr}}, in
  \emph{\bibinfo{booktitle}{Symp. (Int.) Combust.}} (\bibinfo{year}{1990}),
  vol.~\bibinfo{volume}{23}, pp. \bibinfo{pages}{85--90}.

\bibitem[{\citenamefont{Botschwina et~al.}(1995)\citenamefont{Botschwina,
  Oswald, Fl\"ugge, and Horn}}]{Botschwina:ZPC188:29}
\bibinfo{author}{\bibfnamefont{P.}~\bibnamefont{Botschwina}},
  \bibinfo{author}{\bibfnamefont{R.}~\bibnamefont{Oswald}},
  \bibinfo{author}{\bibfnamefont{J.}~\bibnamefont{Fl\"ugge}}, \bibnamefont{and}
  \bibinfo{author}{\bibfnamefont{M.}~\bibnamefont{Horn}}, \bibinfo{journal}{Z.
  Phys. Chem.} \textbf{\bibinfo{volume}{188}}, \bibinfo{pages}{29}
  (\bibinfo{year}{1995}).

\bibitem[{\citenamefont{Botschwina et~al.}(2000)\citenamefont{Botschwina, Horn,
  Oswald, and Schmatz}}]{Botschwina:JElecSpec108:109}
\bibinfo{author}{\bibfnamefont{P.}~\bibnamefont{Botschwina}},
  \bibinfo{author}{\bibfnamefont{M.}~\bibnamefont{Horn}},
  \bibinfo{author}{\bibfnamefont{R.}~\bibnamefont{Oswald}}, \bibnamefont{and}
  \bibinfo{author}{\bibfnamefont{S.}~\bibnamefont{Schmatz}},
  \bibinfo{journal}{J. Electron. Spectrosc.} \textbf{\bibinfo{volume}{108}},
  \bibinfo{pages}{109} (\bibinfo{year}{2000}).

\bibitem[{\citenamefont{Honjou et~al.}(1987)\citenamefont{Honjou, Yoshimine,
  and Pacansky}}]{Honjou:JPC91:4455}
\bibinfo{author}{\bibfnamefont{H.}~\bibnamefont{Honjou}},
  \bibinfo{author}{\bibfnamefont{M.}~\bibnamefont{Yoshimine}},
  \bibnamefont{and} \bibinfo{author}{\bibfnamefont{J.}~\bibnamefont{Pacansky}},
  \bibinfo{journal}{J. Phys. Chem.} \textbf{\bibinfo{volume}{91}},
  \bibinfo{pages}{4455} (\bibinfo{year}{1987}).

\bibitem[{\citenamefont{Hinchcliffe}(1977)}]{Hinchcliffe:JMolStruct37:295}
\bibinfo{author}{\bibfnamefont{A.}~\bibnamefont{Hinchcliffe}},
  \bibinfo{journal}{J. Mol. Struct.} \textbf{\bibinfo{volume}{37}},
  \bibinfo{pages}{295} (\bibinfo{year}{1977}).

\bibitem[{\citenamefont{Wu and Kern}(1987)}]{Wu:JPC91:6291}
\bibinfo{author}{\bibfnamefont{C.~H.} \bibnamefont{Wu}} \bibnamefont{and}
  \bibinfo{author}{\bibfnamefont{R.~D.} \bibnamefont{Kern}},
  \bibinfo{journal}{J. Phys. Chem.} \textbf{\bibinfo{volume}{91}},
  \bibinfo{pages}{6291} (\bibinfo{year}{1987}).

\bibitem[{\citenamefont{Marinov et~al.}(1997)\citenamefont{Marinov, Castaldi,
  Melius, and Tsang}}]{Marinov:CombustSciTech128:295}
\bibinfo{author}{\bibfnamefont{N.~M.} \bibnamefont{Marinov}},
  \bibinfo{author}{\bibfnamefont{M.~J.} \bibnamefont{Castaldi}},
  \bibinfo{author}{\bibfnamefont{C.~F.} \bibnamefont{Melius}},
  \bibnamefont{and} \bibinfo{author}{\bibfnamefont{W.}~\bibnamefont{Tsang}},
  \bibinfo{journal}{Combust. Sci. Technol.} \textbf{\bibinfo{volume}{128}},
  \bibinfo{pages}{295} (\bibinfo{year}{1997}).

\bibitem[{\citenamefont{Glassman}(1988)}]{Glassman:SympCombust22:295}
\bibinfo{author}{\bibfnamefont{I.}~\bibnamefont{Glassman}}, in
  \emph{\bibinfo{booktitle}{Symp. (Int.) Combust.}} (\bibinfo{year}{1988}),
  vol.~\bibinfo{volume}{22}, p. \bibinfo{pages}{295}.

\bibitem[{\citenamefont{Miller}(1996)}]{Miller:SympCombust26:461}
\bibinfo{author}{\bibfnamefont{J.~A.} \bibnamefont{Miller}}, in
  \emph{\bibinfo{booktitle}{Symp. (Int.) Combust.}} (\bibinfo{year}{1996}),
  vol.~\bibinfo{volume}{26}, p. \bibinfo{pages}{461}.

\bibitem[{\citenamefont{Goodings et~al.}(1979)\citenamefont{Goodings, Bohme,
  and Ng}}]{Goodings:CombustFlame36:27}
\bibinfo{author}{\bibfnamefont{J.~M.} \bibnamefont{Goodings}},
  \bibinfo{author}{\bibfnamefont{D.~K.} \bibnamefont{Bohme}}, \bibnamefont{and}
  \bibinfo{author}{\bibfnamefont{C.-W.} \bibnamefont{Ng}},
  \bibinfo{journal}{Combust. Flame} \textbf{\bibinfo{volume}{36}},
  \bibinfo{pages}{27} (\bibinfo{year}{1979}).

\bibitem[{\citenamefont{Olson and Calcote}(1981)}]{Olson:SympCombust18:453}
\bibinfo{author}{\bibfnamefont{D.~B.} \bibnamefont{Olson}} \bibnamefont{and}
  \bibinfo{author}{\bibfnamefont{H.~F.} \bibnamefont{Calcote}}, in
  \emph{\bibinfo{booktitle}{Symp. (Int.) Combust.}} (\bibinfo{year}{1981}),
  vol.~\bibinfo{volume}{18}, p. \bibinfo{pages}{453}.

\bibitem[{\citenamefont{Tanaka et~al.}(1995)\citenamefont{Tanaka, Harada,
  Sakaguchi, Harada, and Tanaka}}]{Tanaka:JCP103:6450}
\bibinfo{author}{\bibfnamefont{K.}~\bibnamefont{Tanaka}},
  \bibinfo{author}{\bibfnamefont{T.}~\bibnamefont{Harada}},
  \bibinfo{author}{\bibfnamefont{K.}~\bibnamefont{Sakaguchi}},
  \bibinfo{author}{\bibfnamefont{K.}~\bibnamefont{Harada}}, \bibnamefont{and}
  \bibinfo{author}{\bibfnamefont{T.}~\bibnamefont{Tanaka}},
  \bibinfo{journal}{J. Chem. Phys.} \textbf{\bibinfo{volume}{103}},
  \bibinfo{pages}{6450} (\bibinfo{year}{1995}).

\bibitem[{\citenamefont{Yuan et~al.}(1998)\citenamefont{Yuan, DeSain, and
  Curl}}]{Yuan:JMolSpec187:102}
\bibinfo{author}{\bibfnamefont{L.}~\bibnamefont{Yuan}},
  \bibinfo{author}{\bibfnamefont{J.}~\bibnamefont{DeSain}}, \bibnamefont{and}
  \bibinfo{author}{\bibfnamefont{R.~F.} \bibnamefont{Curl}},
  \bibinfo{journal}{J. Mol. Spec.} \textbf{\bibinfo{volume}{187}},
  \bibinfo{pages}{102} (\bibinfo{year}{1998}).

\bibitem[{\citenamefont{Tanaka et~al.}(1997)\citenamefont{Tanaka, Sumiyoshi,
  Ohshima, Endo, and Kawaguchi}}]{Tanaka:JCP107:2728}
\bibinfo{author}{\bibfnamefont{K.}~\bibnamefont{Tanaka}},
  \bibinfo{author}{\bibfnamefont{Y.}~\bibnamefont{Sumiyoshi}},
  \bibinfo{author}{\bibfnamefont{Y.}~\bibnamefont{Ohshima}},
  \bibinfo{author}{\bibfnamefont{Y.}~\bibnamefont{Endo}}, \bibnamefont{and}
  \bibinfo{author}{\bibfnamefont{K.}~\bibnamefont{Kawaguchi}},
  \bibinfo{journal}{J. Chem. Phys.} \textbf{\bibinfo{volume}{107}},
  \bibinfo{pages}{2728} (\bibinfo{year}{1997}).

\bibitem[{\citenamefont{Chapovsky and Hermans}(1999)}]{Chapovsky:ARPC50:315}
\bibinfo{author}{\bibfnamefont{P.~L.} \bibnamefont{Chapovsky}}
  \bibnamefont{and} \bibinfo{author}{\bibfnamefont{L.~J.~F.}
  \bibnamefont{Hermans}}, \bibinfo{journal}{Ann. Rev. Phys. Chem.}
  \textbf{\bibinfo{volume}{50}}, \bibinfo{pages}{315} (\bibinfo{year}{1999}).

\bibitem[{\citenamefont{Fushitani and Momose}(2002)}]{Fushitani:JCP116:10739}
\bibinfo{author}{\bibfnamefont{M.}~\bibnamefont{Fushitani}} \bibnamefont{and}
  \bibinfo{author}{\bibfnamefont{T.}~\bibnamefont{Momose}},
  \bibinfo{journal}{J. Chem. Phys.} \textbf{\bibinfo{volume}{116}},
  \bibinfo{pages}{10739} (\bibinfo{year}{2002}).

\bibitem[{\citenamefont{Stiles et~al.}(2003)\citenamefont{Stiles, Nauta, and
  Miller}}]{Stiles:PRL90:135301}
\bibinfo{author}{\bibfnamefont{P.~L.} \bibnamefont{Stiles}},
  \bibinfo{author}{\bibfnamefont{K.}~\bibnamefont{Nauta}}, \bibnamefont{and}
  \bibinfo{author}{\bibfnamefont{R.~E.} \bibnamefont{Miller}},
  \bibinfo{journal}{Phys. Rev. Lett.} \textbf{\bibinfo{volume}{90}},
  \bibinfo{pages}{135301} (\bibinfo{year}{2003}).

\bibitem[{\citenamefont{Botschwina}(2002)}]{Botschwina:propargyl-dipole:2002}
\bibinfo{author}{\bibfnamefont{P.}~\bibnamefont{Botschwina}},
  \emph{\bibinfo{title}{The zero-point corrected dipole moment of propargyl
  radical}} (\bibinfo{year}{2002}), \bibinfo{note}{private communication}.

\bibitem[{\citenamefont{Lindsay et~al.}(2004)\citenamefont{Lindsay, Lewis, and
  Miller}}]{Lindsay:JCP121:6095}
\bibinfo{author}{\bibfnamefont{C.~M.} \bibnamefont{Lindsay}},
  \bibinfo{author}{\bibfnamefont{W.~K.} \bibnamefont{Lewis}}, \bibnamefont{and}
  \bibinfo{author}{\bibfnamefont{R.~E.} \bibnamefont{Miller}},
  \bibinfo{journal}{J. Chem. Phys.} \textbf{\bibinfo{volume}{121}},
  \bibinfo{pages}{6095} (\bibinfo{year}{2004}).

\bibitem[{\citenamefont{Nauta and
  Miller}(2000{\natexlab{b}})}]{Nauta:JCP113:9466}
\bibinfo{author}{\bibfnamefont{K.}~\bibnamefont{Nauta}} \bibnamefont{and}
  \bibinfo{author}{\bibfnamefont{R.~E.} \bibnamefont{Miller}},
  \bibinfo{journal}{J. Chem. Phys.} \textbf{\bibinfo{volume}{113}},
  \bibinfo{pages}{9466} (\bibinfo{year}{2000}{\natexlab{b}}).

\bibitem[{\citenamefont{Polyakova et~al.}(2006)\citenamefont{Polyakova,
  Stolyarov, and Wittig}}]{Polyakova:JCP124:214308}
\bibinfo{author}{\bibfnamefont{E.}~\bibnamefont{Polyakova}},
  \bibinfo{author}{\bibfnamefont{D.}~\bibnamefont{Stolyarov}},
  \bibnamefont{and} \bibinfo{author}{\bibfnamefont{C.}~\bibnamefont{Wittig}},
  \bibinfo{journal}{J. Chem. Phys.} \textbf{\bibinfo{volume}{124}},
  \bibinfo{pages}{214308} (\bibinfo{year}{2006}).

\bibitem[{\citenamefont{Stienkemeier et~al.}(1995)\citenamefont{Stienkemeier,
  Higgins, Ernst, and Scoles}}]{Stienkemeier:ZPB98:413}
\bibinfo{author}{\bibfnamefont{F.}~\bibnamefont{Stienkemeier}},
  \bibinfo{author}{\bibfnamefont{J.}~\bibnamefont{Higgins}},
  \bibinfo{author}{\bibfnamefont{W.~E.} \bibnamefont{Ernst}}, \bibnamefont{and}
  \bibinfo{author}{\bibfnamefont{G.}~\bibnamefont{Scoles}},
  \bibinfo{journal}{Z. Phys. B} \textbf{\bibinfo{volume}{98}},
  \bibinfo{pages}{413} (\bibinfo{year}{1995}).

\bibitem[{\citenamefont{Reho et~al.}(1997)\citenamefont{Reho, Callegari,
  Higgins, Ernst, Lehmann, and Scoles}}]{Reho:FD1997:161}
\bibinfo{author}{\bibfnamefont{J.}~\bibnamefont{Reho}},
  \bibinfo{author}{\bibfnamefont{C.}~\bibnamefont{Callegari}},
  \bibinfo{author}{\bibfnamefont{J.}~\bibnamefont{Higgins}},
  \bibinfo{author}{\bibfnamefont{W.~E.} \bibnamefont{Ernst}},
  \bibinfo{author}{\bibfnamefont{K.~K.} \bibnamefont{Lehmann}},
  \bibnamefont{and} \bibinfo{author}{\bibfnamefont{G.}~\bibnamefont{Scoles}},
  \bibinfo{journal}{Faraday Disc.} \textbf{\bibinfo{volume}{108}},
  \bibinfo{pages}{161} (\bibinfo{year}{1997}).

\bibitem[{\citenamefont{Br\"uhl et~al.}(2001)\citenamefont{Br\"uhl, Trasca, and
  Ernst}}]{Bruehl:JCP115:10220}
\bibinfo{author}{\bibfnamefont{F.~R.} \bibnamefont{Br\"uhl}},
  \bibinfo{author}{\bibfnamefont{R.~A.} \bibnamefont{Trasca}},
  \bibnamefont{and} \bibinfo{author}{\bibfnamefont{W.~E.} \bibnamefont{Ernst}},
  \bibinfo{journal}{J. Chem. Phys.} \textbf{\bibinfo{volume}{115}},
  \bibinfo{pages}{10220} (\bibinfo{year}{2001}).

\bibitem[{\citenamefont{Lugovoj et~al.}(2000)\citenamefont{Lugovoj, Toennies,
  and Vilesov}}]{Lugovoj:JCP112:8217}
\bibinfo{author}{\bibfnamefont{E.}~\bibnamefont{Lugovoj}},
  \bibinfo{author}{\bibfnamefont{J.~P.} \bibnamefont{Toennies}},
  \bibnamefont{and} \bibinfo{author}{\bibfnamefont{A.}~\bibnamefont{Vilesov}},
  \bibinfo{journal}{The Journal of Chemical Physics}
  \textbf{\bibinfo{volume}{112}}, \bibinfo{pages}{8217} (\bibinfo{year}{2000}).

\bibitem[{\citenamefont{Herbst and W.}(1973)}]{Herbst:AJ185:505}
\bibinfo{author}{\bibfnamefont{E.}~\bibnamefont{Herbst}} \bibnamefont{and}
  \bibinfo{author}{\bibfnamefont{K.}~\bibnamefont{W.}},
  \bibinfo{journal}{Astron. J.} \textbf{\bibinfo{volume}{185}},
  \bibinfo{pages}{505} (\bibinfo{year}{1973}).

\bibitem[{\citenamefont{Smith}(1992)}]{Smith:CR92:1473}
\bibinfo{author}{\bibfnamefont{D.}~\bibnamefont{Smith}},
  \bibinfo{journal}{Chem. Rev.} \textbf{\bibinfo{volume}{92}},
  \bibinfo{pages}{1473} (\bibinfo{year}{1992}).

\bibitem[{\citenamefont{Neumark}(2002)}]{Neumark:PCC5:76}
\bibinfo{author}{\bibfnamefont{D.~M.} \bibnamefont{Neumark}},
  \bibinfo{journal}{Phys. Chem. Comm} \textbf{\bibinfo{volume}{5}},
  \bibinfo{pages}{76} (\bibinfo{year}{2002}).

\bibitem[{\citenamefont{Varandas et~al.}(2006)\citenamefont{Varandas, Caridade,
  Zhang, Cui, and Han}}]{Varandas:JCP125:064312}
\bibinfo{author}{\bibfnamefont{A.~J.~C.} \bibnamefont{Varandas}},
  \bibinfo{author}{\bibfnamefont{P.~J. S.~B.} \bibnamefont{Caridade}},
  \bibinfo{author}{\bibfnamefont{J.~Z.~H.} \bibnamefont{Zhang}},
  \bibinfo{author}{\bibfnamefont{Q.}~\bibnamefont{Cui}}, \bibnamefont{and}
  \bibinfo{author}{\bibfnamefont{K.~L.} \bibnamefont{Han}},
  \bibinfo{journal}{J. Chem. Phys.} \textbf{\bibinfo{volume}{125}},
  \bibinfo{pages}{064312} (\bibinfo{year}{2006}).

\bibitem[{\citenamefont{Merritt et~al.}(2006)\citenamefont{Merritt, K\"upper,
  and Miller}}]{Merritt:PCCP:inprep}
\bibinfo{author}{\bibfnamefont{J.~M.} \bibnamefont{Merritt}},
  \bibinfo{author}{\bibfnamefont{J.}~\bibnamefont{K\"upper}}, \bibnamefont{and}
  \bibinfo{author}{\bibfnamefont{R.~E.} \bibnamefont{Miller}},
  \bibinfo{journal}{Phys. Chem. Chem. Phys.}  (\bibinfo{year}{2006}),
  \bibinfo{note}{submitted}.

\bibitem[{\citenamefont{Merritt et~al.}(2005)\citenamefont{Merritt, K\"upper,
  and Miller}}]{Merritt:PCCP7:67}
\bibinfo{author}{\bibfnamefont{J.~M.} \bibnamefont{Merritt}},
  \bibinfo{author}{\bibfnamefont{J.}~\bibnamefont{K\"upper}}, \bibnamefont{and}
  \bibinfo{author}{\bibfnamefont{R.~E.} \bibnamefont{Miller}},
  \bibinfo{journal}{Phys. Chem. Chem. Phys.} \textbf{\bibinfo{volume}{7}},
  \bibinfo{pages}{67} (\bibinfo{year}{2005}).

\bibitem[{\citenamefont{Dubernet and
  Hutson}(1994{\natexlab{a}})}]{Dubernet:JCP101:1939}
\bibinfo{author}{\bibfnamefont{M.~L.} \bibnamefont{Dubernet}} \bibnamefont{and}
  \bibinfo{author}{\bibfnamefont{J.~M.} \bibnamefont{Hutson}},
  \bibinfo{journal}{J. Chem. Phys.} \textbf{\bibinfo{volume}{101}},
  \bibinfo{pages}{1939} (\bibinfo{year}{1994}{\natexlab{a}}).

\bibitem[{\citenamefont{Maierle et~al.}(1997)\citenamefont{Maierle, Schatz,
  Gordon, Mccabe, and Connor}}]{Maierle:JCSFT93:709}
\bibinfo{author}{\bibfnamefont{C.~S.} \bibnamefont{Maierle}},
  \bibinfo{author}{\bibfnamefont{G.~C.} \bibnamefont{Schatz}},
  \bibinfo{author}{\bibfnamefont{M.~S.} \bibnamefont{Gordon}},
  \bibinfo{author}{\bibfnamefont{P.}~\bibnamefont{Mccabe}}, \bibnamefont{and}
  \bibinfo{author}{\bibfnamefont{J.~N.} \bibnamefont{Connor}},
  \bibinfo{journal}{J. Chem. Soc. -- Faraday Trans.}
  \textbf{\bibinfo{volume}{93}}, \bibinfo{pages}{709} (\bibinfo{year}{1997}).

\bibitem[{\citenamefont{Jungwirth et~al.}(1998)\citenamefont{Jungwirth,
  Zdanska, and Schmidt}}]{Jungwirth:JPCA102:7241}
\bibinfo{author}{\bibfnamefont{P.}~\bibnamefont{Jungwirth}},
  \bibinfo{author}{\bibfnamefont{P.}~\bibnamefont{Zdanska}}, \bibnamefont{and}
  \bibinfo{author}{\bibfnamefont{B.}~\bibnamefont{Schmidt}},
  \bibinfo{journal}{J. Phys. Chem. A} \textbf{\bibinfo{volume}{102}},
  \bibinfo{pages}{7241} (\bibinfo{year}{1998}).

\bibitem[{\citenamefont{Bittererova and
  Biskupic}(1999)}]{Bittererova:CPL299:145}
\bibinfo{author}{\bibfnamefont{M.}~\bibnamefont{Bittererova}} \bibnamefont{and}
  \bibinfo{author}{\bibfnamefont{S.}~\bibnamefont{Biskupic}},
  \bibinfo{journal}{Chem. Phys. Lett.} \textbf{\bibinfo{volume}{299}},
  \bibinfo{pages}{145} (\bibinfo{year}{1999}).

\bibitem[{\citenamefont{Meuwly and
  Hutson}(2000{\natexlab{a}})}]{Meuwly:PCCP2:441}
\bibinfo{author}{\bibfnamefont{M.}~\bibnamefont{Meuwly}} \bibnamefont{and}
  \bibinfo{author}{\bibfnamefont{J.~M.} \bibnamefont{Hutson}},
  \bibinfo{journal}{Phys. Chem. Chem. Phys.} \textbf{\bibinfo{volume}{2}},
  \bibinfo{pages}{441} (\bibinfo{year}{2000}{\natexlab{a}}).

\bibitem[{\citenamefont{Meuwly and
  Hutson}(2000{\natexlab{b}})}]{Meuwly:JCP112:592}
\bibinfo{author}{\bibfnamefont{M.}~\bibnamefont{Meuwly}} \bibnamefont{and}
  \bibinfo{author}{\bibfnamefont{J.~M.} \bibnamefont{Hutson}},
  \bibinfo{journal}{J. Chem. Phys.} \textbf{\bibinfo{volume}{112}},
  \bibinfo{pages}{592} (\bibinfo{year}{2000}{\natexlab{b}}).

\bibitem[{\citenamefont{Meuwly and Hutson}(2003)}]{Meuwly:JCP119:8873}
\bibinfo{author}{\bibfnamefont{M.}~\bibnamefont{Meuwly}} \bibnamefont{and}
  \bibinfo{author}{\bibfnamefont{J.~M.} \bibnamefont{Hutson}},
  \bibinfo{journal}{J. Chem. Phys.} \textbf{\bibinfo{volume}{119}},
  \bibinfo{pages}{8873} (\bibinfo{year}{2003}).

\bibitem[{\citenamefont{K{\l}os et~al.}(2001)\citenamefont{K{\l}os,
  Cha{\l}asi\'nski, Szcz\c{e}\'sniak, and Werner}}]{Klos:JCP115:3085}
\bibinfo{author}{\bibfnamefont{J.~A.} \bibnamefont{K{\l}os}},
  \bibinfo{author}{\bibfnamefont{G.}~\bibnamefont{Cha{\l}asi\'nski}},
  \bibinfo{author}{\bibfnamefont{M.~M.} \bibnamefont{Szcz\c{e}\'sniak}},
  \bibnamefont{and} \bibinfo{author}{\bibfnamefont{H.-J.}
  \bibnamefont{Werner}}, \bibinfo{journal}{J. Chem. Phys.}
  \textbf{\bibinfo{volume}{115}}, \bibinfo{pages}{3085} (\bibinfo{year}{2001}).

\bibitem[{\citenamefont{Zeimen et~al.}(2003)\citenamefont{Zeimen, K{\l}os,
  Groenenboom, and van~der Avoird}}]{Zeimen:JPCA107:5110}
\bibinfo{author}{\bibfnamefont{W.~B.} \bibnamefont{Zeimen}},
  \bibinfo{author}{\bibfnamefont{J.~A.} \bibnamefont{K{\l}os}},
  \bibinfo{author}{\bibfnamefont{G.~C.} \bibnamefont{Groenenboom}},
  \bibnamefont{and} \bibinfo{author}{\bibfnamefont{A.}~\bibnamefont{van~der
  Avoird}}, \bibinfo{journal}{J. Phys. Chem. A} \textbf{\bibinfo{volume}{107}},
  \bibinfo{pages}{5110} (\bibinfo{year}{2003}), \bibinfo{note}{see erratum:
  \cite{Zeimen:JPCA108:9319}}.

\bibitem[{\citenamefont{Neumark}(1992)}]{Neumark:ARPC43:153}
\bibinfo{author}{\bibfnamefont{D.~M.} \bibnamefont{Neumark}},
  \bibinfo{journal}{Ann. Rev. Phys. Chem.} \textbf{\bibinfo{volume}{43}},
  \bibinfo{pages}{153} (\bibinfo{year}{1992}).

\bibitem[{\citenamefont{Deskevich et~al.}(2006)\citenamefont{Deskevich, Hayes,
  Takahashi, Skodje, and Nesbitt}}]{Deskevich:JCP124:224303}
\bibinfo{author}{\bibfnamefont{M.~P.} \bibnamefont{Deskevich}},
  \bibinfo{author}{\bibfnamefont{M.~Y.} \bibnamefont{Hayes}},
  \bibinfo{author}{\bibfnamefont{K.}~\bibnamefont{Takahashi}},
  \bibinfo{author}{\bibfnamefont{R.~T.} \bibnamefont{Skodje}},
  \bibnamefont{and} \bibinfo{author}{\bibfnamefont{D.~J.}
  \bibnamefont{Nesbitt}}, \bibinfo{journal}{J. Chem. Phys.}
  \textbf{\bibinfo{volume}{124}}, \bibinfo{pages}{224303}
  (\bibinfo{year}{2006}).

\bibitem[{\citenamefont{Fishchuk
  et~al.}(2006{\natexlab{a}})\citenamefont{Fishchuk, Wormer, and van~der
  Avoird}}]{Fishchuk:JPCA110:5273}
\bibinfo{author}{\bibfnamefont{A.~V.} \bibnamefont{Fishchuk}},
  \bibinfo{author}{\bibfnamefont{P.~E.~S.} \bibnamefont{Wormer}},
  \bibnamefont{and} \bibinfo{author}{\bibfnamefont{A.}~\bibnamefont{van~der
  Avoird}}, \bibinfo{journal}{Journal Of Physical Chemistry A}
  \textbf{\bibinfo{volume}{110}}, \bibinfo{pages}{5273}
  (\bibinfo{year}{2006}{\natexlab{a}}).

\bibitem[{\citenamefont{Fishchuk
  et~al.}(2006{\natexlab{b}})\citenamefont{Fishchuk, Groenenboom, and van~der
  Avoird}}]{Fishchuk:JPCA110:5280}
\bibinfo{author}{\bibfnamefont{A.~V.} \bibnamefont{Fishchuk}},
  \bibinfo{author}{\bibfnamefont{G.~C.} \bibnamefont{Groenenboom}},
  \bibnamefont{and} \bibinfo{author}{\bibfnamefont{A.}~\bibnamefont{van~der
  Avoird}}, \bibinfo{journal}{Journal Of Physical Chemistry A}
  \textbf{\bibinfo{volume}{110}}, \bibinfo{pages}{5280}
  (\bibinfo{year}{2006}{\natexlab{b}}).

\bibitem[{\citenamefont{Dubernet and
  Hutson}(1994{\natexlab{b}})}]{Dubernet:JPC98:5844}
\bibinfo{author}{\bibfnamefont{M.~L.} \bibnamefont{Dubernet}} \bibnamefont{and}
  \bibinfo{author}{\bibfnamefont{J.}~\bibnamefont{Hutson}},
  \bibinfo{journal}{J. Phys. Chem.} \textbf{\bibinfo{volume}{98}},
  \bibinfo{pages}{5844} (\bibinfo{year}{1994}{\natexlab{b}}).

\bibitem[{\citenamefont{Liu et~al.}(1999)\citenamefont{Liu, Kolessov, Partin,
  Bezel, and Wittig}}]{Liu:CPL299:374}
\bibinfo{author}{\bibfnamefont{K.}~\bibnamefont{Liu}},
  \bibinfo{author}{\bibfnamefont{A.}~\bibnamefont{Kolessov}},
  \bibinfo{author}{\bibfnamefont{J.~W.} \bibnamefont{Partin}},
  \bibinfo{author}{\bibfnamefont{I.}~\bibnamefont{Bezel}}, \bibnamefont{and}
  \bibinfo{author}{\bibfnamefont{C.}~\bibnamefont{Wittig}},
  \bibinfo{journal}{Chem. Phys. Lett.} \textbf{\bibinfo{volume}{299}},
  \bibinfo{pages}{374} (\bibinfo{year}{1999}).

\bibitem[{\citenamefont{Imura et~al.}(2000)\citenamefont{Imura, Ohoyama,
  Naaman, Che, Hashinokuchi, and Kasai}}]{Imura:JMolStruct552:137}
\bibinfo{author}{\bibfnamefont{K.}~\bibnamefont{Imura}},
  \bibinfo{author}{\bibfnamefont{H.}~\bibnamefont{Ohoyama}},
  \bibinfo{author}{\bibfnamefont{R.}~\bibnamefont{Naaman}},
  \bibinfo{author}{\bibfnamefont{D.~C.} \bibnamefont{Che}},
  \bibinfo{author}{\bibfnamefont{M.}~\bibnamefont{Hashinokuchi}},
  \bibnamefont{and} \bibinfo{author}{\bibfnamefont{T.}~\bibnamefont{Kasai}},
  \bibinfo{journal}{J. Mol. Struct.} \textbf{\bibinfo{volume}{552}},
  \bibinfo{pages}{137} (\bibinfo{year}{2000}).

\bibitem[{\citenamefont{Mestdagh et~al.}(2003)\citenamefont{Mestdagh, Soep,
  Gaveau, and Visticot}}]{Mestdagh:IRPC22:285}
\bibinfo{author}{\bibfnamefont{J.~M.} \bibnamefont{Mestdagh}},
  \bibinfo{author}{\bibfnamefont{B.}~\bibnamefont{Soep}},
  \bibinfo{author}{\bibfnamefont{M.~A.} \bibnamefont{Gaveau}},
  \bibnamefont{and} \bibinfo{author}{\bibfnamefont{J.~P.}
  \bibnamefont{Visticot}}, \bibinfo{journal}{Int. Rev. Phys. Chem.}
  \textbf{\bibinfo{volume}{22}}, \bibinfo{pages}{285} (\bibinfo{year}{2003}).

\bibitem[{\citenamefont{Andrews and Hunt}(1988)}]{Andrews:JCP89:3502}
\bibinfo{author}{\bibfnamefont{L.}~\bibnamefont{Andrews}} \bibnamefont{and}
  \bibinfo{author}{\bibfnamefont{R.~D.} \bibnamefont{Hunt}},
  \bibinfo{journal}{J. Chem. Phys.} \textbf{\bibinfo{volume}{89}},
  \bibinfo{pages}{3502} (\bibinfo{year}{1988}).

\bibitem[{\citenamefont{Hunt and
  Andrews}(1988{\natexlab{a}})}]{Hunt:JCP88:3599}
\bibinfo{author}{\bibfnamefont{R.~D.} \bibnamefont{Hunt}} \bibnamefont{and}
  \bibinfo{author}{\bibfnamefont{L.}~\bibnamefont{Andrews}},
  \bibinfo{journal}{J. Chem. Phys.} \textbf{\bibinfo{volume}{88}},
  \bibinfo{pages}{3599} (\bibinfo{year}{1988}{\natexlab{a}}).

\bibitem[{\citenamefont{Hunt and
  Andrews}(1988{\natexlab{b}})}]{Hunt:JPC92:3769}
\bibinfo{author}{\bibfnamefont{R.~D.} \bibnamefont{Hunt}} \bibnamefont{and}
  \bibinfo{author}{\bibfnamefont{L.}~\bibnamefont{Andrews}},
  \bibinfo{journal}{J. Phys. Chem.} \textbf{\bibinfo{volume}{92}},
  \bibinfo{pages}{3769} (\bibinfo{year}{1988}{\natexlab{b}}).

\bibitem[{\citenamefont{Ding et~al.}(1973)\citenamefont{Ding, Kirsch, Perry,
  Polanyi, and Schreiber}}]{Ding:FDCS55:252}
\bibinfo{author}{\bibfnamefont{A.~M.~G.} \bibnamefont{Ding}},
  \bibinfo{author}{\bibfnamefont{L.~J.} \bibnamefont{Kirsch}},
  \bibinfo{author}{\bibfnamefont{D.~S.} \bibnamefont{Perry}},
  \bibinfo{author}{\bibfnamefont{J.~C.} \bibnamefont{Polanyi}},
  \bibnamefont{and} \bibinfo{author}{\bibfnamefont{J.~L.}
  \bibnamefont{Schreiber}}, \bibinfo{journal}{Faraday Disc. Chem. Soc}
  \textbf{\bibinfo{volume}{55}}, \bibinfo{pages}{252} (\bibinfo{year}{1973}).

\bibitem[{\citenamefont{W\"urzberg et~al.}(1978)\citenamefont{W\"urzberg,
  Grimley, and Houston}}]{Wuerzberg:CPL57:373}
\bibinfo{author}{\bibfnamefont{E.}~\bibnamefont{W\"urzberg}},
  \bibinfo{author}{\bibfnamefont{A.~J.} \bibnamefont{Grimley}},
  \bibnamefont{and} \bibinfo{author}{\bibfnamefont{P.~L.}
  \bibnamefont{Houston}}, \bibinfo{journal}{Chem. Phys. Lett.}
  \textbf{\bibinfo{volume}{57}}, \bibinfo{pages}{373} (\bibinfo{year}{1978}).

\bibitem[{\citenamefont{W\"urzberg and Houston}(1980)}]{Wuerzberg:JCP72:5915}
\bibinfo{author}{\bibfnamefont{E.}~\bibnamefont{W\"urzberg}} \bibnamefont{and}
  \bibinfo{author}{\bibfnamefont{P.~L.} \bibnamefont{Houston}},
  \bibinfo{journal}{J. Chem. Phys.} \textbf{\bibinfo{volume}{72}},
  \bibinfo{pages}{5915} (\bibinfo{year}{1980}).

\bibitem[{\citenamefont{Moore et~al.}(1994)\citenamefont{Moore, Smith, and
  Stewart}}]{Moore:IJCK26:813}
\bibinfo{author}{\bibfnamefont{C.~M.} \bibnamefont{Moore}},
  \bibinfo{author}{\bibfnamefont{I.~W.~M.} \bibnamefont{Smith}},
  \bibnamefont{and} \bibinfo{author}{\bibfnamefont{D.~W.~A.}
  \bibnamefont{Stewart}}, \bibinfo{journal}{Int. J. Chem. Kin.}
  \textbf{\bibinfo{volume}{26}}, \bibinfo{pages}{813} (\bibinfo{year}{1994}).

\bibitem[{\citenamefont{Tamagake et~al.}(1980)\citenamefont{Tamagake, Setser,
  and Sung}}]{Tamagake:JCP73:2203}
\bibinfo{author}{\bibfnamefont{K.}~\bibnamefont{Tamagake}},
  \bibinfo{author}{\bibfnamefont{D.~W.} \bibnamefont{Setser}},
  \bibnamefont{and} \bibinfo{author}{\bibfnamefont{J.~P.} \bibnamefont{Sung}},
  \bibinfo{journal}{J. Chem. Phys.} \textbf{\bibinfo{volume}{73}},
  \bibinfo{pages}{2203} (\bibinfo{year}{1980}).

\bibitem[{\citenamefont{Zolot and Nesbitt}()}]{Zolot:F+HCl:inprep}
\bibinfo{author}{\bibfnamefont{A.}~\bibnamefont{Zolot}} \bibnamefont{and}
  \bibinfo{author}{\bibfnamefont{D.~J.} \bibnamefont{Nesbitt}},
  \bibinfo{note}{manuscript in preparation}.

\bibitem[{\citenamefont{Metz et~al.}(1994)\citenamefont{Metz, Pfeiffer,
  Thoemke, and Crim}}]{Metz:CPL221:347}
\bibinfo{author}{\bibfnamefont{R.~B.} \bibnamefont{Metz}},
  \bibinfo{author}{\bibfnamefont{J.~M.} \bibnamefont{Pfeiffer}},
  \bibinfo{author}{\bibfnamefont{J.~D.} \bibnamefont{Thoemke}},
  \bibnamefont{and} \bibinfo{author}{\bibfnamefont{F.~F.} \bibnamefont{Crim}},
  \bibinfo{journal}{Chem. Phys. Lett.} \textbf{\bibinfo{volume}{221}},
  \bibinfo{pages}{347} (\bibinfo{year}{1994}).

\bibitem[{\citenamefont{Kreher et~al.}(1996)\citenamefont{Kreher, Theinl, and
  Gericke}}]{Kreher:JCP104:4481}
\bibinfo{author}{\bibfnamefont{C.}~\bibnamefont{Kreher}},
  \bibinfo{author}{\bibfnamefont{R.}~\bibnamefont{Theinl}}, \bibnamefont{and}
  \bibinfo{author}{\bibfnamefont{K.~H.} \bibnamefont{Gericke}},
  \bibinfo{journal}{J. Chem. Phys.} \textbf{\bibinfo{volume}{104}},
  \bibinfo{pages}{4481} (\bibinfo{year}{1996}).

\bibitem[{\citenamefont{Decker et~al.}(2001)\citenamefont{Decker, He, Tokue,
  and Macdonald}}]{Decker:JPCA105:5759}
\bibinfo{author}{\bibfnamefont{B.~K.} \bibnamefont{Decker}},
  \bibinfo{author}{\bibfnamefont{G.}~\bibnamefont{He}},
  \bibinfo{author}{\bibfnamefont{I.}~\bibnamefont{Tokue}}, \bibnamefont{and}
  \bibinfo{author}{\bibfnamefont{R.~G.} \bibnamefont{Macdonald}},
  \bibinfo{journal}{J. Phys. Chem. A} \textbf{\bibinfo{volume}{105}},
  \bibinfo{pages}{5759} (\bibinfo{year}{2001}).

\bibitem[{\citenamefont{Hayes et~al.}(2006)\citenamefont{Hayes, Deskevich,
  Nesbitt, Takahashi, and Skodje}}]{Hayes:JPCA110:436}
\bibinfo{author}{\bibfnamefont{M.~Y.} \bibnamefont{Hayes}},
  \bibinfo{author}{\bibfnamefont{M.~P.} \bibnamefont{Deskevich}},
  \bibinfo{author}{\bibfnamefont{D.~J.} \bibnamefont{Nesbitt}},
  \bibinfo{author}{\bibfnamefont{K.}~\bibnamefont{Takahashi}},
  \bibnamefont{and} \bibinfo{author}{\bibfnamefont{R.~T.}
  \bibnamefont{Skodje}}, \bibinfo{journal}{J. Phys. Chem. A}
  \textbf{\bibinfo{volume}{110}}, \bibinfo{pages}{436} (\bibinfo{year}{2006}).

\bibitem[{\citenamefont{Krogh and Pimentel}(1977)}]{Krogh:JCP67:2993}
\bibinfo{author}{\bibfnamefont{O.~D.} \bibnamefont{Krogh}} \bibnamefont{and}
  \bibinfo{author}{\bibfnamefont{G.~C.} \bibnamefont{Pimentel}},
  \bibinfo{journal}{J. Chem. Phys.} \textbf{\bibinfo{volume}{67}},
  \bibinfo{pages}{2993} (\bibinfo{year}{1977}).

\bibitem[{\citenamefont{Loesch and Remscheid}(1990)}]{Loesch:JCP93:4779}
\bibinfo{author}{\bibfnamefont{H.~J.} \bibnamefont{Loesch}} \bibnamefont{and}
  \bibinfo{author}{\bibfnamefont{A.}~\bibnamefont{Remscheid}},
  \bibinfo{journal}{J. Chem. Phys.} \textbf{\bibinfo{volume}{93}},
  \bibinfo{pages}{4779} (\bibinfo{year}{1990}).

\bibitem[{\citenamefont{Rost et~al.}(1992)\citenamefont{Rost, Griffin,
  Friedrich, and Herschbach}}]{Rost:PRL68:1299}
\bibinfo{author}{\bibfnamefont{J.~M.} \bibnamefont{Rost}},
  \bibinfo{author}{\bibfnamefont{J.~C.} \bibnamefont{Griffin}},
  \bibinfo{author}{\bibfnamefont{B.}~\bibnamefont{Friedrich}},
  \bibnamefont{and} \bibinfo{author}{\bibfnamefont{D.~R.}
  \bibnamefont{Herschbach}}, \bibinfo{journal}{Phys. Rev. Lett.}
  \textbf{\bibinfo{volume}{68}}, \bibinfo{pages}{1299} (\bibinfo{year}{1992}).

\bibitem[{\citenamefont{Merritt}(2006)}]{Merritt:thesis:2006}
\bibinfo{author}{\bibfnamefont{J.~M.} \bibnamefont{Merritt}},
  \bibinfo{type}{{Ph.\,D.} thesis}, \bibinfo{school}{University of North
  Carolina at Chapel Hill}, \bibinfo{address}{Chapel Hill, NC, USA}
  (\bibinfo{year}{2006}).

\bibitem[{\citenamefont{Hunt and Andrews}(1987)}]{Hunt:IC26:3051}
\bibinfo{author}{\bibfnamefont{R.~D.} \bibnamefont{Hunt}} \bibnamefont{and}
  \bibinfo{author}{\bibfnamefont{L.}~\bibnamefont{Andrews}},
  \bibinfo{journal}{Inorg. Chem.} \textbf{\bibinfo{volume}{26}},
  \bibinfo{pages}{3051} (\bibinfo{year}{1987}).

\bibitem[{\citenamefont{Goldschleger et~al.}(2001)\citenamefont{Goldschleger,
  Akimov, Misochko, and Wight}}]{Goldschleger:MC2001:43}
\bibinfo{author}{\bibfnamefont{I.~U.} \bibnamefont{Goldschleger}},
  \bibinfo{author}{\bibfnamefont{A.~V.} \bibnamefont{Akimov}},
  \bibinfo{author}{\bibfnamefont{E.~Y.} \bibnamefont{Misochko}},
  \bibnamefont{and} \bibinfo{author}{\bibfnamefont{C.~A.} \bibnamefont{Wight}},
  \bibinfo{journal}{Mendeleev Comm.} pp. \bibinfo{pages}{43--45}
  (\bibinfo{year}{2001}).

\bibitem[{\citenamefont{Schneider et~al.}(1997)\citenamefont{Schneider,
  Kramper, Schiller, and Mlynek}}]{Schneider:OL22:1293}
\bibinfo{author}{\bibfnamefont{K.}~\bibnamefont{Schneider}},
  \bibinfo{author}{\bibfnamefont{P.}~\bibnamefont{Kramper}},
  \bibinfo{author}{\bibfnamefont{S.}~\bibnamefont{Schiller}}, \bibnamefont{and}
  \bibinfo{author}{\bibfnamefont{J.}~\bibnamefont{Mlynek}},
  \bibinfo{journal}{Opt. Lett.} \textbf{\bibinfo{volume}{22}},
  \bibinfo{pages}{1293} (\bibinfo{year}{1997}).

\bibitem[{\citenamefont{van Herpen et~al.}(2002)\citenamefont{van Herpen,
  te~Lintel~Hekkert, Bisson, and Harren}}]{Herpen:OL27:640}
\bibinfo{author}{\bibfnamefont{M.}~\bibnamefont{van Herpen}},
  \bibinfo{author}{\bibfnamefont{S.}~\bibnamefont{te~Lintel~Hekkert}},
  \bibinfo{author}{\bibfnamefont{S.~E.} \bibnamefont{Bisson}},
  \bibnamefont{and} \bibinfo{author}{\bibfnamefont{F.~J.~M.}
  \bibnamefont{Harren}}, \bibinfo{journal}{Opt. Lett.}
  \textbf{\bibinfo{volume}{27}}, \bibinfo{pages}{640} (\bibinfo{year}{2002}).

\bibitem[{\citenamefont{Merritt and Miller}()}]{Merritt:inprep:HCN-Ga}
\bibinfo{author}{\bibfnamefont{J.~M.} \bibnamefont{Merritt}} \bibnamefont{and}
  \bibinfo{author}{\bibfnamefont{R.~E.} \bibnamefont{Miller}},
  \bibinfo{note}{in preparation}.

\bibitem[{\citenamefont{Douberly et~al.}()\citenamefont{Douberly, Stiles, and
  Miller}}]{Douberly:metals}
\bibinfo{author}{\bibfnamefont{G.~E.} \bibnamefont{Douberly}},
  \bibinfo{author}{\bibfnamefont{P.~L.} \bibnamefont{Stiles}},
  \bibnamefont{and} \bibinfo{author}{\bibfnamefont{R.~E.}
  \bibnamefont{Miller}}, \bibinfo{note}{to be published}.

\bibitem[{\citenamefont{Douberly}(2006)}]{Douberly:thesis:2006}
\bibinfo{author}{\bibfnamefont{G.~E.} \bibnamefont{Douberly}},
  \bibinfo{type}{{Ph.\,D.} thesis}, \bibinfo{school}{University of North
  Carolina at Chapel Hill}, \bibinfo{address}{Chapel Hill, NC, USA}
  (\bibinfo{year}{2006}).

\bibitem[{\citenamefont{Stiles and Miller}(2006)}]{Stiles:JPCA110:5620}
\bibinfo{author}{\bibfnamefont{P.~L.} \bibnamefont{Stiles}} \bibnamefont{and}
  \bibinfo{author}{\bibfnamefont{R.~E.} \bibnamefont{Miller}},
  \bibinfo{journal}{J. Phys. Chem. A} \textbf{\bibinfo{volume}{110}},
  \bibinfo{pages}{5620} (\bibinfo{year}{2006}).

\bibitem[{\citenamefont{Nauta et~al.}(2001)\citenamefont{Nauta, Moore, Stiles,
  and Miller}}]{Nauta:Science292:481}
\bibinfo{author}{\bibfnamefont{K.}~\bibnamefont{Nauta}},
  \bibinfo{author}{\bibfnamefont{D.~T.} \bibnamefont{Moore}},
  \bibinfo{author}{\bibfnamefont{P.~L.} \bibnamefont{Stiles}},
  \bibnamefont{and} \bibinfo{author}{\bibfnamefont{R.~E.}
  \bibnamefont{Miller}}, \bibinfo{journal}{Science}
  \textbf{\bibinfo{volume}{292}}, \bibinfo{pages}{481} (\bibinfo{year}{2001}).

\bibitem[{\citenamefont{Stiles et~al.}(2004)\citenamefont{Stiles, Moore, and
  Miller}}]{Stiles:JCP121:3130}
\bibinfo{author}{\bibfnamefont{P.~L.} \bibnamefont{Stiles}},
  \bibinfo{author}{\bibfnamefont{D.~T.} \bibnamefont{Moore}}, \bibnamefont{and}
  \bibinfo{author}{\bibfnamefont{R.~E.} \bibnamefont{Miller}},
  \bibinfo{journal}{J. Chem. Phys.} \textbf{\bibinfo{volume}{121}},
  \bibinfo{pages}{3130} (\bibinfo{year}{2004}).

\bibitem[{\citenamefont{Moore and
  Miller}(2004{\natexlab{a}})}]{Moore:JPCA108:9908}
\bibinfo{author}{\bibfnamefont{D.~T.} \bibnamefont{Moore}} \bibnamefont{and}
  \bibinfo{author}{\bibfnamefont{R.~E.} \bibnamefont{Miller}},
  \bibinfo{journal}{J. Phys. Chem. A} \textbf{\bibinfo{volume}{108}},
  \bibinfo{pages}{9908} (\bibinfo{year}{2004}{\natexlab{a}}).

\bibitem[{\citenamefont{Dong and Miller}(2004)}]{Dong:JPCA108:2181}
\bibinfo{author}{\bibfnamefont{F.}~\bibnamefont{Dong}} \bibnamefont{and}
  \bibinfo{author}{\bibfnamefont{R.~E.} \bibnamefont{Miller}},
  \bibinfo{journal}{J. Phys. Chem. A} \textbf{\bibinfo{volume}{108}},
  \bibinfo{pages}{2181} (\bibinfo{year}{2004}).

\bibitem[{\citenamefont{Shiu et~al.}(2004)\citenamefont{Shiu, Lin, and
  Liu}}]{Shiu:PRL92:103201}
\bibinfo{author}{\bibfnamefont{W.}~\bibnamefont{Shiu}},
  \bibinfo{author}{\bibfnamefont{J.~J.} \bibnamefont{Lin}}, \bibnamefont{and}
  \bibinfo{author}{\bibfnamefont{K.}~\bibnamefont{Liu}},
  \bibinfo{journal}{Phys. Rev. Lett.} \textbf{\bibinfo{volume}{92}},
  \bibinfo{eid}{103201} (\bibinfo{year}{2004}).

\bibitem[{\citenamefont{Chu et~al.}(2006)\citenamefont{Chu, Zhang, Ju, Yao,
  Han, Wang, and Zhang}}]{Chu:CPL424:243}
\bibinfo{author}{\bibfnamefont{T.~S.} \bibnamefont{Chu}},
  \bibinfo{author}{\bibfnamefont{X.}~\bibnamefont{Zhang}},
  \bibinfo{author}{\bibfnamefont{L.~P.} \bibnamefont{Ju}},
  \bibinfo{author}{\bibfnamefont{L.}~\bibnamefont{Yao}},
  \bibinfo{author}{\bibfnamefont{K.~L.} \bibnamefont{Han}},
  \bibinfo{author}{\bibfnamefont{M.~L.} \bibnamefont{Wang}}, \bibnamefont{and}
  \bibinfo{author}{\bibfnamefont{J.~Z.~H.} \bibnamefont{Zhang}},
  \bibinfo{journal}{Chem. Phys. Lett.} \textbf{\bibinfo{volume}{424}},
  \bibinfo{pages}{243} (\bibinfo{year}{2006}).

\bibitem[{\citenamefont{Johnson and Andrews}(1980)}]{Johnson:JACS102:5736}
\bibinfo{author}{\bibfnamefont{G.~L.} \bibnamefont{Johnson}} \bibnamefont{and}
  \bibinfo{author}{\bibfnamefont{L.}~\bibnamefont{Andrews}},
  \bibinfo{journal}{J. Am. Chem. Soc.} \textbf{\bibinfo{volume}{102}},
  \bibinfo{pages}{5736} (\bibinfo{year}{1980}).

\bibitem[{\citenamefont{Misochko et~al.}(1995)\citenamefont{Misochko,
  Benderskii, Goldschleger, Akimov, and Shestakov}}]{Misochko:JACS117:11997}
\bibinfo{author}{\bibfnamefont{E.~Y.} \bibnamefont{Misochko}},
  \bibinfo{author}{\bibfnamefont{V.~A.} \bibnamefont{Benderskii}},
  \bibinfo{author}{\bibfnamefont{A.~U.} \bibnamefont{Goldschleger}},
  \bibinfo{author}{\bibfnamefont{A.~V.} \bibnamefont{Akimov}},
  \bibnamefont{and} \bibinfo{author}{\bibfnamefont{A.~F.}
  \bibnamefont{Shestakov}}, \bibinfo{journal}{J. Am. Chem. Soc.}
  \textbf{\bibinfo{volume}{117}}, \bibinfo{pages}{11997}
  (\bibinfo{year}{1995}).

\bibitem[{\citenamefont{Misochko et~al.}(1996)\citenamefont{Misochko,
  Benderskii, Goldschleger, Akimov, Benderskii, and
  Wight}}]{Misochko:JCP106:3146}
\bibinfo{author}{\bibfnamefont{E.~Y.} \bibnamefont{Misochko}},
  \bibinfo{author}{\bibfnamefont{V.~A.} \bibnamefont{Benderskii}},
  \bibinfo{author}{\bibfnamefont{A.~U.} \bibnamefont{Goldschleger}},
  \bibinfo{author}{\bibfnamefont{A.~V.} \bibnamefont{Akimov}},
  \bibinfo{author}{\bibfnamefont{A.~V.} \bibnamefont{Benderskii}},
  \bibnamefont{and} \bibinfo{author}{\bibfnamefont{C.~A.} \bibnamefont{Wight}},
  \bibinfo{journal}{J. Chem. Phys.} \textbf{\bibinfo{volume}{106}},
  \bibinfo{pages}{3146} (\bibinfo{year}{1996}).

\bibitem[{\citenamefont{Liu et~al.}(2000)\citenamefont{Liu, Gomez, and
  Neumark}}]{Liu:CPL332:65}
\bibinfo{author}{\bibfnamefont{Z.}~\bibnamefont{Liu}},
  \bibinfo{author}{\bibfnamefont{H.}~\bibnamefont{Gomez}}, \bibnamefont{and}
  \bibinfo{author}{\bibfnamefont{D.~M.} \bibnamefont{Neumark}},
  \bibinfo{journal}{Chem. Phys. Lett.} \textbf{\bibinfo{volume}{332}},
  \bibinfo{pages}{65} (\bibinfo{year}{2000}).

\bibitem[{\citenamefont{Rudi\'c et~al.}(2006)\citenamefont{Rudi\'c, Merritt,
  and Miller}}]{Rudic:JCP124:104305}
\bibinfo{author}{\bibfnamefont{S.}~\bibnamefont{Rudi\'c}},
  \bibinfo{author}{\bibfnamefont{J.~M.} \bibnamefont{Merritt}},
  \bibnamefont{and} \bibinfo{author}{\bibfnamefont{R.~E.}
  \bibnamefont{Miller}}, \bibinfo{journal}{J. Chem. Phys.}
  \textbf{\bibinfo{volume}{124}}, \bibinfo{eid}{104305} (\bibinfo{year}{2006}).

\bibitem[{\citenamefont{Liu}(2001)}]{Liu:ARPC52:139}
\bibinfo{author}{\bibfnamefont{K.~P.} \bibnamefont{Liu}},
  \bibinfo{journal}{Ann. Rev. Phys. Chem.} \textbf{\bibinfo{volume}{52}},
  \bibinfo{pages}{139} (\bibinfo{year}{2001}).

\bibitem[{\citenamefont{Gilbert et~al.}(1999)\citenamefont{Gilbert, Grebner,
  Fischer, and Chen}}]{Gilbert:JCP110:5485}
\bibinfo{author}{\bibfnamefont{T.}~\bibnamefont{Gilbert}},
  \bibinfo{author}{\bibfnamefont{T.~L.} \bibnamefont{Grebner}},
  \bibinfo{author}{\bibfnamefont{I.}~\bibnamefont{Fischer}}, \bibnamefont{and}
  \bibinfo{author}{\bibfnamefont{P.}~\bibnamefont{Chen}},
  \bibinfo{journal}{Journal Of Chemical Physics}
  \textbf{\bibinfo{volume}{110}}, \bibinfo{pages}{5485} (\bibinfo{year}{1999}).

\bibitem[{\citenamefont{Castiglioni et~al.}(2005)\citenamefont{Castiglioni,
  Bach, and Chen}}]{Castiglioni:JPCA109:962}
\bibinfo{author}{\bibfnamefont{L.}~\bibnamefont{Castiglioni}},
  \bibinfo{author}{\bibfnamefont{A.}~\bibnamefont{Bach}}, \bibnamefont{and}
  \bibinfo{author}{\bibfnamefont{P.}~\bibnamefont{Chen}}, \bibinfo{journal}{J.
  Phys. Chem. A} \textbf{\bibinfo{volume}{109}}, \bibinfo{pages}{962}
  (\bibinfo{year}{2005}).

\bibitem[{\citenamefont{Blanksby and Ellison}(2003)}]{Blanksby:ACR36:255}
\bibinfo{author}{\bibfnamefont{S.~J.} \bibnamefont{Blanksby}} \bibnamefont{and}
  \bibinfo{author}{\bibfnamefont{G.~B.} \bibnamefont{Ellison}},
  \bibinfo{journal}{Acct. Chem. Res.} \textbf{\bibinfo{volume}{36}},
  \bibinfo{pages}{255} (\bibinfo{year}{2003}).

\bibitem[{\citenamefont{Curran et~al.}(1998)\citenamefont{Curran, Gaffuri,
  Pitz, and Westbrook}}]{Curran:CombustFlame114:149}
\bibinfo{author}{\bibfnamefont{H.~J.} \bibnamefont{Curran}},
  \bibinfo{author}{\bibfnamefont{P.}~\bibnamefont{Gaffuri}},
  \bibinfo{author}{\bibfnamefont{W.~J.} \bibnamefont{Pitz}}, \bibnamefont{and}
  \bibinfo{author}{\bibfnamefont{C.~K.} \bibnamefont{Westbrook}},
  \bibinfo{journal}{Combust. Flame} \textbf{\bibinfo{volume}{114}},
  \bibinfo{pages}{149} (\bibinfo{year}{1998}).

\bibitem[{\citenamefont{Nielsen and
  Wallington}(1997)}]{Nielsen:PeroxyRadicals:UV}
\bibinfo{author}{\bibfnamefont{O.~J.} \bibnamefont{Nielsen}} \bibnamefont{and}
  \bibinfo{author}{\bibfnamefont{T.~J.} \bibnamefont{Wallington}},
  \emph{\bibinfo{title}{Ultraviolet Absorption Spectra of Peroxy Radicals in
  the Gas Phase}} (\bibinfo{publisher}{John Wiley \& Sons},
  \bibinfo{address}{New York, NY, USA}, \bibinfo{year}{1997}), pp.
  \bibinfo{pages}{72--73}.

\bibitem[{\citenamefont{Pushkarsky et~al.}(2000)\citenamefont{Pushkarsky,
  Zalyubovsky, and Miller}}]{Pushkarsky:JCP112:10695}
\bibinfo{author}{\bibfnamefont{M.~B.} \bibnamefont{Pushkarsky}},
  \bibinfo{author}{\bibfnamefont{S.~J.} \bibnamefont{Zalyubovsky}},
  \bibnamefont{and} \bibinfo{author}{\bibfnamefont{T.~A.}
  \bibnamefont{Miller}}, \bibinfo{journal}{J. Chem. Phys.}
  \textbf{\bibinfo{volume}{112}}, \bibinfo{pages}{10695}
  (\bibinfo{year}{2000}).

\bibitem[{\citenamefont{Zalyubovsky et~al.}(2003)\citenamefont{Zalyubovsky,
  Glover, and Miller}}]{Zalyubovsky:JPCA107:7704}
\bibinfo{author}{\bibfnamefont{S.~J.} \bibnamefont{Zalyubovsky}},
  \bibinfo{author}{\bibfnamefont{B.~G.} \bibnamefont{Glover}},
  \bibnamefont{and} \bibinfo{author}{\bibfnamefont{T.~A.}
  \bibnamefont{Miller}}, \bibinfo{journal}{J. Phys. Chem. A}
  \textbf{\bibinfo{volume}{107}}, \bibinfo{pages}{7704} (\bibinfo{year}{2003}).

\bibitem[{\citenamefont{Zalyubovsky et~al.}(2005)\citenamefont{Zalyubovsky,
  Glover, Miller, Hayes, Merle, and Hadad}}]{Zalyubovsky:JPCA109:1308}
\bibinfo{author}{\bibfnamefont{S.~J.} \bibnamefont{Zalyubovsky}},
  \bibinfo{author}{\bibfnamefont{B.~G.} \bibnamefont{Glover}},
  \bibinfo{author}{\bibfnamefont{T.~A.} \bibnamefont{Miller}},
  \bibinfo{author}{\bibfnamefont{C.}~\bibnamefont{Hayes}},
  \bibinfo{author}{\bibfnamefont{J.~K.} \bibnamefont{Merle}}, \bibnamefont{and}
  \bibinfo{author}{\bibfnamefont{C.~M.} \bibnamefont{Hadad}},
  \bibinfo{journal}{J. Phys. Chem. A} \textbf{\bibinfo{volume}{109}},
  \bibinfo{pages}{1308} (\bibinfo{year}{2005}).

\bibitem[{\citenamefont{Glover and Miller}(2005)}]{Glover:JPCA109:11191}
\bibinfo{author}{\bibfnamefont{B.~G.} \bibnamefont{Glover}} \bibnamefont{and}
  \bibinfo{author}{\bibfnamefont{T.~A.} \bibnamefont{Miller}},
  \bibinfo{journal}{J. Phys. Chem. A} \textbf{\bibinfo{volume}{109}},
  \bibinfo{pages}{11191} (\bibinfo{year}{2005}).

\bibitem[{\citenamefont{Tarczay et~al.}(2005)\citenamefont{Tarczay,
  Zalyubovsky, and Miller}}]{Tarczay:CPL406:81}
\bibinfo{author}{\bibfnamefont{G.}~\bibnamefont{Tarczay}},
  \bibinfo{author}{\bibfnamefont{S.~J.} \bibnamefont{Zalyubovsky}},
  \bibnamefont{and} \bibinfo{author}{\bibfnamefont{T.~A.}
  \bibnamefont{Miller}}, \bibinfo{journal}{Chem. Phys. Lett.}
  \textbf{\bibinfo{volume}{406}}, \bibinfo{pages}{81} (\bibinfo{year}{2005}).

\bibitem[{\citenamefont{Rienstra-Kiracofe
  et~al.}(2000)\citenamefont{Rienstra-Kiracofe, Allen, and
  Schaefer}}]{Rienstra-Kiracofe:JPCA104:9823}
\bibinfo{author}{\bibfnamefont{J.}~\bibnamefont{Rienstra-Kiracofe}},
  \bibinfo{author}{\bibfnamefont{W.}~\bibnamefont{Allen}}, \bibnamefont{and}
  \bibinfo{author}{\bibfnamefont{H.}~\bibnamefont{Schaefer},
  \bibfnamefont{III}}, \bibinfo{journal}{J. Phys. Chem. A}
  \textbf{\bibinfo{volume}{104}}, \bibinfo{pages}{9823} (\bibinfo{year}{2000}).

\bibitem[{\citenamefont{Fishchuk
  et~al.}(2006{\natexlab{c}})\citenamefont{Fishchuk, Merritt, and van~der
  Avoird}}]{Fishchuk:HCN-Br:inprep}
\bibinfo{author}{\bibfnamefont{A.}~\bibnamefont{Fishchuk}},
  \bibinfo{author}{\bibfnamefont{J.~M.} \bibnamefont{Merritt}},
  \bibnamefont{and} \bibinfo{author}{\bibfnamefont{A.}~\bibnamefont{van~der
  Avoird}} (\bibinfo{year}{2006}{\natexlab{c}}), \bibinfo{note}{in
  preparation}.

\bibitem[{\citenamefont{Fawzy et~al.}(1990)\citenamefont{Fawzy, Fraser, Hougen,
  and Pine}}]{Fawzy:JCP93:2992}
\bibinfo{author}{\bibfnamefont{W.~M.} \bibnamefont{Fawzy}},
  \bibinfo{author}{\bibfnamefont{G.~T.} \bibnamefont{Fraser}},
  \bibinfo{author}{\bibfnamefont{J.~T.} \bibnamefont{Hougen}},
  \bibnamefont{and} \bibinfo{author}{\bibfnamefont{A.~S.} \bibnamefont{Pine}},
  \bibinfo{journal}{J. Chem. Phys.} \textbf{\bibinfo{volume}{93}},
  \bibinfo{pages}{2992} (\bibinfo{year}{1990}).

\bibitem[{\citenamefont{Western}()}]{Western:pgopher}
\bibinfo{author}{\bibfnamefont{C.~M.} \bibnamefont{Western}},
  \emph{\bibinfo{title}{Pgopher, a program for simulating rotational
  structure}}, \bibinfo{note}{{U}niversity of {B}ristol, {B}ristol, {GB}}.

\bibitem[{\citenamefont{Goyal et~al.}(1992)\citenamefont{Goyal, Schutt, and
  Scoles}}]{Goyal:PRL69:933}
\bibinfo{author}{\bibfnamefont{S.}~\bibnamefont{Goyal}},
  \bibinfo{author}{\bibfnamefont{D.~L.} \bibnamefont{Schutt}},
  \bibnamefont{and} \bibinfo{author}{\bibfnamefont{G.}~\bibnamefont{Scoles}},
  \bibinfo{journal}{Phys. Rev. Lett.} \textbf{\bibinfo{volume}{69}},
  \bibinfo{pages}{933} (\bibinfo{year}{1992}), \bibinfo{note}{see erratum
  \cite{Goyal:PRL73:2512}}.

\bibitem[{\citenamefont{Nauta and
  Miller}(2001{\natexlab{b}})}]{Nauta:JCP115:10254}
\bibinfo{author}{\bibfnamefont{K.}~\bibnamefont{Nauta}} \bibnamefont{and}
  \bibinfo{author}{\bibfnamefont{R.~E.} \bibnamefont{Miller}},
  \bibinfo{journal}{J. Chem. Phys.} \textbf{\bibinfo{volume}{115}},
  \bibinfo{pages}{10254} (\bibinfo{year}{2001}{\natexlab{b}}).

\bibitem[{\citenamefont{Lee et~al.}(1999)\citenamefont{Lee, Farrelly, and
  Whaley}}]{Lee:PRL83:3812}
\bibinfo{author}{\bibfnamefont{E.}~\bibnamefont{Lee}},
  \bibinfo{author}{\bibfnamefont{D.}~\bibnamefont{Farrelly}}, \bibnamefont{and}
  \bibinfo{author}{\bibfnamefont{K.~B.} \bibnamefont{Whaley}},
  \bibinfo{journal}{Phys. Rev. Lett.} \textbf{\bibinfo{volume}{83}},
  \bibinfo{pages}{3812} (\bibinfo{year}{1999}).

\bibitem[{\citenamefont{Viel and Whaley}(2001)}]{Viel:JCP115:10186}
\bibinfo{author}{\bibfnamefont{A.}~\bibnamefont{Viel}} \bibnamefont{and}
  \bibinfo{author}{\bibfnamefont{K.~B.} \bibnamefont{Whaley}},
  \bibinfo{journal}{J. Chem. Phys.} \textbf{\bibinfo{volume}{115}},
  \bibinfo{pages}{10186} (\bibinfo{year}{2001}).

\bibitem[{\citenamefont{Blinov et~al.}(2004)\citenamefont{Blinov, Song, and
  Roy}}]{Blinov:JCP120:5916}
\bibinfo{author}{\bibfnamefont{N.}~\bibnamefont{Blinov}},
  \bibinfo{author}{\bibfnamefont{X.~G.} \bibnamefont{Song}}, \bibnamefont{and}
  \bibinfo{author}{\bibfnamefont{P.~N.} \bibnamefont{Roy}},
  \bibinfo{journal}{J. Chem. Phys.} \textbf{\bibinfo{volume}{120}},
  \bibinfo{pages}{5916} (\bibinfo{year}{2004}).

\bibitem[{\citenamefont{Jucks and Miller}(1988)}]{Jucks:JCP88:2196}
\bibinfo{author}{\bibfnamefont{K.~W.} \bibnamefont{Jucks}} \bibnamefont{and}
  \bibinfo{author}{\bibfnamefont{R.~E.} \bibnamefont{Miller}},
  \bibinfo{journal}{J. Chem. Phys.} \textbf{\bibinfo{volume}{88}},
  \bibinfo{pages}{2196} (\bibinfo{year}{1988}).

\bibitem[{\citenamefont{Anex et~al.}(1988)\citenamefont{Anex, Davidson,
  Douketis, and Ewing}}]{Anex:JPC92:2913}
\bibinfo{author}{\bibfnamefont{D.~S.} \bibnamefont{Anex}},
  \bibinfo{author}{\bibfnamefont{E.~R.} \bibnamefont{Davidson}},
  \bibinfo{author}{\bibfnamefont{C.}~\bibnamefont{Douketis}}, \bibnamefont{and}
  \bibinfo{author}{\bibfnamefont{G.~E.} \bibnamefont{Ewing}},
  \bibinfo{journal}{J. Phys. Chem.} \textbf{\bibinfo{volume}{92}},
  \bibinfo{pages}{2913} (\bibinfo{year}{1988}).

\bibitem[{\citenamefont{Quack and Suhm}(1991)}]{Quack:JCP95:28}
\bibinfo{author}{\bibfnamefont{M.}~\bibnamefont{Quack}} \bibnamefont{and}
  \bibinfo{author}{\bibfnamefont{M.~A.} \bibnamefont{Suhm}},
  \bibinfo{journal}{J. Chem. Phys.} \textbf{\bibinfo{volume}{95}},
  \bibinfo{pages}{28} (\bibinfo{year}{1991}).

\bibitem[{\citenamefont{Howard et~al.}(1984)\citenamefont{Howard, Dyke, and
  Klemperer}}]{Howard:JCP81:5417}
\bibinfo{author}{\bibfnamefont{B.~J.} \bibnamefont{Howard}},
  \bibinfo{author}{\bibfnamefont{T.~R.} \bibnamefont{Dyke}}, \bibnamefont{and}
  \bibinfo{author}{\bibfnamefont{W.}~\bibnamefont{Klemperer}},
  \bibinfo{journal}{J. Chem. Phys.} \textbf{\bibinfo{volume}{81}},
  \bibinfo{pages}{5417} (\bibinfo{year}{1984}).

\bibitem[{\citenamefont{Kuhn et~al.}(1999)\citenamefont{Kuhn, Rizzo, Luckhaus,
  Quack, and Suhm}}]{Kuhn:JCP111:2565}
\bibinfo{author}{\bibfnamefont{B.}~\bibnamefont{Kuhn}},
  \bibinfo{author}{\bibfnamefont{T.~R.} \bibnamefont{Rizzo}},
  \bibinfo{author}{\bibfnamefont{D.}~\bibnamefont{Luckhaus}},
  \bibinfo{author}{\bibfnamefont{M.}~\bibnamefont{Quack}}, \bibnamefont{and}
  \bibinfo{author}{\bibfnamefont{M.~A.} \bibnamefont{Suhm}},
  \bibinfo{journal}{J. Chem. Phys.} \textbf{\bibinfo{volume}{111}},
  \bibinfo{pages}{2565} (\bibinfo{year}{1999}).

\bibitem[{\citenamefont{Douberly and Miller}(2003)}]{Douberly:JPCB107:4500}
\bibinfo{author}{\bibfnamefont{G.~E.} \bibnamefont{Douberly}} \bibnamefont{and}
  \bibinfo{author}{\bibfnamefont{R.~E.} \bibnamefont{Miller}},
  \bibinfo{journal}{J. Phys. Chem. B} \textbf{\bibinfo{volume}{107}},
  \bibinfo{pages}{4500} (\bibinfo{year}{2003}).

\bibitem[{\citenamefont{Blume et~al.}(1996)\citenamefont{Blume, Lewerenz,
  Huisken, and Kaloudis}}]{Blume:JCP105:8666}
\bibinfo{author}{\bibfnamefont{D.}~\bibnamefont{Blume}},
  \bibinfo{author}{\bibfnamefont{M.}~\bibnamefont{Lewerenz}},
  \bibinfo{author}{\bibfnamefont{F.}~\bibnamefont{Huisken}}, \bibnamefont{and}
  \bibinfo{author}{\bibfnamefont{M.}~\bibnamefont{Kaloudis}},
  \bibinfo{journal}{J. Chem. Phys.} \textbf{\bibinfo{volume}{105}},
  \bibinfo{pages}{8666} (\bibinfo{year}{1996}).

\bibitem[{\citenamefont{Huisken
  et~al.}(1995{\natexlab{a}})\citenamefont{Huisken, Tarakanova, Vigasin, and
  Yukhnevich}}]{Huisken:CPL245:319}
\bibinfo{author}{\bibfnamefont{F.}~\bibnamefont{Huisken}},
  \bibinfo{author}{\bibfnamefont{E.~G.} \bibnamefont{Tarakanova}},
  \bibinfo{author}{\bibfnamefont{A.~A.} \bibnamefont{Vigasin}},
  \bibnamefont{and} \bibinfo{author}{\bibfnamefont{G.~V.}
  \bibnamefont{Yukhnevich}}, \bibinfo{journal}{Chem. Phys. Lett.}
  \textbf{\bibinfo{volume}{245}}, \bibinfo{pages}{319}
  (\bibinfo{year}{1995}{\natexlab{a}}).

\bibitem[{\citenamefont{Huisken
  et~al.}(1995{\natexlab{b}})\citenamefont{Huisken, Kaloudis, Kulcke, Laush,
  and Lisy}}]{Huisken:JCP103:5366}
\bibinfo{author}{\bibfnamefont{F.}~\bibnamefont{Huisken}},
  \bibinfo{author}{\bibfnamefont{M.}~\bibnamefont{Kaloudis}},
  \bibinfo{author}{\bibfnamefont{A.}~\bibnamefont{Kulcke}},
  \bibinfo{author}{\bibfnamefont{C.}~\bibnamefont{Laush}}, \bibnamefont{and}
  \bibinfo{author}{\bibfnamefont{J.~M.} \bibnamefont{Lisy}},
  \bibinfo{journal}{J. Chem. Phys.} \textbf{\bibinfo{volume}{103}},
  \bibinfo{pages}{5366} (\bibinfo{year}{1995}{\natexlab{b}}).

\bibitem[{\citenamefont{Moore and
  Miller}(2004{\natexlab{b}})}]{Moore:JPCA108:1930}
\bibinfo{author}{\bibfnamefont{D.~T.} \bibnamefont{Moore}} \bibnamefont{and}
  \bibinfo{author}{\bibfnamefont{R.~E.} \bibnamefont{Miller}},
  \bibinfo{journal}{J. Phys. Chem. A} \textbf{\bibinfo{volume}{108}},
  \bibinfo{pages}{1930} (\bibinfo{year}{2004}{\natexlab{b}}).

\bibitem[{\citenamefont{Moore and
  Miller}(2003{\natexlab{a}})}]{Moore:JCP118:9629}
\bibinfo{author}{\bibfnamefont{D.~T.} \bibnamefont{Moore}} \bibnamefont{and}
  \bibinfo{author}{\bibfnamefont{R.~E.} \bibnamefont{Miller}},
  \bibinfo{journal}{J. Chem. Phys.} \textbf{\bibinfo{volume}{118}},
  \bibinfo{pages}{9629} (\bibinfo{year}{2003}{\natexlab{a}}).

\bibitem[{\citenamefont{Moore and
  Miller}(2003{\natexlab{b}})}]{Moore:JCP119:4713}
\bibinfo{author}{\bibfnamefont{D.~T.} \bibnamefont{Moore}} \bibnamefont{and}
  \bibinfo{author}{\bibfnamefont{R.~E.} \bibnamefont{Miller}},
  \bibinfo{journal}{J. Chem. Phys.} \textbf{\bibinfo{volume}{119}},
  \bibinfo{pages}{4713} (\bibinfo{year}{2003}{\natexlab{b}}).

\bibitem[{\citenamefont{Moore and
  Miller}(2003{\natexlab{c}})}]{Moore:JPCA107:10805}
\bibinfo{author}{\bibfnamefont{D.~T.} \bibnamefont{Moore}} \bibnamefont{and}
  \bibinfo{author}{\bibfnamefont{R.~E.} \bibnamefont{Miller}},
  \bibinfo{journal}{J. Phys. Chem. A} \textbf{\bibinfo{volume}{107}},
  \bibinfo{pages}{10805} (\bibinfo{year}{2003}{\natexlab{c}}).

\bibitem[{\citenamefont{Lide}(1990)}]{CRC:HandbookChemPhys71}
\bibinfo{editor}{\bibfnamefont{D.~R.} \bibnamefont{Lide}}, ed.,
  \emph{\bibinfo{title}{CRC Handbook of Chemistry and Physics}},
  vol.~\bibinfo{volume}{71} (\bibinfo{publisher}{CRC Press},
  \bibinfo{address}{Boca Raton}, \bibinfo{year}{1990}).

\bibitem[{\citenamefont{Strazisar et~al.}(2000)\citenamefont{Strazisar, Lin,
  and Davis}}]{Strazisar:Science290:958}
\bibinfo{author}{\bibfnamefont{B.~R.} \bibnamefont{Strazisar}},
  \bibinfo{author}{\bibfnamefont{C.}~\bibnamefont{Lin}}, \bibnamefont{and}
  \bibinfo{author}{\bibfnamefont{H.~F.} \bibnamefont{Davis}},
  \bibinfo{journal}{Science} \textbf{\bibinfo{volume}{290}},
  \bibinfo{pages}{958} (\bibinfo{year}{2000}).

\bibitem[{\citenamefont{Smith and Crim}(2002)}]{Smith:PCCP4:3543}
\bibinfo{author}{\bibfnamefont{I.~W.~M.} \bibnamefont{Smith}} \bibnamefont{and}
  \bibinfo{author}{\bibfnamefont{F.~F.} \bibnamefont{Crim}},
  \bibinfo{journal}{Phys. Chem. Chem. Phys.} \textbf{\bibinfo{volume}{4}},
  \bibinfo{pages}{3543} (\bibinfo{year}{2002}).

\bibitem[{\citenamefont{Zhang et~al.}(2005)\citenamefont{Zhang, Morokuma, and
  Wodtke}}]{Zhang:JCP122:014106}
\bibinfo{author}{\bibfnamefont{P.}~\bibnamefont{Zhang}},
  \bibinfo{author}{\bibfnamefont{K.}~\bibnamefont{Morokuma}}, \bibnamefont{and}
  \bibinfo{author}{\bibfnamefont{A.~M.} \bibnamefont{Wodtke}},
  \bibinfo{journal}{J. Chem. Phys.} \textbf{\bibinfo{volume}{122}},
  \bibinfo{pages}{014106} (\bibinfo{year}{2005}).

\bibitem[{\citenamefont{Bittererova et~al.}(2001)\citenamefont{Bittererova,
  Ostmark, and Brinck}}]{Bittererova:CPL347:220}
\bibinfo{author}{\bibfnamefont{M.}~\bibnamefont{Bittererova}},
  \bibinfo{author}{\bibfnamefont{H.}~\bibnamefont{Ostmark}}, \bibnamefont{and}
  \bibinfo{author}{\bibfnamefont{T.}~\bibnamefont{Brinck}},
  \bibinfo{journal}{Chem. Phys. Lett.} \textbf{\bibinfo{volume}{347}},
  \bibinfo{pages}{220} (\bibinfo{year}{2001}).

\bibitem[{\citenamefont{Bittererova et~al.}(2002)\citenamefont{Bittererova,
  Ostmark, and Brinck}}]{Bittererova:JCP116:9740}
\bibinfo{author}{\bibfnamefont{M.}~\bibnamefont{Bittererova}},
  \bibinfo{author}{\bibfnamefont{H.}~\bibnamefont{Ostmark}}, \bibnamefont{and}
  \bibinfo{author}{\bibfnamefont{T.}~\bibnamefont{Brinck}},
  \bibinfo{journal}{J. Chem. Phys.} \textbf{\bibinfo{volume}{116}},
  \bibinfo{pages}{9740} (\bibinfo{year}{2002}).

\bibitem[{\citenamefont{Hansen and Wodtke}(2003)}]{Hansen:JPCA107_10608}
\bibinfo{author}{\bibfnamefont{N.}~\bibnamefont{Hansen}} \bibnamefont{and}
  \bibinfo{author}{\bibfnamefont{A.~M.} \bibnamefont{Wodtke}},
  \bibinfo{journal}{J. Phys. Chem. A} \textbf{\bibinfo{volume}{107}},
  \bibinfo{pages}{10608} (\bibinfo{year}{2003}).

\bibitem[{\citenamefont{Cacace et~al.}(2002)\citenamefont{Cacace, de~Petris,
  and Troiani}}]{Cacace:Science295:480}
\bibinfo{author}{\bibfnamefont{F.}~\bibnamefont{Cacace}},
  \bibinfo{author}{\bibfnamefont{G.}~\bibnamefont{de~Petris}},
  \bibnamefont{and} \bibinfo{author}{\bibfnamefont{A.}~\bibnamefont{Troiani}},
  \bibinfo{journal}{Science} \textbf{\bibinfo{volume}{295}},
  \bibinfo{pages}{480} (\bibinfo{year}{2002}).

\bibitem[{\citenamefont{Gordon et~al.}(1978)\citenamefont{Gordon,
  Mezhov-Deglin, Pugachev, and Khmelenko}}]{Gordon:CPL54:282}
\bibinfo{author}{\bibfnamefont{E.~B.} \bibnamefont{Gordon}},
  \bibinfo{author}{\bibfnamefont{L.~P.} \bibnamefont{Mezhov-Deglin}},
  \bibinfo{author}{\bibfnamefont{O.~F.} \bibnamefont{Pugachev}},
  \bibnamefont{and} \bibinfo{author}{\bibfnamefont{V.~V.}
  \bibnamefont{Khmelenko}}, \bibinfo{journal}{Chem. Phys. Lett.}
  \textbf{\bibinfo{volume}{54}}, \bibinfo{pages}{282} (\bibinfo{year}{1978}).

\bibitem[{\citenamefont{Gordon et~al.}(1993)\citenamefont{Gordon, Khmelenko,
  Pelmenev, Popov, Pugachev, and Shestakov}}]{Gordon:CP170:411}
\bibinfo{author}{\bibfnamefont{E.~B.} \bibnamefont{Gordon}},
  \bibinfo{author}{\bibfnamefont{V.~V.} \bibnamefont{Khmelenko}},
  \bibinfo{author}{\bibfnamefont{A.~A.} \bibnamefont{Pelmenev}},
  \bibinfo{author}{\bibfnamefont{E.~A.} \bibnamefont{Popov}},
  \bibinfo{author}{\bibfnamefont{O.~F.} \bibnamefont{Pugachev}},
  \bibnamefont{and} \bibinfo{author}{\bibfnamefont{A.~F.}
  \bibnamefont{Shestakov}}, \bibinfo{journal}{Chem. Phys.}
  \textbf{\bibinfo{volume}{170}}, \bibinfo{pages}{411} (\bibinfo{year}{1993}).

\bibitem[{\citenamefont{Boltnev et~al.}(1994)\citenamefont{Boltnev, Gordon,
  Khmelenko, Krushinskaya, Martynenko, Pelmenev, Popov, and
  Shestakov}}]{Boltnev:CP189:367}
\bibinfo{author}{\bibfnamefont{R.~E.} \bibnamefont{Boltnev}},
  \bibinfo{author}{\bibfnamefont{E.~B.} \bibnamefont{Gordon}},
  \bibinfo{author}{\bibfnamefont{V.~V.} \bibnamefont{Khmelenko}},
  \bibinfo{author}{\bibfnamefont{I.~N.} \bibnamefont{Krushinskaya}},
  \bibinfo{author}{\bibfnamefont{M.~V.} \bibnamefont{Martynenko}},
  \bibinfo{author}{\bibfnamefont{A.~A.} \bibnamefont{Pelmenev}},
  \bibinfo{author}{\bibfnamefont{E.~A.} \bibnamefont{Popov}}, \bibnamefont{and}
  \bibinfo{author}{\bibfnamefont{A.~F.} \bibnamefont{Shestakov}},
  \bibinfo{journal}{Chem. Phys.} \textbf{\bibinfo{volume}{189}},
  \bibinfo{pages}{367} (\bibinfo{year}{1994}).

\bibitem[{\citenamefont{Gordon et~al.}(1989)\citenamefont{Gordon, Khmelenko,
  Pelmenev, Popov, and Pugachev}}]{Gordon:CPL155:301}
\bibinfo{author}{\bibfnamefont{E.~B.} \bibnamefont{Gordon}},
  \bibinfo{author}{\bibfnamefont{V.~V.} \bibnamefont{Khmelenko}},
  \bibinfo{author}{\bibfnamefont{A.~A.} \bibnamefont{Pelmenev}},
  \bibinfo{author}{\bibfnamefont{E.~A.} \bibnamefont{Popov}}, \bibnamefont{and}
  \bibinfo{author}{\bibfnamefont{O.~F.} \bibnamefont{Pugachev}},
  \bibinfo{journal}{Chem. Phys. Lett.} \textbf{\bibinfo{volume}{155}},
  \bibinfo{pages}{301} (\bibinfo{year}{1989}).

\bibitem[{\citenamefont{Gordon}(2000)}]{Gordon:Plunger:2000}
\bibinfo{author}{\bibfnamefont{E.~B.} \bibnamefont{Gordon}},
  \bibinfo{howpublished}{private communication with R.\,E.\ Miller}
  (\bibinfo{year}{2000}).

\bibitem[{\citenamefont{Boltnev et~al.}(2000)}]{Boltnev:OSU2000:WI13}
\bibinfo{author}{\bibfnamefont{R.~E.} \bibnamefont{Boltnev}}
  \bibnamefont{et~al.}, in \emph{\bibinfo{booktitle}{55th International
  Symposium on Molecular Spectroscopy}} (\bibinfo{publisher}{Ohio State
  University}, \bibinfo{address}{Columbus, OH, USA}, \bibinfo{year}{2000}),
  \bibinfo{number}{WI13}, p. \bibinfo{pages}{183}.

\bibitem[{\citenamefont{Popov et~al.}(2000)}]{Popov:OSU2000:WI14}
\bibinfo{author}{\bibfnamefont{E.~A.} \bibnamefont{Popov}}
  \bibnamefont{et~al.}, in \emph{\bibinfo{booktitle}{55th International
  Symposium on Molecular Spectroscopy}} (\bibinfo{publisher}{Ohio State
  University}, \bibinfo{address}{Columbus, OH, USA}, \bibinfo{year}{2000}),
  \bibinfo{number}{WI14}, p. \bibinfo{pages}{183}.

\bibitem[{\citenamefont{Gordon}(2004)}]{Gordon:LTP30:756}
\bibinfo{author}{\bibfnamefont{E.~B.} \bibnamefont{Gordon}},
  \bibinfo{journal}{Low. Temp. Phys.} \textbf{\bibinfo{volume}{30}},
  \bibinfo{pages}{756} (\bibinfo{year}{2004}).

\bibitem[{\citenamefont{Etters et~al.}(1975)\citenamefont{Etters, John
  V.~Dugan, and Palmer}}]{Etters:JCP62:313}
\bibinfo{author}{\bibfnamefont{R.~D.} \bibnamefont{Etters}},
  \bibinfo{author}{\bibfnamefont{J.}~\bibnamefont{John V.~Dugan}},
  \bibnamefont{and} \bibinfo{author}{\bibfnamefont{R.~W.}
  \bibnamefont{Palmer}}, \bibinfo{journal}{The Journal of Chemical Physics}
  \textbf{\bibinfo{volume}{62}}, \bibinfo{pages}{313} (\bibinfo{year}{1975}).

\bibitem[{\citenamefont{Silvera and Walraven}(1980)}]{Silvera:PRL44:164}
\bibinfo{author}{\bibfnamefont{I.~F.} \bibnamefont{Silvera}} \bibnamefont{and}
  \bibinfo{author}{\bibfnamefont{J.~T.~M.} \bibnamefont{Walraven}},
  \bibinfo{journal}{Phys. Rev. Lett.} \textbf{\bibinfo{volume}{44}},
  \bibinfo{pages}{164} (\bibinfo{year}{1980}).

\bibitem[{\citenamefont{Meyer et~al.}(1994)\citenamefont{Meyer, Zhao, Mester,
  and Silvera}}]{Meyer:PRB50:9339}
\bibinfo{author}{\bibfnamefont{E.~S.} \bibnamefont{Meyer}},
  \bibinfo{author}{\bibfnamefont{Z.}~\bibnamefont{Zhao}},
  \bibinfo{author}{\bibfnamefont{J.~C.} \bibnamefont{Mester}},
  \bibnamefont{and} \bibinfo{author}{\bibfnamefont{I.~F.}
  \bibnamefont{Silvera}}, \bibinfo{journal}{Phys. Rev. B}
  \textbf{\bibinfo{volume}{50}}, \bibinfo{pages}{9339} (\bibinfo{year}{1994}).

\bibitem[{\citenamefont{Cornell and Wieman}(2002)}]{Cornell:RMP74:875}
\bibinfo{author}{\bibfnamefont{E.~A.} \bibnamefont{Cornell}} \bibnamefont{and}
  \bibinfo{author}{\bibfnamefont{C.~E.} \bibnamefont{Wieman}},
  \bibinfo{journal}{Rev. Mod. Phys.} \textbf{\bibinfo{volume}{74}},
  \bibinfo{pages}{875} (\bibinfo{year}{2002}).

\bibitem[{\citenamefont{Ketterle}(2002)}]{Ketterle:RMP74:1131}
\bibinfo{author}{\bibfnamefont{W.}~\bibnamefont{Ketterle}},
  \bibinfo{journal}{Rev. Mod. Phys.} \textbf{\bibinfo{volume}{74}},
  \bibinfo{pages}{1131} (\bibinfo{year}{2002}).

\bibitem[{\citenamefont{Nieto et~al.}(2003)\citenamefont{Nieto, Holzscheiter,
  and Phillips}}]{Nieto:JOB5:S547}
\bibinfo{author}{\bibfnamefont{M.~M.} \bibnamefont{Nieto}},
  \bibinfo{author}{\bibfnamefont{M.~H.} \bibnamefont{Holzscheiter}},
  \bibnamefont{and} \bibinfo{author}{\bibfnamefont{T.~J.}
  \bibnamefont{Phillips}}, \bibinfo{journal}{J. Opt. B}
  \textbf{\bibinfo{volume}{5}}, \bibinfo{pages}{S547} (\bibinfo{year}{2003}).

\bibitem[{\citenamefont{Kozlov and
  Derevianko}(2006)}]{Kozlov:arXiv:physics/0602111}
\bibinfo{author}{\bibfnamefont{M.~G.} \bibnamefont{Kozlov}} \bibnamefont{and}
  \bibinfo{author}{\bibfnamefont{A.}~\bibnamefont{Derevianko}},
  \bibinfo{journal}{arXiv} \textbf{\bibinfo{volume}{physics}},
  \bibinfo{pages}{0602111} (\bibinfo{year}{2006}), \bibinfo{note}{accepted for
  \emph{Phys.\ Rev.\ Lett.}}

\bibitem[{\citenamefont{van~de Meerakker et~al.}(2005)\citenamefont{van~de
  Meerakker, Smeets, Vanhaecke, Jongma, and Meijer}}]{Meerakker:PRL94:023004}
\bibinfo{author}{\bibfnamefont{S.~Y.~T.} \bibnamefont{van~de Meerakker}},
  \bibinfo{author}{\bibfnamefont{P.~H.~M.} \bibnamefont{Smeets}},
  \bibinfo{author}{\bibfnamefont{N.}~\bibnamefont{Vanhaecke}},
  \bibinfo{author}{\bibfnamefont{R.~T.} \bibnamefont{Jongma}},
  \bibnamefont{and} \bibinfo{author}{\bibfnamefont{G.}~\bibnamefont{Meijer}},
  \bibinfo{journal}{Phys. Rev. Lett.} \textbf{\bibinfo{volume}{94}},
  \bibinfo{pages}{023004} (\bibinfo{year}{2005}).

\bibitem[{\citenamefont{Avdeenkov and Bohn}(2003)}]{Avdeenkov:PRL90:043006}
\bibinfo{author}{\bibfnamefont{A.~V.} \bibnamefont{Avdeenkov}}
  \bibnamefont{and} \bibinfo{author}{\bibfnamefont{J.~L.} \bibnamefont{Bohn}},
  \bibinfo{journal}{Phys. Rev. Lett.} \textbf{\bibinfo{volume}{90}},
  \bibinfo{pages}{043006} (\bibinfo{year}{2003}).

\bibitem[{\citenamefont{Avdeenkov and Bohn}(2002)}]{Avdeenkov:PRA66:052718}
\bibinfo{author}{\bibfnamefont{A.~V.} \bibnamefont{Avdeenkov}}
  \bibnamefont{and} \bibinfo{author}{\bibfnamefont{J.~L.} \bibnamefont{Bohn}},
  \bibinfo{journal}{Phys. Rev. A} \textbf{\bibinfo{volume}{66}},
  \bibinfo{pages}{052718} (\bibinfo{year}{2002}).

\bibitem[{\citenamefont{Stienkemeier et~al.}(1999)\citenamefont{Stienkemeier,
  Meier, Hagele, Lutz, Schreiber, Schulz, and
  Hertel}}]{Stienkemeier:PRL83:2320}
\bibinfo{author}{\bibfnamefont{F.}~\bibnamefont{Stienkemeier}},
  \bibinfo{author}{\bibfnamefont{F.}~\bibnamefont{Meier}},
  \bibinfo{author}{\bibfnamefont{A.}~\bibnamefont{Hagele}},
  \bibinfo{author}{\bibfnamefont{H.~O.} \bibnamefont{Lutz}},
  \bibinfo{author}{\bibfnamefont{E.}~\bibnamefont{Schreiber}},
  \bibinfo{author}{\bibfnamefont{C.~P.} \bibnamefont{Schulz}},
  \bibnamefont{and} \bibinfo{author}{\bibfnamefont{I.~V.}
  \bibnamefont{Hertel}}, \bibinfo{journal}{Phys. Rev. Lett.}
  \textbf{\bibinfo{volume}{83}}, \bibinfo{pages}{2320} (\bibinfo{year}{1999}).

\bibitem[{\citenamefont{Schulz et~al.}(2001)\citenamefont{Schulz, Claas, and
  Stienkemeier}}]{Schulz:PRL87:153401}
\bibinfo{author}{\bibfnamefont{C.~P.} \bibnamefont{Schulz}},
  \bibinfo{author}{\bibfnamefont{P.}~\bibnamefont{Claas}}, \bibnamefont{and}
  \bibinfo{author}{\bibfnamefont{F.}~\bibnamefont{Stienkemeier}},
  \bibinfo{journal}{Phys. Rev. Lett.} \textbf{\bibinfo{volume}{8715}},
  \bibinfo{pages}{153401} (\bibinfo{year}{2001}).

\bibitem[{\citenamefont{Droppelmann et~al.}(2004)\citenamefont{Droppelmann,
  Bunermann, Schulz, and Stienkemeier}}]{Droppelmann:PRL93:023402}
\bibinfo{author}{\bibfnamefont{G.}~\bibnamefont{Droppelmann}},
  \bibinfo{author}{\bibfnamefont{O.}~\bibnamefont{Bunermann}},
  \bibinfo{author}{\bibfnamefont{C.~P.} \bibnamefont{Schulz}},
  \bibnamefont{and}
  \bibinfo{author}{\bibfnamefont{F.}~\bibnamefont{Stienkemeier}},
  \bibinfo{journal}{Phys. Rev. Lett.} \textbf{\bibinfo{volume}{93}},
  \bibinfo{pages}{023402} (\bibinfo{year}{2004}).

\bibitem[{\citenamefont{Scoles}(1988)}]{_Scoles:MolBeam:1}
\bibinfo{editor}{\bibfnamefont{G.}~\bibnamefont{Scoles}}, ed.,
  \emph{\bibinfo{title}{Atomic and molecular beam methods}},
  vol.~\bibinfo{volume}{1} (\bibinfo{publisher}{Oxford University Press},
  \bibinfo{address}{New York, NY, USA}, \bibinfo{year}{1988}).

\bibitem[{\citenamefont{Zeimen et~al.}(2004)\citenamefont{Zeimen, K{\l}os,
  Groenenboom, and van~der Avoird}}]{Zeimen:JPCA108:9319}
\bibinfo{author}{\bibfnamefont{W.~B.} \bibnamefont{Zeimen}},
  \bibinfo{author}{\bibfnamefont{J.~A.} \bibnamefont{K{\l}os}},
  \bibinfo{author}{\bibfnamefont{G.~C.} \bibnamefont{Groenenboom}},
  \bibnamefont{and} \bibinfo{author}{\bibfnamefont{A.}~\bibnamefont{van~der
  Avoird}}, \bibinfo{journal}{J. Phys. Chem. A} \textbf{\bibinfo{volume}{108}},
  \bibinfo{pages}{9319} (\bibinfo{year}{2004}).

\bibitem[{\citenamefont{Goyal et~al.}(2004)\citenamefont{Goyal, Schutt, and
  Scoles}}]{Goyal:PRL73:2512}
\bibinfo{author}{\bibfnamefont{S.}~\bibnamefont{Goyal}},
  \bibinfo{author}{\bibfnamefont{D.~L.} \bibnamefont{Schutt}},
  \bibnamefont{and} \bibinfo{author}{\bibfnamefont{G.}~\bibnamefont{Scoles}},
  \bibinfo{journal}{Phys. Rev. Lett.} \textbf{\bibinfo{volume}{73}},
  \bibinfo{pages}{2512} (\bibinfo{year}{2004}).

\bibitem[{\citenamefont{C\^ot\'e et~al.}(2002)\citenamefont{C\^ot\'e,
  Kharchenko, and Lukin}}]{Cote:PRL89:093001}
\bibinfo{author}{\bibfnamefont{R.}~\bibnamefont{C\^ot\'e}},
  \bibinfo{author}{\bibfnamefont{V.}~\bibnamefont{Kharchenko}},
  \bibnamefont{and} \bibinfo{author}{\bibfnamefont{M.~D.} \bibnamefont{Lukin}},
  \bibinfo{journal}{Phys. Rev. Lett.} \textbf{\bibinfo{volume}{89}},
  \bibinfo{pages}{093001} (\bibinfo{year}{2002}).

\bibitem[{\citenamefont{Mudrich et~al.}(2004)\citenamefont{Mudrich, Bunermann,
  Stienkemeier, Dulieu, and Weidem\"uller}}]{Mudrich:EPJD31:291}
\bibinfo{author}{\bibfnamefont{N.}~\bibnamefont{Mudrich}},
  \bibinfo{author}{\bibfnamefont{O.}~\bibnamefont{Bunermann}},
  \bibinfo{author}{\bibfnamefont{F.}~\bibnamefont{Stienkemeier}},
  \bibinfo{author}{\bibfnamefont{O.}~\bibnamefont{Dulieu}}, \bibnamefont{and}
  \bibinfo{author}{\bibfnamefont{M.}~\bibnamefont{Weidem\"uller}},
  \bibinfo{journal}{Eur. Phys. J. D} \textbf{\bibinfo{volume}{31}},
  \bibinfo{pages}{291} (\bibinfo{year}{2004}).

\end{thebibliography}
\bibliographystyle{apsrev-nourl}%



\end{document}